\documentclass[11pt, 3p,fleqn,times]{elsarticle}
\usepackage[british]{babel}
\usepackage[utf8]{inputenc}
\usepackage{graphics}
\usepackage{graphicx}
\usepackage{amsmath,amssymb,esint}
\usepackage{lineno}
\usepackage{multirow}
\usepackage{subfig}
\usepackage{natbib}
\usepackage{eucal}
\usepackage{mathtools}
\usepackage{placeins}
\usepackage{enumerate}
\usepackage[usenames,dvipsnames]{xcolor}
\usepackage[colorlinks]{hyperref}
\usepackage{nicefrac}
\usepackage{setspace}
\usepackage{tikz}
\usepackage{wasysym}
\usepackage{booktabs}
\usepackage{adjustbox}
\usepackage{array}
\usepackage{soul}
\DeclareFontEncoding{LS1}{}{}
\DeclareFontSubstitution{LS1}{stix}{m}{n}
\DeclareMathAlphabet{\mathscr}{LS1}{stixscr}{m}{n}
\SetMathAlphabet{\mathscr}{bold}{LS1}{stixscr}{b}{n}
\let\oldemph\emph
\renewcommand{\vec}[1]{\textbf{\oldemph{#1}}}

\usetikzlibrary{calc}
\usetikzlibrary{shapes.geometric, arrows}
\usetikzlibrary{arrows.meta}
\tikzset{>={Latex[width=1mm,length=1mm]}}

\tikzstyle{process} = [rectangle, minimum width=2.5cm, minimum height=1cm, text
centered, text width = 3cm, draw=black]
\tikzstyle{decision} = [diamond, minimum width=2.5cm, minimum height=1cm,
aspect=2,inner sep=-0.5ex,text centered, text width = 2.5cm, draw=black]
\tikzstyle{arrow} = [thick,->,>=stealth]
\tikzstyle{line} = [draw, -latex']

\tikzset{
    cross/.pic = {
    \draw[rotate = 45] (-#1,0) -- (#1,0);
    \draw[rotate = 45] (0,-#1) -- (0, #1);
    }
}

\biboptions{sort&compress}
\graphicspath{{./}{Figures/}}

\makeatletter
\def\ps@pprintTitle{%
    \let\@oddhead\@empty
    \let\@evenhead\@empty
    \def\@oddfoot{\footnotesize\itshape
         {} \hfill {}}%
    \let\@evenfoot\@oddfoot
    }
\makeatother

\newcommand{\vect}[1]{\boldsymbol{#1}}

\newcommand{\tens}[1]{\overline{\overline{#1}}}

\newcommand\Rep{$Re_\textup{p}$}
\newcommand\cd{$C_\textup{D}$}
\newcommand\cl{$C_\textup{L}$}
\newcommand\ct{$C_\textup{T}$}

\newcommand*{\addheight}[2][.5ex]{%
  \raisebox{0pt}[\dimexpr\height+(#1)\relax]{#2}%
}

\newcommand{\reviewerII}[1]{{#1}}

\begin{document}

\begin{frontmatter}

\title{Drag, lift and torque correlations for axi-symmetric {rod-like} non-spherical particles in locally {linear shear} flows}%

\author{Victor Ch\'{e}ron}
\author{Fabien Evrard\corref{cor2}}
\author{Berend van Wachem\corref{cor1}}
\cortext[cor2]{also at: Sibley School of Mechanical and Aerospace Engineering, Cornell University,\\ Ithaca, NY 14853, United States of America}
\cortext[cor1]{Corresponding author}\ead{Berend.van.Wachem@gmail.com}
\address{Lehrstuhl f\"ur Mechanische Verfahrenstechnik, Otto-von-Guericke-Universit\"at Magdeburg, \mbox{Universit\"atsplatz 2, 39106 Magdeburg, Germany}}

\begin{abstract}

This paper presents new correlations to predict the drag, lift and torque coefficients of axi-symmetric non-spherical rod-like particles for several fluid flow regimes and velocity profiles.
The fluid velocity profiles considered are locally uniform flow and locally linear shear flow.
The novel correlations for the drag, lift and torque coefficients depend on the particle Reynolds number \Rep, the orientation of the particle with respect to the main fluid direction $\theta$, the aspect ratio of the rod-like particle $\alpha$, and the dimensionless local shear rate $\tilde{G}$.
The effect of the linear shear flow on the hydrodynamic forces is modeled as an additional component for the resultant of forces acting on a particle in a locally uniform flow, hence the independent expressions for the drag, lift and torque coefficients of axi-symmetric particles in a locally uniform flow are also provided in this work.
The data provided to fit the coefficient in the new correlation are generated using available analytical expressions in the viscous regime, and performing direct numerical simulations (DNS) of the flow past the axi-symmetric particles at finite particle Reynolds number.
The DNS are performed using the direct-forcing immersed boundary method.
\reviewerII{The coefficients in the proposed drag, lift and torque correlations are determined with a high degree of accuracy, where the mean error in the prediction lies below $2\%$ for the locally uniform flow correlations, and below $1.67\%$, $5.35\%$, $6.78\%$ for the correlations accounting for the change in the drag, lift, and torque coefficients in case of a linear shear flow, respectively.}
The proposed correlations for the drag, lift and torque coefficients can be used in large-scale simulations performed in the Eulerian-Lagrangian framework with locally uniform and non-uniform flows.

\end{abstract}


\end{frontmatter}


\section{Introduction}


\reviewerII{In most large-scale numerical studies of particle-laden flows, two predominant frameworks can be employed, namely the Eulerian-Lagrangian (EL) or Lagrangian Particle Tracking (LPT) framework~\citep{Hilton2011,Li2013c,Mallouppas2013b,Zhao2013a,Zhou2014a,Ren2014,vanWachem2015,Kuerten2016,Fitzgerald2021,Mema2021,Markauskas2022}, and the Eulerian-Eulerian (EE) framework~\citep{vanWachem2001a,Gupta2013,Nigmetova2022}.
The EL framework entails tracking individual particle trajectories within a continuous fluid, relying on analytical, empirical, or semi-empirical correlations to estimate momentum exchange between the fluid and particles. Conversely, the EE framework treats both the fluid and particle phases as continuous and interpenetrating media, using distinct sets of equations to describe their behaviors and interactions. In this context, the correlations to estimate momentum exchange between the fluid and particles play a crucial role in the EL framework, as they determine the accuracy of momentum exchange estimations based on local fluid properties. Similarly, in the Euler-Euler framework, appropriate correlations are crucial to establish the interplay between the two phases and the resulting interaction forces in addition to accurate closures~\citep{Balachandar2020a}. The effectiveness of the EL and EE approaches hinges on the validity of these correlations, ensuring accurate results for specific particle-laden applications.}
Although recent EL simulations have been used to predict the behavior of particle-laden flows with non-spherical particles~\citep{Marchioli2010,Zhao2013a,vanWachem2015}, assuming particles to be spherical simplifies the dynamic equations that need to be solved to predict the behavior of the particles.
{For example, for spherical particles, the moment of inertia is a constant scalar. For non-spherical particles, it is an orientation dependent second order tensor. The implications of this are, that predicting the rotation of a non-spherical particle is significantly more complex than for a spherical one~\cite{Zhao2013a}.}
The assumption that particles are spherical also enables using several correlations describing the interactions between spherical particles and the properties of the local fluid flow, to accurately describe the transport of spherical particles in the EL framework~\citep{Schiller1933,Clift1971,Tenneti2011,Tang2015}.


However, {many realistic applications involve the transport of non-spherical particles, such as pneumatic conveying~\citep{Hilton2011}, fluidization bed mixing~\citep{Hilton2010}, or sedimenting particles~\citep{Concha2002}, to name just a few.}
Considering non-spherical particles increases the complexity of the equations governing the particle-laden flow, as the description of non-spherical particles relies upon multiple parameters~\citep{Chhabra1999}.
The main parameters are the shape and the orientation of the particle with respect to the main local flow direction.
To reduce the number of parameters, non-spherical particles are often assumed axi-symmetric about the major axis or the minor axis of the particle.
Then, a rotation of the co-ordinate system can be used to simplify the description of the orientation {between the main direction of the flow and the orientation of the particle} from a full 3D to a 2D one.
This drastically reduces the number of orientation angles to study in order to model the hydrodynamic forces for the specific particle.

Assuming the particle is axi-symmetric,~\citet{Brenner1963} derives an expression to model the hydrodynamic forces of inertial particles in the viscous regime.
~\citet{Brenner1963} demonstrates that the evolution of the drag force as a function of the orientation of the particle follows a so-called `sinesquare' profile.
As a result, the drag coefficient of the particle at any orientation can be determined from the two extreme values of the drag coefficients, namely the drag coefficient of the particle with its major axis in the direction of the local flow, and its major axis perpendicular to the local flow.
The `sinesquare' form of the drag force can be extended, to some degree, to finite particle Reynolds number~\citep{Sanjeevi2017}.
In addition to the drag force, the axi-symmetric particles are also subject to a so-called shape-induced lift force.
The magnitude of the shape-induced lift force varies with the orientation of the particle, and is often considered proportional to the drag force through the cross-flow principle~\citep{Hoerner1965}.
Along with the drag and the shape-induced lift forces, the effects of the torques also have to be considered at finite particle Reynolds number.

\reviewerII{Modeling the forces and torque coefficients for non-spherical particles in fluid flows is complex due to the numerous parameters involved, such as the shape or eventhe aspect ratio of the particle. Until recently, predictions have relied mainly on semi-empirical correlations using just the sphericity factor~\citep{Haider1989}. However, a key limitation of this approach is that the forces and torque coefficients are averaged over multiple instances of the same particle shape and do not take in account the orientation of the specific particle.
This is shown to be a significant limitation for these correlations, since the forces and torque can significantly differ, depending on the orientation of the particle with respect to the local fluid flow~\citep{Brenner1963}, thus more recent correlations showed the importance of including the orientation of the particle in the correlations~\citep{Ganser1993,Holzer2008}.}

With the aim to investigate and derive accurate correlations to model the fluid-particle interaction forces,~\citet{Holzer2008} performed a series of direct numerical simulations (DNS) of flow around non-spherical particles, varying the shape of the particle, the flow regime and the orientation angle.
The results of these simulations were used to derive an empirical expression for the drag force. 
In~\citet{Zastawny2012c}, DNS was also used to model the shape-induced lift force, the hydrodynamic torque and the rotational torque.
These coefficients have later been investigated by other authors, who have varied the shape of the particle~\reviewerII{\citep{Ouchene2015,Ouchene2016,Frohlich2020,Sanjeevi2022,Feng2023}}, or have studied the particle behavior at a higher particle Reynolds numbers~\citep{Sanjeevi2018}.

Although correlations describing the interactions between axi-symmetric particles and the fluid flow enables the transport of axi-symmetric particles in the EL framework, the effects of non-uniformity in the flow at the particle scale at a finite particle Reynolds number have, so far, been neglected.
The main reason for this is that, even for a simple flow configuration such as an unbounded linear shear flow, it is complex to extend the correlations to predict the drag and lift coefficients derived in the viscous regime for the unbounded linear shear flow profile to a finite particle Reynolds number~\citep{Kleinstreuer2013,Harper1968,Fan1995,Cui2018}, {as done for the hydrodynamic torque coefficient by~\citet{Dabade2016}}.

In an effort to derive empirical models to transport axi-symmetric particles subject to a locally non-uniform flow, we perform a series of DNS of rod-like particles subject to a uniform flow and an unbounded linear shear flow.
From the DNS results, new models for the drag, lift and torque coefficients are proposed.
The parameters varied in the simulations are the aspect ratio of the rod-like particle, the orientation angle between the main axis of the particle and the main local flow direction in the direction of the shearing of the flow, the local fluid shear rate and the flow regime.
The three rod-like particles considered in this work are shown in table~\ref{table:fibersshapes}.
From a series of DNS, empirical correlations for the drag and lift forces, as well as the torque are derived.
These correlations include the effect of the particle Reynolds number, the orientation angle, the aspect ratio and the local shear rate of the flow.
\reviewerII{The analytical results of~\citet{Brenner1963}, as well as the results of~\citet{Harper1968} and~\citet{Jeffery1922}, are used to close our correlations for uniform flow in the viscous regime for the drag and lift coefficients, and the torque coefficients in the viscous regime in case of a linear shear flow, respectively.
In the present manuscript, the term viscous regime is characterized by a particle Reynolds number equal or lower than \Rep = $\nicefrac{\vert\tilde{u}\vert D_\textup{eq}}{\nu_\textup{f}}$ = 0.1, where $\nu_\textup{f}$ is the viscosity of the fluid, $D_\textup{eq}$ the volume-based-equivalent diameter of the particle, and $\tilde{u}$ the relative velocity between the particle and the fluid flow.
To make sure the correct limits are achieved for the correlations proposed in this work, data from the correlation of~\citet{Schiller1933} is added to the dataset, representing the drag coefficient of a sphere as a function of the particle Reynolds number in case of uniform flow.
The DNS results of~\citet{Kurose1999} are included to the data to construct the fitted models for the drag and the lift coefficients in case of linear shear flow. However, it should be pointed out that the correlations proposed in this work have not been specifically validated for rod-like particles with an aspect ratio of less than 2.5.}

\begin{table}
\begin{tabular}{rccc}
       &
      \addheight{\includegraphics[width=0.25\columnwidth, trim={400 20 400 20},clip]{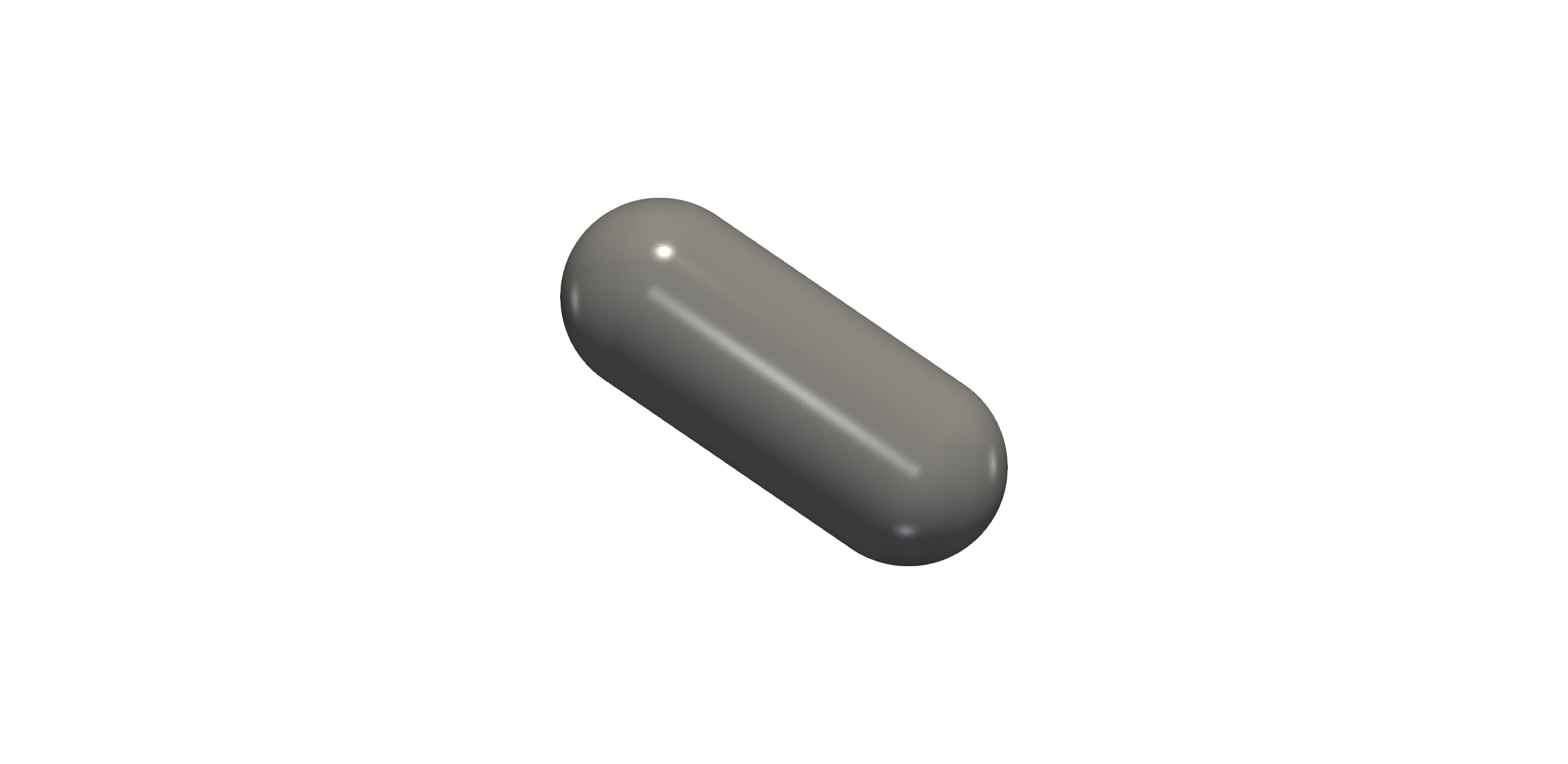}} &
      \addheight{\includegraphics[width=0.25\columnwidth, trim={400 20 400 20},clip]{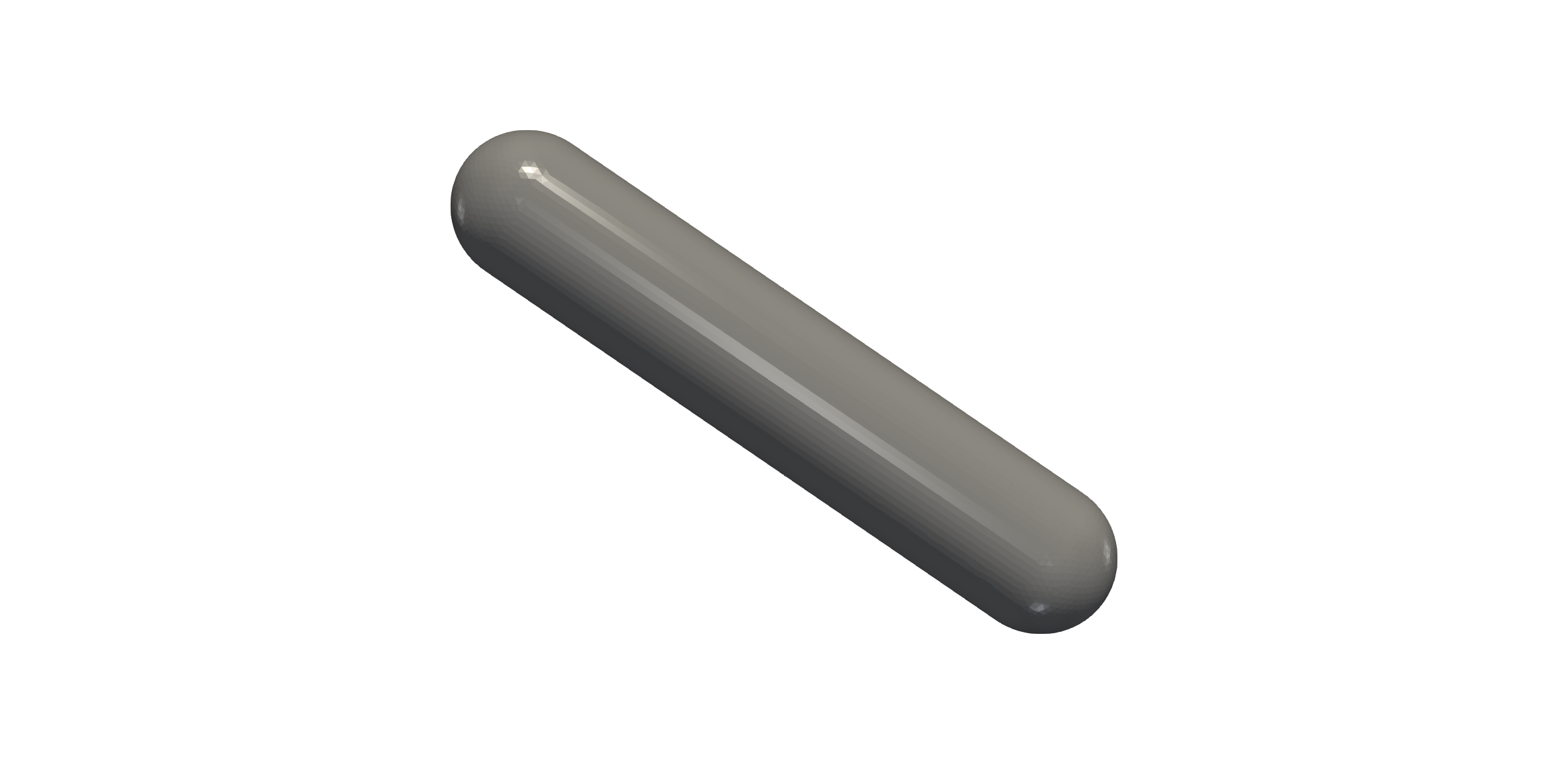}} &
      \addheight{\includegraphics[width=0.25\columnwidth, trim={400 10 400 10},clip]{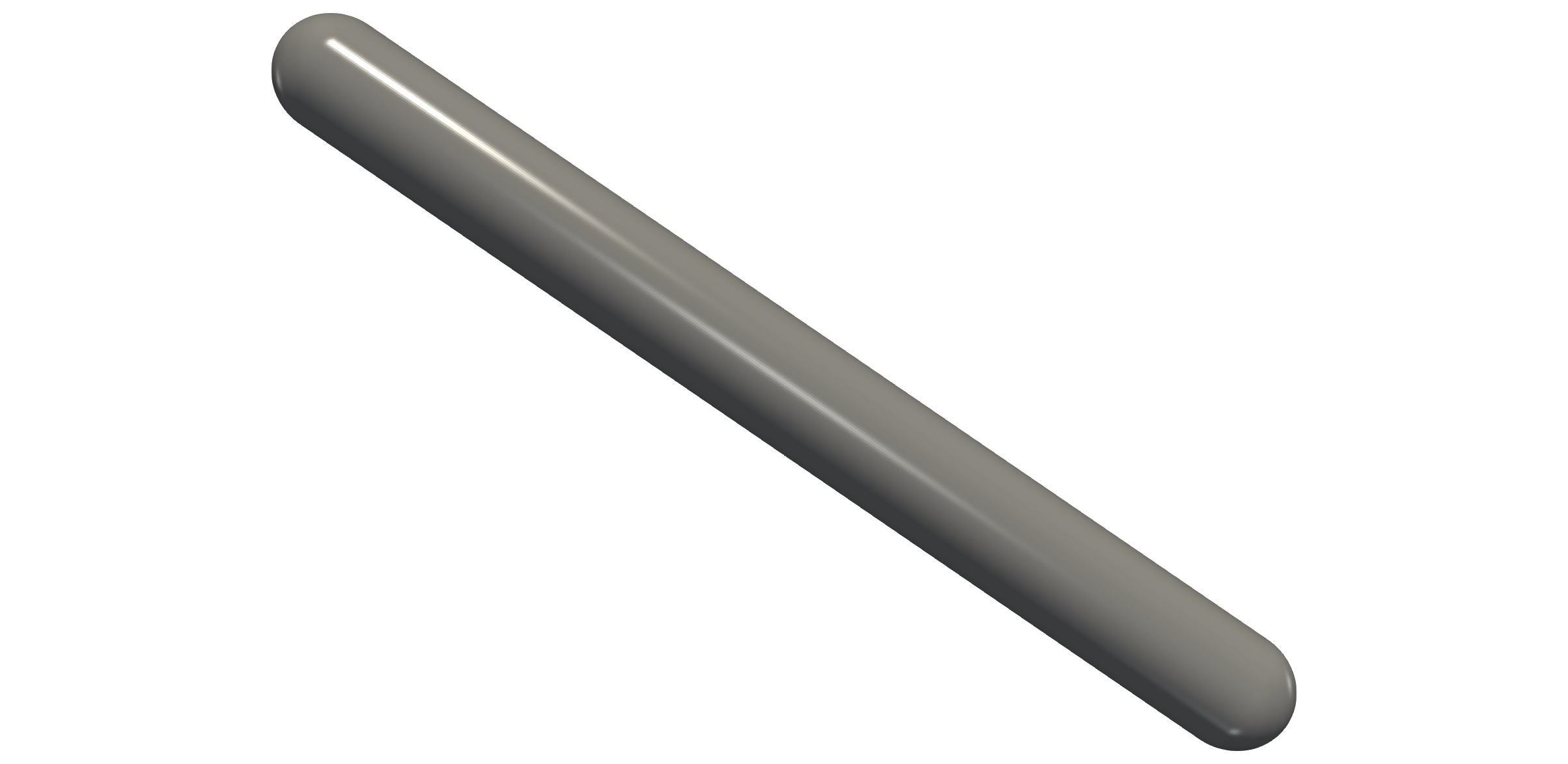}} \\
      \small Aspect ratio $\alpha=\frac{a}{b}$ & 2.5 & 5. & 10. \\
      \small Sphericity $\Phi=\frac{A_{\textup{S}}}{A_\textup{p}}$ & 0.878 & 0.737 & 0.596 \\
\end{tabular}
     \caption{The axi-symmetric rod-like particles considered in this work and their shape coefficients. $\Phi$ is the ratio of the surface of the volume-based equivalent sphere $A_\textup{S}$ and the surface of the particle $A_\textup{p}$. $\alpha$ is the ratio between the semi-major, $a$, and semi-minor $b$, axes of the particle.
     }\label{table:fibersshapes}
\end{table}

This paper is organized as follows: Section~\ref{sec:forcesonparticle} describes the main forces used to model the transport of a non-spherical particle, Section~\ref{sec:numericalframework} briefly discusses the numerical method employed to perform the DNS study. Section~\ref{sec:description} presents the numerical configuration. The correlations derived from the analytical solutions and the numerical simulations are discussed in Section~\ref{sec:results}. The conclusions of this work are summarized in Section~\ref{sec:conclusions}.

\section{Lagrangian equations of motion\label{sec:forcesonparticle}}

The rigid body dynamics of a non-spherical particle in the Lagrangian framework is governed by the Newton's second law, given as~\citep{Cundall1979}
\begin{equation}
\label{eq:NewtonSecondLaw-one}
\rho_{\textup{p}} V_{\textup{p}} \vect{a}_{\textup{p}} =
\sum \vect{F}_{\textup{p}}\, ,
\end{equation}
where $V_{\textup{p}}$ and $\rho_{\textup{p}}$ are the volume and the density of the particle, and $\vec{a}_{\textup{p}}$ and $\sum \vect{F}_{\textup{p}}$ are the acceleration of the particle and the resultant of the forces acting on the particle in the Lagrangian framework.
The resultant of the forces acting on the particle can be expressed as
\begin{equation}
\label{eq:NewtonSecondLaw-two}
\rho_{\textup{p}} V_{\textup{p}} \vect{a}_{\textup{p}} = \vect{F}_{\textup{p,ext}} + \vect{F}_{\textup{p,fluid}}\, , 
\end{equation}
where $\vect{F}_{\textup{p,ext}}$ is the resultant of the external forces acting on the particle, such as the gravity force or collisions, and $\vect{F}_{\textup{p,fluid}}$ is the resultant of the fluid forces acting on the particle, in theory obtained by integrating the fluid stresses over the surface of the particle.
For a small particle affecting only its local surroundings, the resultant of the fluid forces can be split in several contributions, considering steady and unsteady forces~\citep{Maxey1983}.
\reviewerII{In the scope of our work, only steady forces are considered.
Thus, the force balance over a non-spherical particle reduces to the drag and lift forces and gravitational effects, given as}
\begin{equation}
\label{eq:NewtonSecondLaw-three}
\rho_\textup{p} V_\textup{p} \vect{a}_\textup{p} = V_\textup{p} \left( \rho_\textup{p} - \rho_\textup{f}\right)\vect{g} + \vect{F}_\textup{D} +\vect{F}_\textup{L}\, ,
\end{equation}
where $\vect{g}$ is the gravitational acceleration, $\rho_\textup{f}$ is the fluid density, and $\vect{F}_\textup{D}$ and $\vect{F}_\textup{L}$ are the drag and lift forces acting on the particle.

\begin{figure}[htb]
  \centering
    \parbox[t]{0.25\textwidth}
  {
  \includegraphics[width=0.25\textwidth]{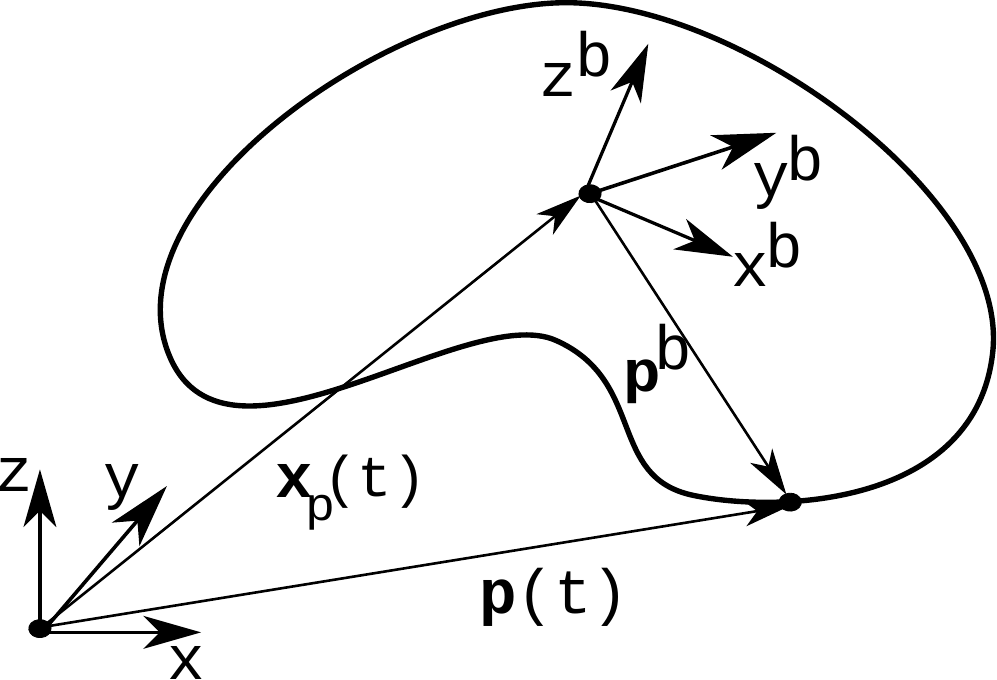}
  \centerline{(a) world space}
  }
  \parbox[t]{0.01\textwidth} {\mbox{}}
  \parbox[t]{0.18\textwidth}
  {
  \includegraphics[width=0.18\textwidth]{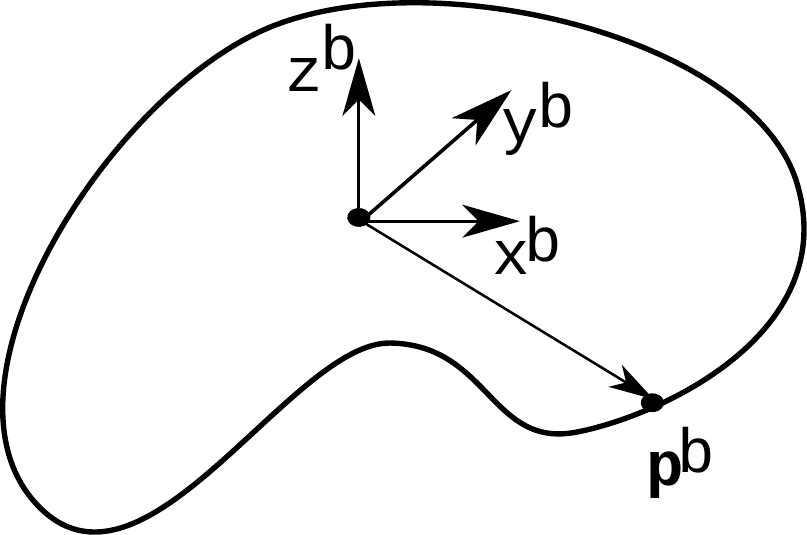}
  \centerline{(b) body space}
  }
  \caption{\label{fig:worldspace} The relation between world space (a) and body
  space (b). The fixed axes of body space, $x^\textup{b}$, $y^\textup{b}$ and $z^\textup{b}$ are indicated in
  both figures. The position of a fixed point in body space, $\pmb{p}^\textup{b}$ is transformed
  to world space, $\pmb{p}(t)$.}
\end{figure}

In the world space system, where the Cartesian co-ordinates are fixed in the origin of the initial Cartesian framework, the inertia tensor of a spherical particle is constant and diagonal.
However, the inertia tensor of a non-spherical particle changes with the orientation of the particle.
For an axi-symmetric rigid particle, it is then convenient to define an additional Cartesian co-ordinate system where the inertia tensor remains constant, the so-called body space.
This Cartesian system is aligned with the axi-symmetric major axis of the particle and is located at the center of mass of the particle.
This co-ordinate system is referred to as the body space reference frame, and is illustrated in figure~\ref{fig:worldspace} along with the Cartesian world space.
Then in body space, the equations governing the rotational motion of an axi-symmetric particle reads
\begin{align}\label{eq:rotation}
        \tens{I}^\textup{b} \dfrac{\textup{d}{\vect{\omega}^\textup{b}}}{\textup{d} t} + {\vect{w}^b \times \left( \tens{I}^\textup{b} \vect{w}^b \right)} =  \vect{T}^\textup{b}\, ,
\end{align}
where $\vect{\omega}^\textup{b}$ is the angular velocity, $\tens{I}^\textup{b}$ the inertia tensor, and $\vect{T}^\textup{b}$ the torques, all defined in the body space reference frame.
A transformation matrix~\citep{Fan1995,Marchioli2010}, or unit quaternion~\citep{Gillissen2008,Zhao2013}, is generally used to convert tensor and vector variables from body space to world space.

\subsection{Drag force}

The viscous drag force on an axi-symmetric particle acts in the direction of the main local flow and is characterized by the dimensionless drag coefficient $C_\textup{D}$, as
\begin{equation}
\text{\cd} = \frac{\vert\vect{F}_\textup{D}\vert}{\frac{1}{2} \rho_\textup{f} \vert\vect{\tilde{v}}\vert^2 \frac{\pi}{4} D_\textup{eq}^2}\, ,
\end{equation}
where $D_\textup{eq}$ is the volume-based equivalent diameter of the particle, and $\vect{\tilde{v}}$ is the relative velocity between the velocity of the particle and the velocity of the locally undisturbed fluid at the center of the particle.
The definition of $D_\textup{eq}$ is orientation and shape independent and is commonly used to characterize the size of axi-symmetric particles~\citep{Chhabra1999}.

In the viscous regime,~\citet{Brenner1963} derives an expression for the hydrodynamic drag force, $\vect{F}_\textup{D}$, for an axi-symmetric particle as
\begin{equation}\label{eq:viscousdragforce-brenner}
\vect{F}_\textup{D} = \pi b \rho_\textup{f} {\nu_\textup{f}} \tens{K} \vect{\tilde{v}}\, ,
\end{equation}
where $b$ is the semi-minor axis of the axi-symmetric particle, {$\nu_\textup{f}$ is the kinematic viscosity of the fluid}, and $\tens{K}$ is the matrix of coefficients of the resistance tensor (also known as the translation dyadic).
In body space, the resistance tensor reduces to a diagonal matrix $\tens{K}^\textup{b}$, whose coefficients vary with respect to the shape of the axi-symmetric particle~\citep{Brenner1963}.
The shape of the particle is characterized by its aspect ratio, $\alpha$, which is the ratio between the semi-major and semi-minor axes of the particle, $\alpha = \nicefrac{a}{b}$.
\reviewerII{Assuming that the major axis of the axi-symmetric particle is aligned with the x axis of the body space, the diagonal coefficients of $\tens{K}^\textup{b}$ is given for a prolate spheroid particle by}
\begin{align}\label{eq:resistance-tensor-K}
K^\textup{b}_\textup{xx} & = \frac{8 \left(\alpha^2 - 1 \right)^{3/2}}{\left[\left( 2\alpha^2 - 1 \right)\ln\left( \alpha + \sqrt{\alpha^2 - 1 }\right) - \alpha\left(\sqrt{\alpha^2 - 1 }\right) \right]}\, ,\\
K^\textup{b}_\textup{yy} & = K^\textup{b}_\textup{zz} = \frac{16 \left(\alpha^2 - 3 \right)^{3/2}}{\left[\left(2\alpha^2 - 3 \right)\ln\left( \alpha + \sqrt{\alpha^2 - 1 }\right) + \alpha\left(\sqrt{\alpha^2 - 1 }\right) \right]}\, .
\end{align}
\reviewerII{As discussed in~\citet{Happel1981} the expression for the drag coefficient of a prolate spheroid is also a good approximation to predict the drag coefficient of spherocylinders or rod-like particles, which are the shapes considered in the present study.
Eq.~\refeq{eq:viscousdragforce-brenner} is then used with the resistance matrix given in Eqs.~\refeq{eq:resistance-tensor-K} to validate our flow simulations, since an analytical solution is not present in the literature for rod-like spherical particles~\citep{Happel1981}.}

At higher particle Reynolds number, empirical correlations are derived for the drag coefficient from available data from the literature~\citep{Haider1989,Ganser1993,Madhav1995,Rosendahl2000,Loth2008a}, or more recently from DNS~\citep{Holzer2009,Zastawny2012c,Sanjeevi2018,Ouchene2020,Frohlich2020,Sanjeevi2022}.
These correlations depend on the orientation of the particle, characterized by the orientation angle $\theta$ between the main axis of the particle and the main local flow direction, and the flow regime, {characterized by the particle Reynolds number $Re_\textup{p}$, given by}
\begin{equation}
{\text{\Rep} = \frac{\vert\tilde{u}\vert D_\textup{eq}}{\nu_\textup{f}}\, .}
\label{eq:Reynolds}
\end{equation}
{The correlations for the drag coefficient} are often expressed by the following expression~\citep{Rosendahl2000}
\begin{equation}\label{eq:sinesquarelaw}
\text{\cd}(\text{\Rep},\theta) = C_\textup{D,$\parallel$}(Re_\textup{p}) + \left(C_\textup{D,$\perp$}(Re_\textup{p}) - C_\textup{D,$\parallel$}(Re_\textup{p})\right)\sin^\textup{k}(\theta)\, ,
\end{equation}
with $C_{\textup{D,$\parallel$}}$ and $C_{\textup{D,$\perp$}}$ being the fitted drag coefficients of the particle with its major axis in the direction of the local flow ($\theta = 0^o$), and its major axis perpendicular to the local flow ($\theta = 90^o$).
{Here $k$ is a constant, and for a flow regime up to a particle Reynolds number of \Rep $= 0.1$,this constant is equal to $k=2$, meaning the drag force follows a `sinesquare' profile~\citep{Happel1981}}.
For a finite particle Reynolds number, the value of the constant $k$ varies in the literature.
In~\citet{Rosendahl2000}, who considers super-ellipsoids, the constant $k$ is set to $k=3$.
In the work of~\citet{Zastawny2012c}, the coefficient $k$ is fitted from the DNS results and varies with respect to the shape of the axi-symmetric particle.
\reviewerII{In~\citet{Feng2023}, who perform DNS of uniform flow past a fixed rod-like particle varying the orientation angle between the main local fluid flow and the particle, the aspect ratio of the particle, and the particle Reynolds number, the coefficient $k$ depends on both the aspect ratio of the rod-like particle and the particle Reynolds number.
Thus, the aspect ratio of the particle can also be directly included to derive a drag model which varies from spherical particle to elongated one~\citep{Ouchene2016,Frohlich2020,Sanjeevi2022,Feng2023}.}

For an unbounded linear shear flow in the viscous regime,~\citet{Harper1968} show that the drag force experienced by an axi-symmetric particle does not vary from the uniform flow configuration.
The DNS results reported in~\citet{Holzer2009} corroborate these conclusions for a prolate of aspect ratio $\alpha=1.5$.
Moreover, \citet{Holzer2009} show a negligible influence of the shear rate on the drag coefficient, up to a particle Reynolds number of \Rep $=90$.
Above this particle Reynolds number value, the drag force of the specific particle is increased in case of linear shear flow compared to uniform flow~\citep{Holzer2009}.
However, the exact effects are not yet known for axi-symmetrical particles.

\subsection{Lift force}

Axi-symmetric particles are subject to a lift force in a locally uniform flow, referred to as the shape-induced lift force.
The shape-induced lift force depends on the orientation of the particle relative to the direction of the local fluid velocity~\citep{Gallily1979}, and acts in the direction perpendicular to the relative local mean flow.
This force is characterized by a lift coefficient, \cl, defined as
\begin{equation}
\text{\cl} = \frac{\vert\vect{F}_\textup{L}\vert}{\frac{1}{2} \rho_\textup{f} \vert\vect{\tilde{v}}\vert^2 \frac{\pi}{4} D_\textup{eq}^2}\, .
\end{equation}
\reviewerII{For $\theta = 0^o$ or $90^o$, the value of the lift force coefficient is equal to $C_\textup{L} = 0$ only, as the particle is symmetrical along these orientations.
At particle Reynolds number \Rep = 0.1 and below, the lift coefficient at any orientation angle can be given as a function of the drag coefficients of the particle with its major axis in the direction of the local flow ($C_\textup{D,$\parallel$}$), and its major axis perpendicular to the local flow ($C_\textup{D,$\perp$}$)~\citep{Happel1981,Sanjeevi2017}:
\begin{equation}
  C_\textup{L}\left(\theta,\text{\Rep} \rightarrow 0 \right) = \left[C_\textup{D,$\perp$}(\text{\Rep} \rightarrow 0) - C_\textup{D,$\parallel$}(\text{\Rep} \rightarrow 0)\right]\sin{\left(\theta\right)}\cos{\left(\theta\right)}\, ,
\end{equation}
which yields to a maximum for the lift coefficient at $\theta = 45^{o}$, and an evolution of the lift coefficient as a function of the orientation angle which follows a cosine-sine profile~\citep{Brenner1963,Sanjeevi2017}.}
Owing to the proportionality between the lift and the drag forces,~\citet{Hoerner1965} proposes to estimate the lift coefficient from the drag coefficient,
\begin{equation}\label{eq:crossflowprinciple}
\frac{\text{\cl}\left(\theta,\text{\Rep}\right)}{\text{\cd}\left(\theta,\text{\Rep}\right)} = \sin^2\left(\theta\right)\cos\left(\theta\right)\, ,
\end{equation}
which is referred to as the cross-flow principle.
However,~\citet{Mando2010} conclude that the cross-flow principle is valid only at very large particle Reynolds number.
Moreover, this expression tends to strongly deviate from the cosine-sine profile of the lift coefficient in the viscous regime~\citep{Happel1981}.
~\citet{Frohlich2020} propose to model the evolution of the lift coefficient from the viscous regime to high particle Reynolds number applying a change of co-ordinate on the orientation angle of the particle.
Thus, the expression accurately predicts the shift of the maximum of the lift coefficient, from an orientation angle of $\theta = 45^{\text{o}}$ in the viscous regime, toward a higher orientation angle for finite particle Reynolds number.
This expression is given by
\begin{equation}\label{eq:shiftlift}
\frac{\text{\cl}}{\max \left( \text{\cl}\right)} = \cos(\varPsi_{\theta}(\text{\Rep},\alpha))\sin(\varPsi_{\theta}(\text{\Rep},\alpha))\, ,
\end{equation}
where $\varPsi_{\theta}(\text{\Rep},\alpha)$ is the function used to shift the maximum of the lift coefficient.

In contrast to uniform flow, in the presence of a locally non-uniform flow, a spherical particle can experience a contribution to its lift force.
For instance, in an unbounded linear shear flow,~\citet{Saffman1965}, who derives an expression for the shear-induced lift force in the viscous regime using an asymptotic method, shows a contribution of the shear-induced lift force to the particle lift force from the lower velocity side to the higher velocity side.
For a spherical particle, the shear-induced lift force is given by
\begin{equation}\label{eq:Saffman1965}
\mathbf{F}_\textup{L} = 6.46 \rho_\textup{f} \left(\frac{D_\textup{eq}}{2}\right)^2 \sqrt{\frac{\nu_\textup{f}}{\vert \vert \nabla \times \mathbf{u}\vert \vert}} \left[\left(\nabla \times \mathbf{u}\right) \times{\vect{\tilde{v}}}\right]\, ,
\end{equation}
where $\nabla \times \mathbf{u}$ is the {vorticity of the fluid at the location of the particle.}
For an unbounded one directional linear shear flow, $\nabla \times \mathbf{u}$ reduces to the velocity gradient in the direction of the linear shear, \reviewerII{in this work $\nicefrac{\partial u_x}{\partial y}$}.
\reviewerII{By extending the derivations of~\citet{Saffman1965}, valid only for spherical particles and in the viscous regime, to particle of axi-symmetric shape based upon the Stokes drag coefficient,~\citet{Harper1968} demonstrate that the lift force of an axi-symmetric particle is increased from the lower velocity side to the higher velocity side in case of linear shear flow compared to uniform flow.
The expression to determine the shear-induced lift force for prolate spheroid particle is given by}
\begin{equation}\label{eq:Fan1995}
\reviewerII{\mathbf{F_\text{L}} = \pi^2 \rho_\text{f} b^2 \sqrt{\frac{\nu_\text{f}}{\vert \vert \nicefrac{\partial u_x}{\partial y}\vert \vert}} \left[\left(\frac{\partial u_x}{\partial y}\right) \tens{K}\text{ }\tens{L}\text{ }\tens{K}  {\vect{\tilde{v}}}\right]}\, ,
\end{equation}
where $\tens{K}$ is the resistance tensor, defined in Eq.~\eqref{eq:resistance-tensor-K}, and $\tens{L}$ is the coefficient matrix of the lift tensor.
\reviewerII{Because the drag expression for the prolate spheroids is a very good assumption for the rod-like particles, Eq.~\eqref{eq:Fan1995} is used in this work to predict the lift coefficient in case of shear flow of the particles.}
The coefficients of the matrix $\tens{L}$ are given in~\citet{Harper1968}; it reads for a linear shear $\partial{u}/\partial{y}$ among the x-component of the fluid velocity vector
\begin{equation}
\tens{L} =
\begin{bmatrix}
5.01 \times 10^{-2}& 3.29 \times 10^{-2}& 0 \\
1.82 \times 10^{-2}& 1.73 \times 10^{-2}& 0 \\
0 & 0 & 3.73 \times 10^{-2}
\end{bmatrix}
\, .
\end{equation}
For a spherical particle, $b=\nicefrac{D_\text{eq}}{2}$ and $\tens{K} = 6 \tens{I}$, ~\citet{Fan1995} demonstrate that Eq.~\eqref{eq:Fan1995} reduces to Eq.~\eqref{eq:Saffman1965}.

Outside the viscous regime, studies of the change in the axi-symmetric particle lift force caused by the linear shear flow are very limited, and are restricted to the DNS results of~\citet{Holzer2009}.
For a prolate of aspect ratio $\alpha = 1.5$,~\citet{Holzer2009} show that above a particle Reynolds number of \Rep $\approx 90$, the lift force of the particle in a linear flow is increased from the higher velocity toward the lower velocity, hence negatively increase the particle lift force.
This phenomenon is also observed for spherical particles~\citep{Kurose1999,Bagchi2002}, and results from the specific flow recirculation at high particle Reynolds number in the wake of the particle, which modifies the stress distribution on the surface of the particle.


\subsection{Hydrodynamic torque}

The torque experienced by an axi-symmetric particle can be split in two dominant contributions, the hydrodynamic or pitching torque, and the rotational torque~\citep{Zastawny2012c}.
The hydrodynamic torque arises from the offset between the center of pressure of the particle in the fluid and the center of mass of the particle.
This offset is caused by the shift of the location of the center of pressure as the center of mass remains constant in the reference frame of a rigid particle.
For a uniform flow configuration, the shift of the center of pressure is observed when the orientation angle of the particle is within the range $0^\text{o} < \theta < 90^\text{o}$.
At the specific orientation angles $\theta = 0^\text{o}$ and $90^\text{o}$, the distribution of the pressure on the surface of the particle is symmetric, and the center of pressure coincides with the center of mass of the particle, hence the particle does not experience a hydrodynamic torque.

The rotational motion of an axi-symmetric particle, resulting from the hydrodynamic torque or external forces, such as collisions, causes a rotational torque.
Within the scope of the current paper, only the hydrodynamic torque is analyzed.
The reader is referred to the work of~\citet{Zastawny2012c} for a discussion of the rotational torque modeling.

The hydrodynamic torque is characterized by a torque coefficient, \ct, given by
\begin{equation}
\text{\ct} = \frac{\vert\vect{T}_\text{P}\vert}{\frac{1}{2} \rho_{\text{f}} \vert\vect{\tilde{v}}\vert^{2} \frac{\pi}{8} D_\text{eq}^3}\, ,
\end{equation}
where $\vect{T_\text{P}}$ is the hydrodynamic torque of the particle.
Several correlations estimate the hydrodynamic torque coefficient from the hydrodynamic forces and the co-ordinates of the center of pressure, but such correlations are typically not accurate~\citep{Rosendahl2000}.
In more recent work, the correlations to estimate the hydrodynamic torque experienced by axi-symmetric particles subject to a locally uniform flow are derived from DNS~\citep{Zastawny2012c,Ouchene2016,Sanjeevi2018,Frohlich2020}.

In contrast to an axi-symmetric particle subject to an uniform flow in the viscous regime, a particle subject to an unbounded linear shear flow receives a contribution to its hydrodynamic torque from the fluid flow.
\citet{Jeffery1922} derives an expression to determine the hydrodynamic torque of an axi-symmetric particles in an infinite linear shear flow in creeping conditions.
It reads, in body space, for each component of the hydrodynamic torque
\begin{align}\label{eq:torquejeffery}
\begin{split}
& \boldsymbol{T}_\textup{x}^\textup{b}=\frac{32 \pi \mu_\textup{f} {b}^3 \alpha}{3\left(\lambda_2+\lambda_3\right)}\left(\Omega_\textup{zy}^\textup{b}-\omega_\textup{x}^\textup{b}\right), \\
& \boldsymbol{T}_\textup{y}^\textup{b}=\frac{16 \pi \mu_\textup{f} {b}^3 \alpha}{3\left(\lambda_3+\alpha^2 \lambda_1\right)}
\left[\left(1-\alpha^2\right) S_{\textup{xz}}^\textup{b}+\left(1+\alpha^2\right)\left(\Omega_{\textup{xz}}^\textup{b}-\omega_\textup{y}^\textup{b}\right)\right], \\
 & \boldsymbol{T}_\textup{z}^\textup{b}=\frac{16 \pi \mu_\textup{f} {b}^3 \alpha}{3\left(\lambda_2+\alpha^2 \lambda_1\right)}\left[\left(\alpha^2-1\right) S_\textup{yx}^\textup{b}+\left(\alpha^2+1\right)\left(\Omega_\textup{yx}^\textup{b}-\omega_\textup{z}^\textup{b}\right)\right]\, ,
\end{split}
\end{align}
with $S^\textup{b}$ and $\Omega^\textup{b}$ the fluid strain and rotation tensors in body space, {given by}
\begin{align}\label{eq:Fluid strain Rate and Rotation tensor}
S^\textup{b} = & \frac{1}{2}\Big[ \nabla \vect{u}^b + \left(\nabla \vect{u}^b\right)^T\Big]\\
\Omega^\textup{b} = & \frac{1}{2}\Big[ \nabla \vect{u}^b - \left(\nabla \vect{u}^b\right)^T\Big]
\end{align}
{where $\nabla \vect{u}^b$ is the fluid velocity gradient in the body space reference frame}~\citep{Zhao2013a}.
In Eq.~\eqref{eq:torquejeffery}, $\mu_{\textup{f}}$ is the viscosity of the fluid, $\alpha$ is the aspect ratio of the axi-symmetric particle, $a$ is the semi-major axis length of the particle, and $\lambda_1$, $\lambda_2$, and $\lambda_3$ are geometric coefficients derived in~\citet{Gallily1979} as
\begin{align}\label{eq:Gallily-lambda}
\begin{split}
& \lambda_1 = -\frac{2}{\alpha^2 -1} - \frac{\alpha}{\left( \alpha^2 - 1\right)^{3/2}}\ln{\left[\frac{\alpha-\left(\alpha^2-1\right)^{1/2}}{\alpha+\left(\alpha^2-1\right)^{1/2}}\right]}\, ,\\
& \reviewerII{\lambda_2 = \lambda_3 = \frac{\alpha^2}{\alpha^2 -1} + \frac{\alpha}{2\left( \alpha^2 - 1\right)^{3/2}}\ln{\left[\frac{\alpha-\left(\alpha^2-1\right)^{1/2}}{\alpha+\left(\alpha^2-1\right)^{1/2}}\right]}\,} .
\end{split}
\end{align}
{In~\citet{Dabade2016}, an expression to predict the hydrodynamic torque coefficient of a \reviewerII{spheroid} particle is analytically derived based on the Stokes flow solution, and is valid up to finite particle Reynolds number of \Rep$\approx \mathcal{O}(1)$}.
For flow regimes up to a particle Reynolds number of \Rep$\approx \mathcal{O}(10^{2})$, the hydrodynamic torque coefficient of an axi-symmetric particle subject to a locally linear shear flow is very different from the results obtained with a locally uniform flow.
The DNS results of~\citet{Holzer2009} show that these differences drastically reduce for a prolate of aspect ratio $\alpha=1.5$ at high particle Reynolds number.
In this paper, we will study the influence of the particle Reynolds number on the torque coefficient for axi-symmetric particles of varying aspect ratio.

\section{Numerical framework\label{sec:numericalframework}}
In this work, particle resolved DNS of a fluid surrounding an axi-symmetric particle is performed with the smooth immersed boundary method (IBM)~\citep{Uhlmann2005}, using the direct-forcing formulation of~\citet{AbdolAzis2019} with the additional correction of~\citet{Cheron2023a}.
The IBM couples the modeling of the fluid domain through an Eulerian framework and the representation of the surface of the particle with a Lagrangian framework.
The Lagrangian framework consists of evenly spaced markers, discretizing the surface of the particle.
The Eulerian framework is based on a finite-volume framework with a collocated variable arrangement.
The incompressible fluid phase, subject to the Navier-Stokes equations, is solved with an implicit pressure-velocity coupling~\citep{Denner2014a}, and the source terms of the momentum equations are discretized to numerically balance the flow pressure gradient~\citep{Bartholomew2018}.

The Eulerian and Lagrangian frameworks being independent, an interpolation/spreading strategy is used to connect the Eulerian and Lagrangian fluid variables~\citep{Peskin2003}.
The interpolation and spreading compact supports are constructed through a mollified moving-least-squares algorithm~\citep{Bale2021}.
The size of the compact support determines the number of fluid cells used for the interpolation and the spreading of the fluid variables.
In this work a five-point spline kernel function is used~\citep{Bao2016}.
The spreading of the IBM source terms toward the source terms of the fluid momentum equations is scaled by a relaxation factor~\citep{Zhou2021}.
This relaxation factor is based on stability condition criterion, and controls the rate at which the no-slip condition is reached as well the magnitude of the no-slip error.
The reader is referred to the work of~\citet{Cheron2023a} for the details of implementation and validation of the present IB method.

\begin{figure}[htbp!]
\centering
\includegraphics[width=0.8\columnwidth, trim={110 120 40 70}, clip]{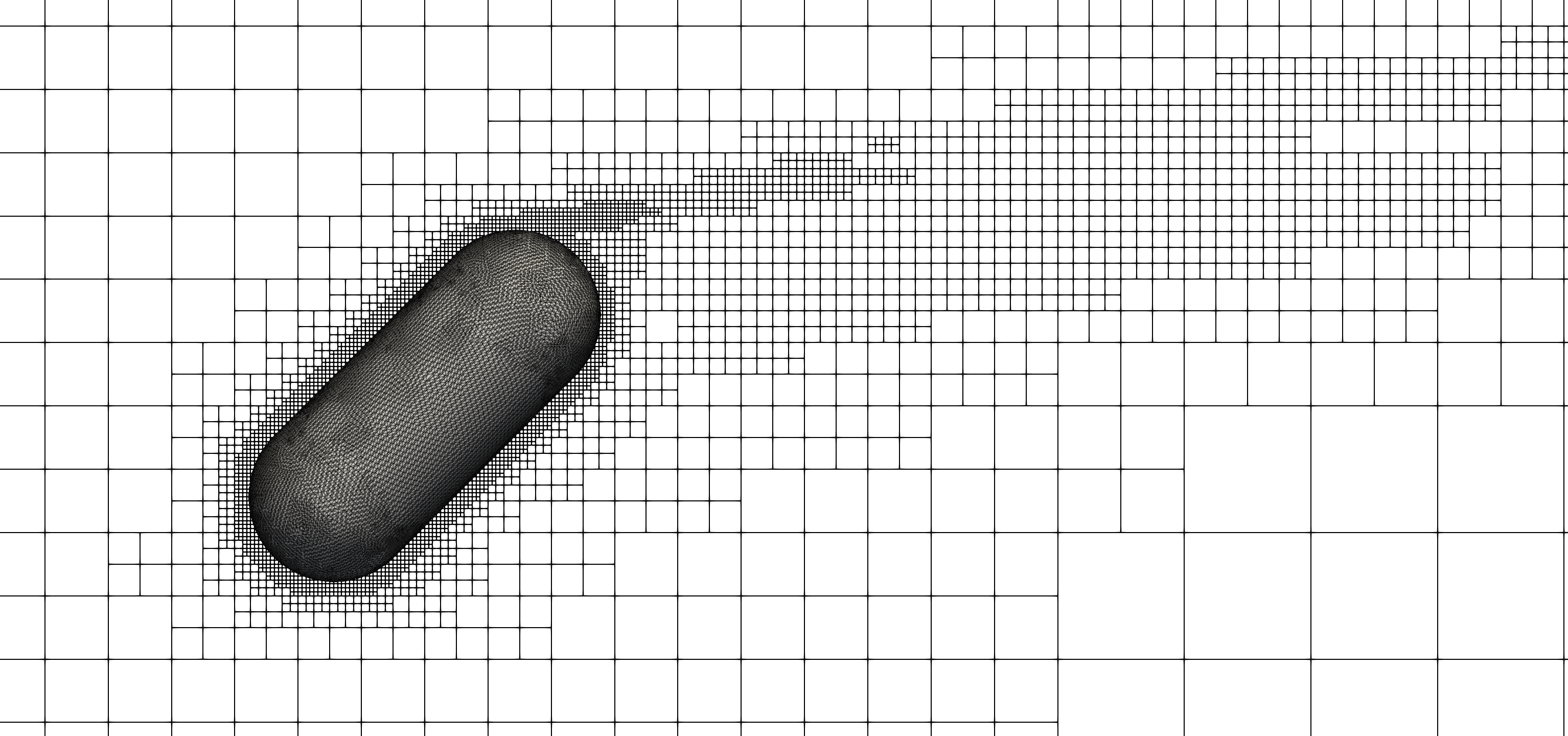}
\caption{Instantaneous snapshot of the Lagrangian and fluid meshes generated in the DNS of a particle of aspect ratio $\alpha=2.5$ subject to an uniform flow at a particle Reynolds number \Rep $=200$, and  orientation angle $\theta = 45^o$.}
\label{fig:mesh}
\end{figure}

To reduce the computational cost, the DNS are performed with a dynamic adaptive mesh refinement (AMR) for the Eulerian fluid mesh in order to focus the computational efforts in the region of interests.
The refinement and coarsening criteria are a distance-based refinement criterion and a vorticity-based criterion, which ensure that the fluid mesh is refined near the surface of the particle and in the wake of the particle.
An instantaneous snapshot of the fluid and the Lagrangian meshes generated in the DNS is shown in figure~\ref{fig:mesh} for a particle Reynolds number of \Rep $=200$.

\section{Numerical configuration\label{sec:description}}

\subsection{Simulation set-up\label{subsec:simu-set-up}}

The numerical framework presented in Section~\ref{sec:numericalframework} is employed to carry out the simulations of the flow around a particle, from which the drag and lift forces and the hydrodynamic torque of the axi-symmetric particles shown in table~\ref{table:fibersshapes} are determined.
A triangulation of the surface of the axi-symmetric particle is used to construct the Lagrangian markers.
The distance between the vertices is set equal to the local fluid mesh spacing.
The computational domain is a cubic box of size $20D_\textup{eq}$ for the high particle Reynolds number simulations (\Rep $\geq20$).
{This ensures domain independent results, since the length in the rear of the particle is of at least $10D_\textup{eq}$ for the high particle Reynolds number (\Rep $>20$) simulations, as prescribed in the domain independence study of~\citet{Sanjeevi2018}}.
{For lower particle Reynolds number (\Rep$<20$) simulations, a cubic box of size $40D_\textup{eq}$ is used. This computational domain size is twice larger than prescribed in~\citet{Sanjeevi2018}, which reduces confinement effects.}
{The center of co-ordinates of the particle is located at the center of the computational domain}.
{At large particle Reynolds number, the hydrodynamic coefficients may slightly fluctuate for non-spherical particles~\citep{Zastawny2012c,Sanjeevi2018}, therefore all DNS are performed over a long enough physical time to ensure time independent averaged results.}

To simulate the uniform flow and the shear flow DNS, two set of boundary conditions are used on the faces of the computational domain.
For the uniform flow simulation, the numerical configuration of~\citet{Zastawny2012c} is used.
{For the velocity,} a Dirichlet and a Neumann boundary conditions are used in the direction of the flow, the $x$ co-ordinate, with a constant inlet velocity $u_\textup{in}$ set at the inlet, {and a velocity gradient set to 0 at the outlet}.
A full slip boundary condition is applied on the top and bottom boundaries of the domain, the $y$ co-ordinate.
A periodic boundary condition is used for the boundaries in the direction of the $z$ co-ordinate.
{The pressure is set to 0 at the outlet and the pressure gradient is set to 0 for the remaining non-periodic boundaries.}

\begin{figure}[htbp!]
\centering
\includegraphics[width=0.9\columnwidth, trim={110 120 40 50}, clip]{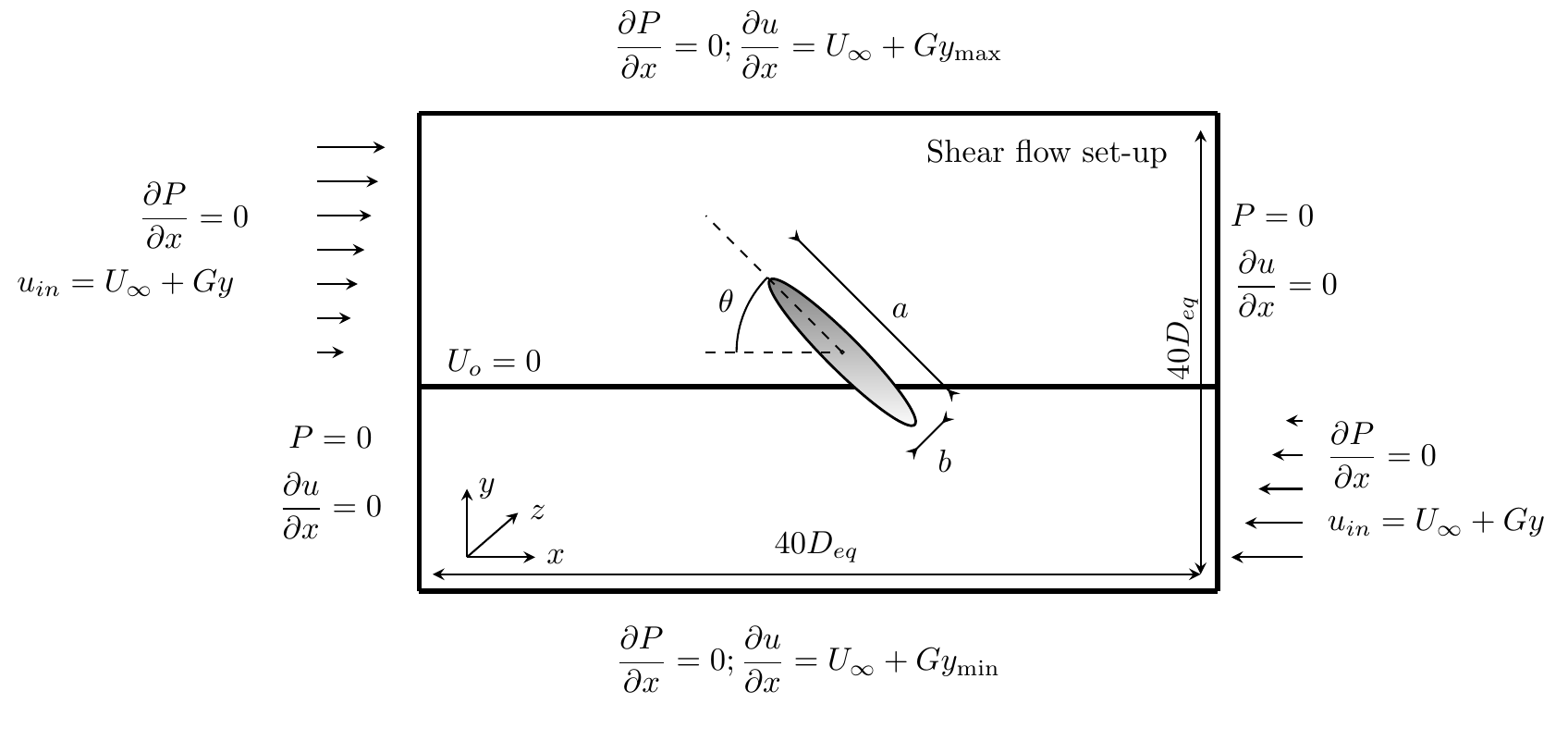}
    \caption{A sketch of the simulation setup and boundary conditions used to simulate the shear flow past a rod-like axi-symmetric particle.}\label{fig:introductionproblem}
\end{figure}

For the unbounded linear shear flow, the inlet velocity profile is prescribed by
$$
u_\textup{in} = U_{\infty} + G y\, ,
$$
where $U_{\infty}$ is a reference velocity, $G$ is the dimensional shear rate, and $y$ is the co-ordinate along the $y$ axis.
To derive a correlation based on the shear rate of the fluid flow, it is convenient to define a dimensionless number based on the dimensional shear rate $G$.
~\citet{Kurose1999} provide the following expression for the dimensionless shear rate $\tilde{G}$
\begin{equation}\label{eq:dimensionless-shear-rate}
\tilde{G} = \frac{D_\textup{eq} G}{U_{\infty}}\, ,
\end{equation}
which is also used in this work.
Thus, the inlet velocity profile can also be written in dimensionless form
$$
U = 1 + \tilde{G} Y\, ,
$$
where $U=u_\textup{in}/U_{\infty}$ and $Y = y/D_\textup{eq}$.
For large computational domains and large dimensionless shear rates, the definition provided in Eq.~\eqref{eq:dimensionless-shear-rate} yields to a negative velocity in the inlet profile.
Therefore, the unbounded linear shear flow configuration requires a modification of the boundary conditions along the $x$ co-ordinate.
{The solution adopted, is to split the inlet and outlet boundaries along the velocity node $U_0$ into both an outlet/inlet boundary, as shown in the sketch of the simulation setup, figure~\ref{fig:introductionproblem}.
For the left boundary in figure~\ref{fig:introductionproblem}, this results in a Dirichlet condition for the velocity and Neumann condition for the pressure for the upper part, and a Dirichlet condition for the pressure and Neumann condition for the velocity for the lower part of the left boundary.
The same methodology is applied to the right boundary, as shown in figure~\ref{fig:introductionproblem}}.
Other boundaries are left unchanged.

{To solve the discretised equations governing the fluid flow, the DNSs are performed with second order spatial and temporal accuracy using a Laplacian discretization of the diffusion term, a Minmod TVD scheme for the discretization of the advection term,
and a second order backward Euler temporal scheme for the transient terms~\citep{Denner2020}}.
The time-step is fixed and based on the convection and viscosity criteria for the temporal derivatives, {as given in~\citet{Kang2000}, with a fixed CFL criterium set to 0.05}.
{This low CFL number is required to accurately enforce the no-slip conditions on the particle surface in the IBM framework}.
The number of fluid cells in the DNS is based on a spatial convergence study, see Section~\ref{subsec:spatial-convergence}.
In this convergence study, the number of Lagrangian markers is adapted, so that the spacing between them equals the Eulerian fluid mesh length.

\subsection{Parameters of influence\label{subsec:param-influence}}

Four parameters are varied to derive the models to predict the drag, lift and torque coefficients.
These parameters are the dimensionless shear rate $\tilde{G}$, defined in Eq.~\eqref{eq:dimensionless-shear-rate},
the orientation angle $\theta$ between the main axis of the particle and the main local flow direction, which always vary in the same plane that the direction of the local shear flow,
the aspect ratio of the particle, $\alpha=a/b$, where $\alpha$ is the aspect ratio and $a$ and $b$ are the semi-major and the semi-minor axes of the particle, respectively,
and the particle Reynolds number $Re_\textup{p}$ { using the characteristic velocity of the undisturbed fluid velocity interpolated to the center of the particle,  $u_\textup{f@p}$.}
Here $D_\textup{eq}$ is constant for all the particles considered in this study, see table~\ref{table:fibersshapes}, and $u_\textup{f@p}$ is given by
\begin{equation}\label{eq:VolumeRestriction}
  u_{\text{f@p}} =
    \begin{cases}
      U_{\infty} & \text{uniform flow}\\
      U_{\infty} + G y_\text{p} & \text{linear shear flow}
    \end{cases}
\end{equation}
where $y_\textup{p}$ indicates the $y$ component of the particle center co-ordinates.
This set of dimensionless parameters is illustrated in figure~\ref{fig:introductionproblem}, and the matrix of parameters varied in this study is given in table~\ref{table:matrix-simulation}.

\begin{table}
\centering
\begin{tabular}{ {r}||{l}|{l}  }
 & \textbf{Uniform flow}& \textbf{Linear shear flow}\\
 \hline
\hline
 $\alpha$ $[-]$   & 2.5, 5, 10 & 2.5, 5, 10\\
 \hline
 $\text{\Rep}$ $[-]$ & 2, 10, 25, 50, 100, 200, 300 & 2, 10, 25, 50, 100, 200, 300\\
 \hline
 $\tilde{G}$ $[-]$  & - & 0.1, 0.2\\
\hline
 $\theta$ $[^\textup{o}]$  & 0, 30, 45, 60, 90   & 0, 30, 45, 60, 90, 120, 135, 150\\

\end{tabular}
     \caption{Matrix of parameters varied for the DNS of the uniform flow and unbounded linear shear flow configurations, $\alpha$ is the aspect ratio of the axi-symmetric particle, \Rep is the particle Reynolds number, $\tilde{G}$ is the dimensionless shear rate, and $\theta$ is the orientation angle between the particle and the local main flow. The orientation angle of the particle always vary in the same plane that the direction of the local shear flow.}\label{table:matrix-simulation}
\end{table}

\subsection{Spatial convergence analysis\label{subsec:spatial-convergence}}

\begin{figure}[ht]
\centering
\includegraphics[width=0.48\textwidth]{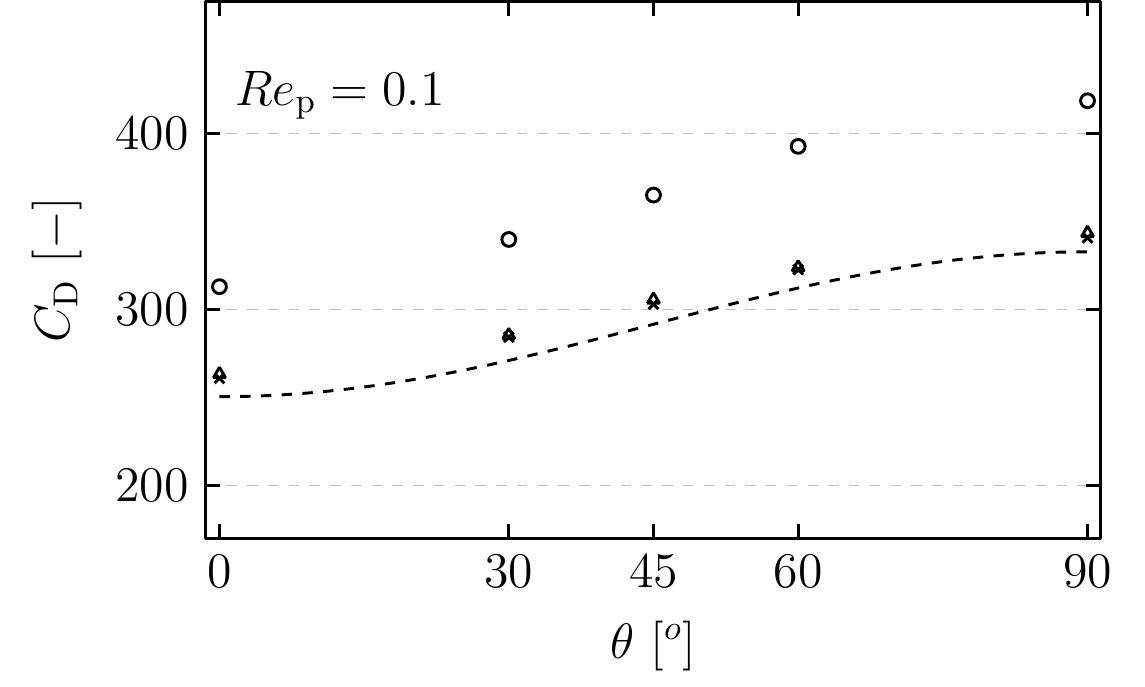}
\includegraphics[width=0.475\textwidth]{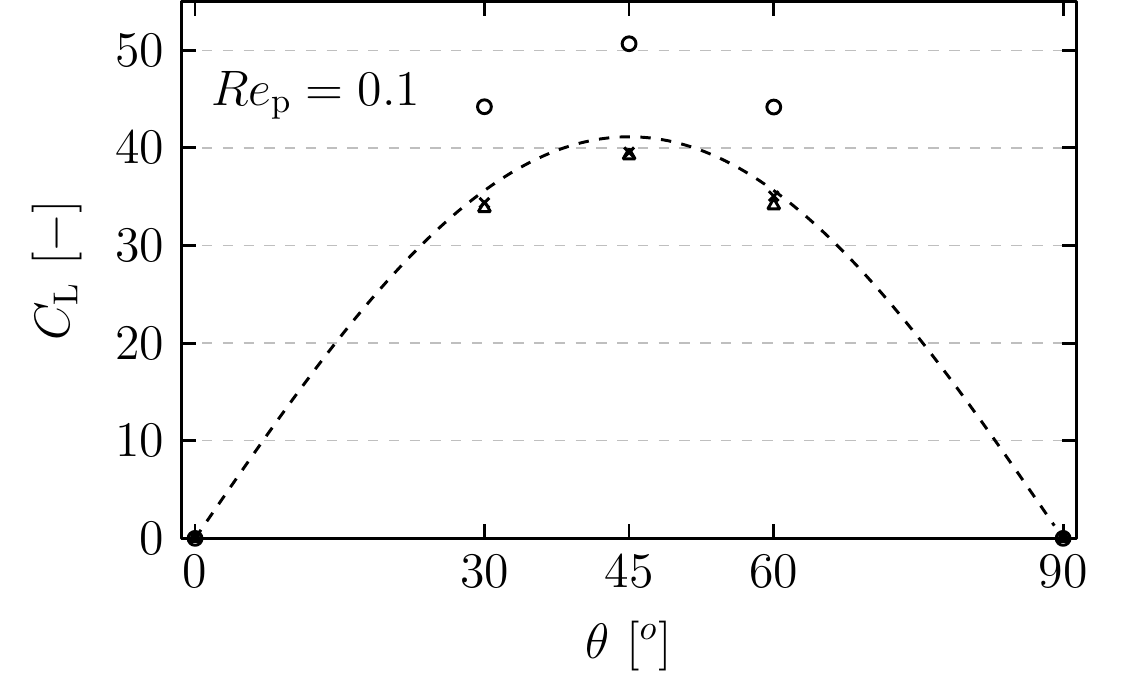}
    \caption{The drag coefficient \cd (left), and the lift coefficient \cl (right) as a function of the orientation angle, for an axi-symmetric particle of aspect ratio $\alpha=5$ subject to a uniform flow in the viscous regime. 
    Dashed line:~\citet{Happel1981}.
    Current work, $D_\text{eq}/\Delta x =16$: \protect\tikz{\protect\draw[thick] (0,0) circle (2pt)},
    $D_\text{eq}/\Delta x =32$: $\Delta$,
    $D_\text{eq}/\Delta x =64$:
    {\protect\tikz{\protect\draw pic[black, rotate = 0, line width=1] {cross=3pt};}}.
    }\label{fig:Convergence-ViscousRegime}
\end{figure}

\begin{figure}[ht]
  \centering
  \includegraphics[width=0.325\columnwidth]{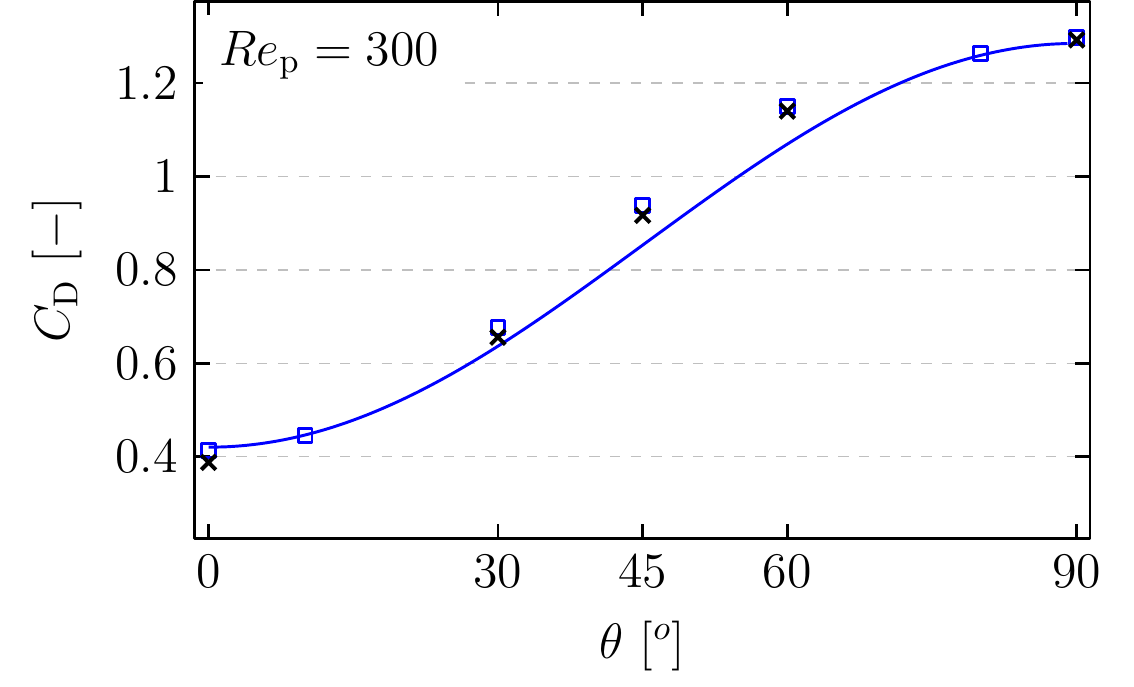}
  \includegraphics[width=0.325\columnwidth]{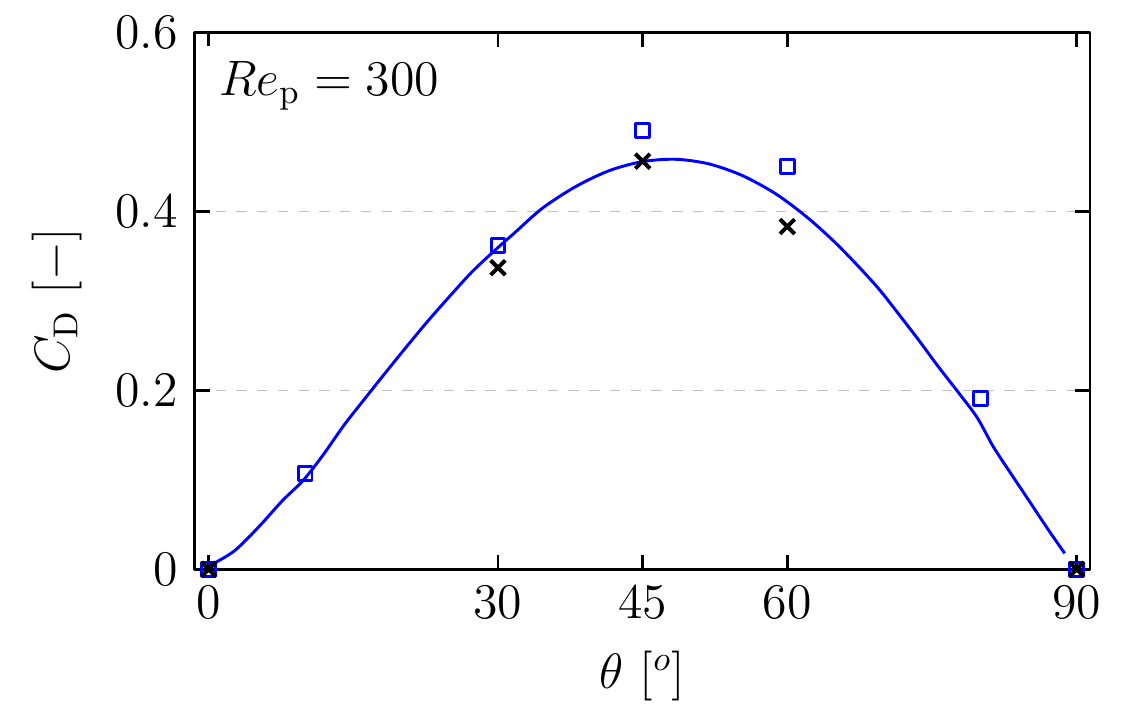}
  \includegraphics[width=0.325\columnwidth]{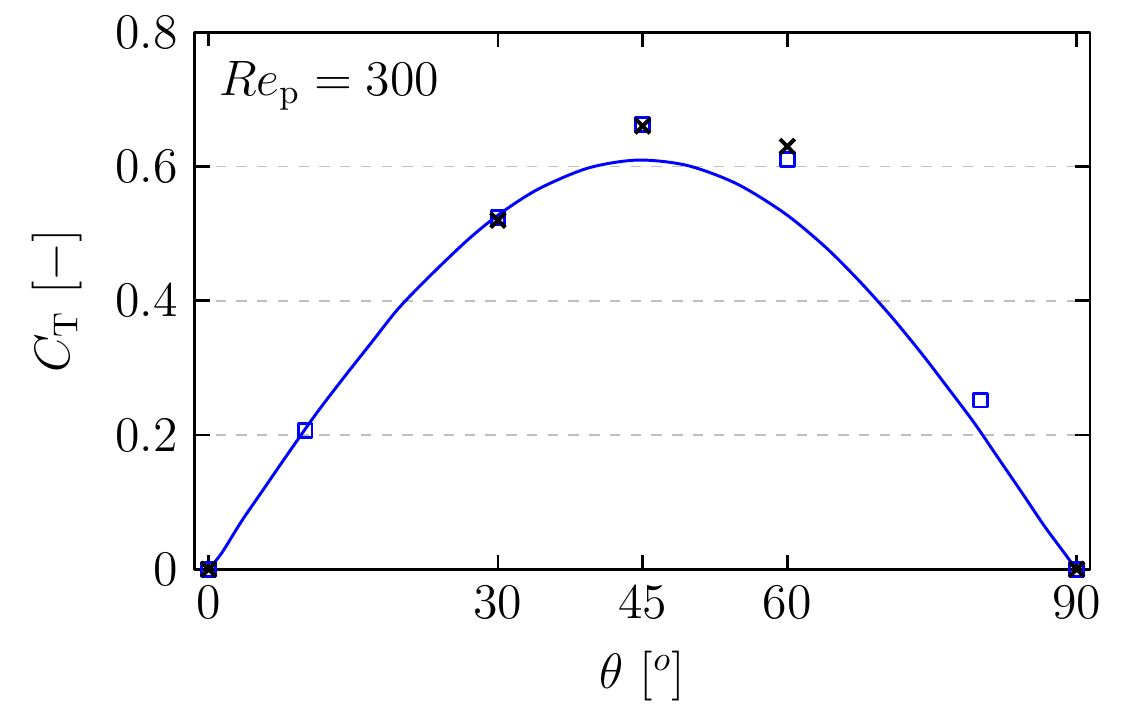}
      \caption{The drag coefficient \cd (left), the lift coefficient \cl (center), and the torque coefficient \ct (right), as a function of the orientation angle, for an axi-symmetric particle of aspect ratio $\alpha=4$ subject to a uniform flow at a particle Reynolds number of $Re_p = 300$.
      Correlation of~\citet{Sanjeevi2018}: solid line, DNS of~\citet{Sanjeevi2018}: $\color{blue}{\square}$.
      Current work $D_\text{eq}/\Delta x =64$:
      {\protect\tikz{\protect\draw pic[black, rotate = 0, line width=1] {cross=3pt};}}.}\label{fig:Convergence-Sanjeevi}
  \end{figure}

In this section, the spatial convergence of the simulation setup is studied.
To achieve this, DNS of the uniform fluid flow configuration past an axi-symmetric particle at a particle Reynolds number of \Rep$=0.1$ and $300$, are performed.
Three numerical resolutions are studied, $D_\textup{eq}/\Delta x = 16, 32$ and $64$, with $\Delta x$ being the fluid mesh spacing.
{At particle Reynolds number \Rep$=300$, an additional numerical resolution is simulated, $D_\textup{eq}/\Delta x = 128$.}
The simulations are performed for the following orientation angles: $\theta = 0, 30, 45, 60$ and $90^\textup{o}$.
In the viscous regime, the results obtained for the drag and the shape-induced lift forces are compared to the analytical solution of~\citet{Happel1981}.
\reviewerII{For a particle Reynolds number of \Rep = 300, the results of~\citet{Zastawny2012c}, who study a rod-like particle with an aspect ratio of $\alpha = 5$ with a numerical resolution of~$D_\text{eq}/\Delta x = 12$, are used as reference results for the convergence study. The results of~\citet{Zastawny2012c} have also been confirmed and used by~\citet{Sanjeevi2018}.}\\
\\
In the viscous regime, the DNS results obtained at $D_\textup{eq}/\Delta x = 32$ and $64$ are in good agreement with the reference results of~\citet{Happel1981}, for both the drag and the shape-induced lift forces, see figure~\ref{fig:Convergence-ViscousRegime}.
\reviewerII{Moreover, the evolution of the drag and the lift forces  as a function of the orientation angle of the particle follows the sinesquare and cosine-sine profile, in which a maxima is reached at $\theta = 90^{o}$ and $\theta=45^{o}$ for the drag and lift coefficients, respectively~\citep{Happel1981}.}
\reviewerII{A maximum relative error of $C_\text{D,DNS}/C_\text{D,Stokes} = 260.485/250.503 = 3.985$\% is observed between the DNS results and the reference results of~\citet{Happel1981} for a numerical resolution of $D_\textup{eq}/\Delta x = 64$ at $\theta = 0^{o}$. This error is attributed to the finite computational domain size.}\\
\\
At high particle Reynolds number, \Rep = 300, and aspect ratio $\alpha = 5$, the spatial convergence study is performed against the DNS of~\citet{Zastawny2012c}; the results for the drag and lift coefficients are reported in figure~\citet{Zastawny2012c}.
At \Rep $=300$, the desired accuracy is reached with a mesh refinement of $D_\textup{eq}/\Delta x = 64$.
The comparison with~\citet{Zastawny2012c} indicates a good agreement for the prediction of the drag coefficient but a slight shift in the prediction of the lift coefficient.\\
\\
\reviewerII{Additional validations for the present solver are included using the converged numerical resolution, $D_\textup{eq}/\Delta x = 64$, by performing DNS of the uniform fluid flow past a particle with an aspect ratio of $\alpha=4$, at particle Reynolds number \Rep = 300, to compare with the DNS results of~\citet{Sanjeevi2018}, obtained with $20$ fluid mesh cells across the minor axis of the particle, $(2 b)/\Delta x = 20$.
The results are shown in figure~\ref{fig:Convergence-Sanjeevi} for the drag, lift and torque coefficients.
The results obtained with the present DNS solver show a very good agreement with the reference results of~\citet{Sanjeevi2018} for both the drag, the lift and the torque coefficients.}
Therefore, all the DNS at a finite particle Reynolds number are performed with the spatial resolution of $D_\textup{eq}/\Delta x = 64$.

\begin{figure}[ht]
\centering
\includegraphics[width=0.475\columnwidth]{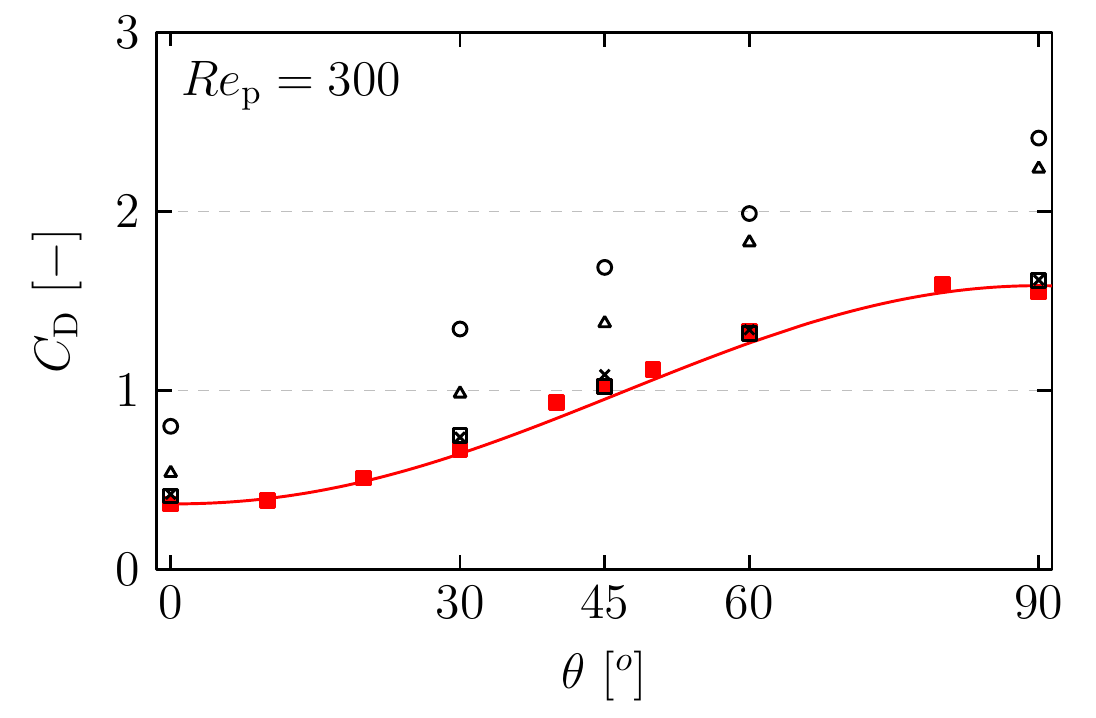}
\includegraphics[width=0.485\columnwidth]{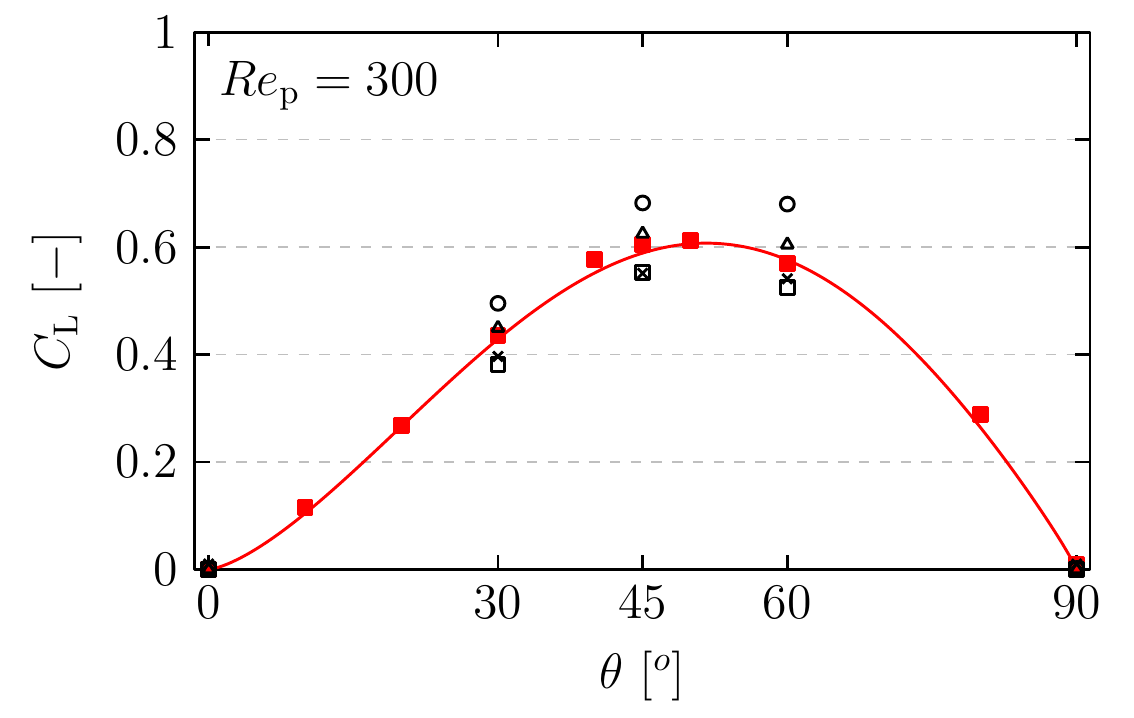}
    \caption{The drag coefficient \cd (left), and the lift coefficient \cl (right) as a function of the orientation angle, for an axi-symmetric particle of aspect ratio $\alpha=5$ subject to a uniform flow at a particle Reynolds number of $Re_p = 300$.
    Correlation of~\citet{Zastawny2012c}: solid line, DNS of~\citet{Zastawny2012c}: $\color{red}{\square}$.
    Current work, $D_\text{eq}/\Delta x =16$: \protect\tikz{\protect\draw[thick] (0,0) circle (2pt)},
    $D_\text{eq}/\Delta x =32$: $\Delta$,
    $D_\text{eq}/\Delta x =64$:
    {\protect\tikz{\protect\draw pic[black, rotate = 0, line width=1] {cross=3pt};}},
    $D_\text{eq}/\Delta x =128$: $\square$.
    }\label{fig:Convergence-HighReynolds}
\end{figure}
\section{Results\label{sec:results}}

The simulations of the flow past the axi-symmetric rod-like particles are performed to extract the drag and lift forces, and the torque.
These results provide the data to derive a model to predict the drag, lift and torque coefficients which depend on the shear rate of the flow, the particle Reynolds number, the orientation angle between the main axis of the particle and the main fluid velocity direction, and the aspect ratio of the rod-like particle.
{The fit parameters in the derived models are \reviewerII{estimated} with curve fitting algorithms based on an optimized mapping function using \reviewerII{the SciPy library~\citep{Virtanen2020}}.}

\begin{figure}[htbp!]
\centering
\includegraphics[width=0.62\columnwidth, trim={0, 100, 0, 100}, clip]{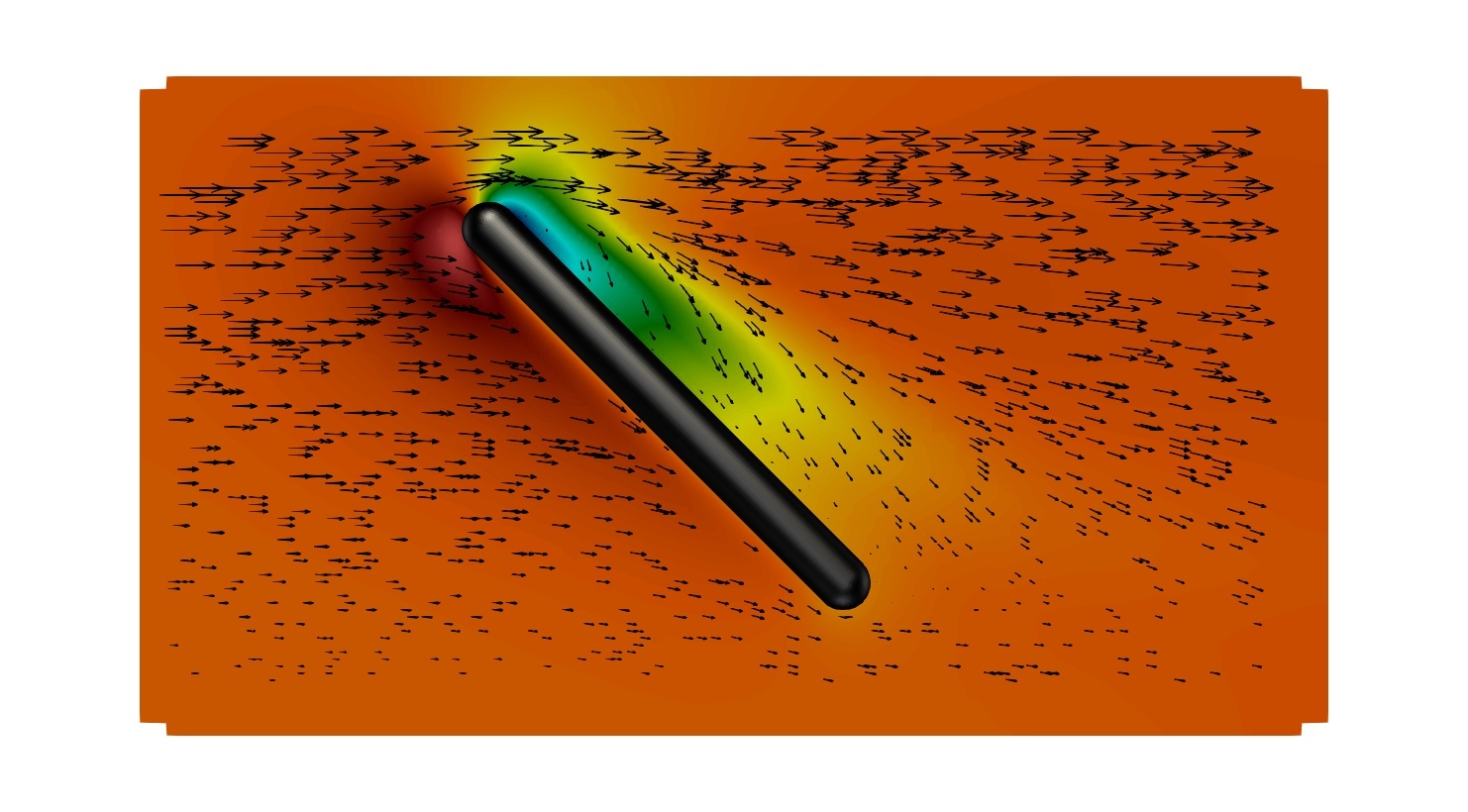}
    \caption{Snapshot of the magnitude of the velocity field (arrows) and pressure (colors), of the linear shear flow past a rod-like particle of aspect ratio $10$, at a particle Reynolds number of~\Rep $=300$, and orientation angle $\theta=135^\text{o}$.}\label{fig:FlowRendering}
\end{figure}

An illustration of the flow field around a particle of aspect ratio $10$ at a particle Reynolds number of \Rep=$300$ subject to a linear shear rate of $\tilde{G} = 0.2$ is shown in figure~\ref{fig:FlowRendering}.
{For this specific simulation, the no-slip error based on the Euclidean norm of the no-slip velocity interpolated at the location of the Lagrangian markers is below~1\%, which ensures that the fluid adapts to the presence of the particle and ensures that the boundary condition is accurately enforced on the surface of the particle.
For all simulation the no-slip error on the surface of the particle is kept below~1\%.}

\subsection{Drag coefficient\label{subsec:dragforceresults}}
\subsubsection{Results and discussion}
\begin{figure}[htbp!]
\centering
    \includegraphics[width=0.995\columnwidth]{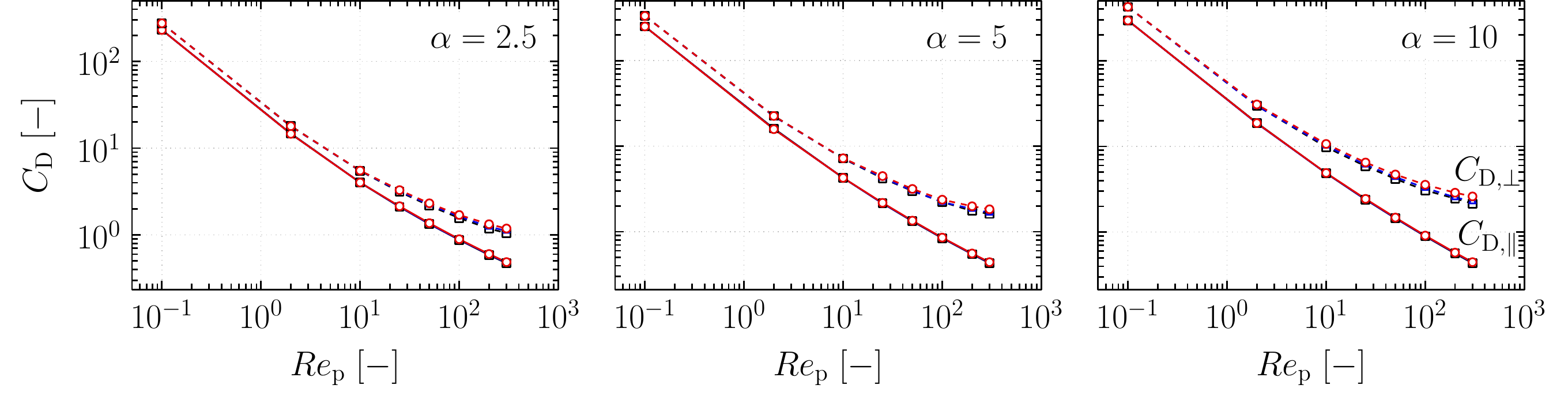}
    \caption{The drag coefficient, \cd, at the orientation angles $\theta = 0^{\text{o}}$ and $90^{\text{o}}$, as a function of the particle Reynolds number, \Rep, varying the shear rate of the flow and the aspect ratios of the particles (from the left figure to the right figure: $\alpha= 2.5$, $5$, and $10$).
    The symbols indicate the shear rate of the flow, $\tilde{G} = 0$: $\square$, $\tilde{G} = 0.1$: {$\triangle$}, $\tilde{G} = 0.2$:
\protect\tikz{\protect\draw[thick, red] (0,0) circle (3pt)}.
Solid line: $C_{\text{D},\parallel}$, dashed line: $C_{\text{D},\perp}$.
    }
    \label{fig:CDvsReynoldsCorrelation}
\end{figure}

In uniform flow, the drag coefficient of an axi-symmetric particle can be expressed for all orientation angles $\theta$ in the drag coefficients at orientation angles $\theta= 0^{\text{o}}$ and $90^{\text{o}}$, $C_{\textup{D},\parallel}$ and $C_{\textup{D},\perp}$~\citep{Frohlich2020,Sanjeevi2022}.
The evolution of the coefficients $C_{\textup{D},\parallel}$ and $C_{\textup{D},\perp}$, as a function of the particle Reynolds number, is shown in figure~\ref{fig:CDvsReynoldsCorrelation} for the three shear rates, $\tilde{G} = 0, 0.1$ and $0.2$, and all the particles studied, $\alpha = 2.5$, 5, and 10.
At an orientation angle of $\theta = 0^o$, the results for all the cases overlap.
At an orientation angle of $\theta = 90^o$, the drag coefficients of the flow configurations at shear rates $\tilde{G} = 0.1$ and $0.2$ are larger than for the uniform flow configuration, for intermediate to high particle Reynolds number.
{The major increase in the drag coefficient at an orientation angle of~$\theta = 90^o$ as compared to $\theta = 0^o$, in case of a linear shear flow, is caused by the larger portion of area exposed to the shearing of the flow in the direction normal to the main local fluid flow.}
This increase is more significant for the most elongated particle, hence the larger in the aspect ratio of the particle, the larger is the increase in the drag force of the particle in case of a linear shear flow compared to uniform flow.

\begin{figure}[htbp!]
\centering
    \includegraphics[width=\columnwidth, trim={30 0 0 0}, clip]{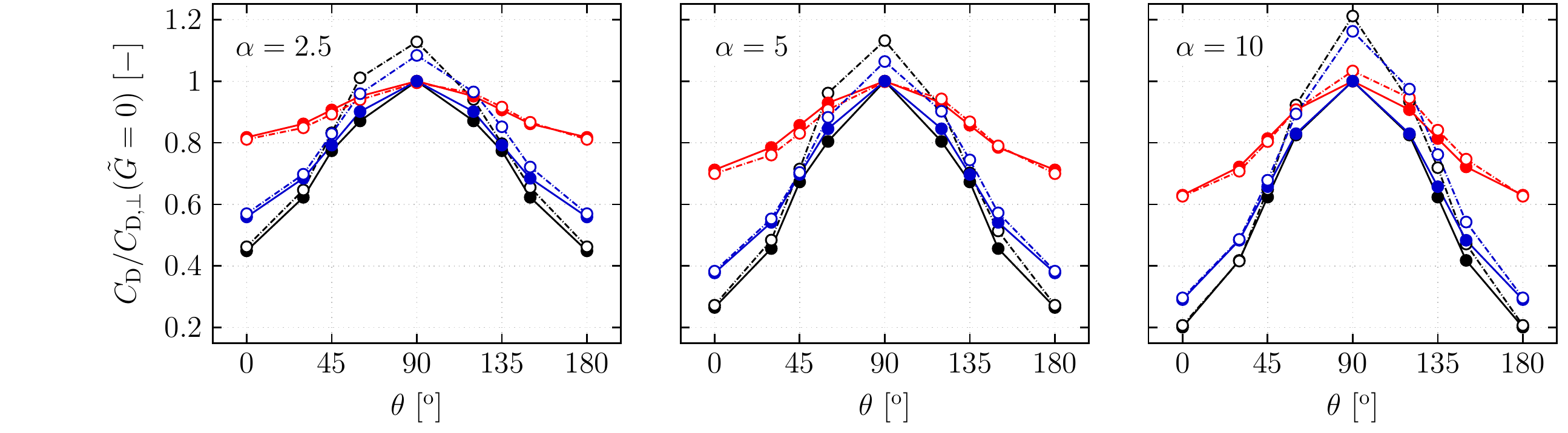}
    \caption{The drag coefficient, $C_\text{D}$, scaled by the maximum uniform flow drag coefficient $C_{\textup{D},\perp}(\tilde{G}=0)$, as a function of the orientation angle $\theta$.
    Three particle Reynolds numbers are shown, \Rep = 2: {red}, \Rep = 100: {blue}, \Rep = 300: black.
    The line style indicates the shear rate, $\tilde{G} = 0$: solid line {with filled markers}, $\tilde{G} = 0.2$: dashed line {with empty markers}. 
    The three particles studied are shown from the left figure to the right figure: $\alpha= 2.5$, $5$, and $10$.}
    \label{fig:CDvsRepvsTheta}
\end{figure}

In figure~\ref{fig:CDvsRepvsTheta}, the drag coefficients for the flow configurations $\tilde{G} = 0$ and $0.2$ are shown for all the studied orientation angles ($\theta = 0^{\text{o}}$ to $180^{\text{o}}$).
The results are shown for particle Reynolds number \Rep  = 2, 100 and 300.
The drag coefficients are scaled by the maximum value of the drag coefficient of the particle in a uniform flow, $C_{\textup{D},\perp}$ at $\tilde{G} = 0$.
The trends observed in figure~\ref{fig:CDvsRepvsTheta} show that at high particle Reynolds number, the drag coefficient of the particle is always increased in case of linear shear flow compared to uniform flow,
This increase is always maximum at an orientation angle of $\theta=90^{\text{o}}$.
As shown in figure~\ref{fig:CDvsReynoldsCorrelation}, the magnitude of the increase also depends on the aspect ratio of the particle.
For instance, at \Rep = 300, the drag coefficient of the particle of aspect ratio $\alpha=2.5$ increases by a factor of $C_{\textup{D},\perp}(\tilde{G}=0.2)/C_{\textup{D},\perp}(\tilde{G}=0) = 1.12$ in case of local shear flow compared to local uniform flow, this increase reaches a factor of $1.22$ for the most elongated particle.

When a particle is in a shear flow, the profile of the dependency of the drag force as a function of the orientation angle changes.
The profile is no longer symmetrical around $\theta = 90^{\text{o}}$.

This can be illustrated by considering the DNS results at for all the particles studied.
For these flow regimes, the increase of the drag coefficient of the particle in case of linear shear flow is always larger in the range of orientation angles varying from $\theta = 90^{\text{o}}$ to $180^{\text{o}}$ than for the orientation angles varying from $\theta = 0^{\text{o}}$ to $90^{\text{o}}$.
At particle Reynolds number \Rep = 300 and aspect ratios $\alpha = 2.5$ and 5, however, the increase of the drag coefficient of the particle in case of linear shear flow is larger in the range of orientation angles varying from $\theta = 0^{\text{o}}$ to $90^{\text{o}}$ than for the orientation angles varying from $\theta = 90^{\text{o}}$ to $180^{\text{o}}$.
For the most elongated particle, however, the major contribution to the drag coefficient remains in the range of orientation angles $\theta = 90^{\text{o}}$ to $180^{\text{o}}$.
Hence, the combination of the shear rate of the flow and the aspect ratio of the particle modifies both the profile of the drag coefficient as a function of the orientation angle, and the maximum value of the drag coefficient.


\subsubsection{Correlation for the drag coefficient}

The model to predict the drag coefficient as a function of the orientation of the particle, the particle Reynolds number, the aspect ratio of the particle, and the shear rate of the fluid, is divided into a term accounting for the uniform flow, and into a term describing the change in the drag force caused by the local shear flow compared to the local uniform flow.
The general correlation is given by
\begin{equation}\label{eq:CD-GeneralExpression}
C_\textup{D}(Re_\textup{p}, \theta, \alpha, \tilde{G}) = C_\text{D}(Re_\textup{p}, \theta, \alpha) + C_{\textup{D},\tilde{G}}(Re_\textup{p}, \theta, \alpha, \tilde{G})\, ,
\end{equation}
with $C_{\textup{D}}(Re_\textup{p}, \theta, \alpha)$ the drag coefficient of the particle in a uniform flow, and $C_{\textup{D},\tilde{G}}(Re_\textup{p}, \theta, \alpha, \tilde{G})$ the drag coefficient to account for the change in the drag caused by the linear shear flow compared to the uniform flow.
The derivation of the correlation for the drag coefficient considering uniform flow follows the work of~\citet{Frohlich2020} and~\citet{Sanjeevi2022}, who propose a correlation based on the sinesquare law, Eq.~\eqref{eq:sinesquarelaw}.
\reviewerII{In this work, the drag coefficient of a spherical particle, given by the correlation of~\citet{Schiller1933}, is included to the set of data for all the particle Reynolds number studied. This provides a lower bound to the correlation when $\alpha \rightarrow 1$.}
\reviewerII{However, the accuracy of the proposed correlations for both the uniform drag coefficient and the change in the drag coefficient in case of linear shear flow is calculated only in the range between $\alpha = 2.5$ to $\alpha = 10$, and particle Reynolds number ranging from \Rep = 0.1 to \Rep = 300.}
\textbf{VC come back why against answer to reviewer ?}
The correlation is given by
\begin{align}\label{eq:CD-uniform-fit}
& C_{\textup{D}}(Re_\textup{p}, \theta, \alpha) =  C_{\text{D},\parallel} + \left[ C_{\text{D},\perp} - C_{\text{D},\parallel}\right]\sin(\theta)^2
\end{align}
where $C_{\text{D},\parallel}$ and $C_{\text{D},\perp}$ are given by
\reviewerII{
\begin{align}
& C_{\text{D},\parallel} = \left[C_{\text{D},\text{Stokes},\parallel} + \cfrac{c_{\text{d},1} + c_{\text{d},2} (\alpha - 1)^{c_{\text{d},3}}}{Re_\text{p}^{\beta_{\parallel}}} \right]\exp{(- c_{\text{d},7}Re_\text{p}^{c_{\text{d},8}})}\, ,\\
& C_{\text{D},\perp} = \left[C_{\text{D},\text{Stokes},\perp} + \cfrac{c_{\text{d},1} + c_{\text{d},2} (\alpha - 1)^{c_{\text{d},3}}}{Re_\text{p}^{\beta_{\perp}}} \right]\exp{(- c_{\text{d},7}\text{\Rep}^{c_{\text{d},8}})}\, ,
\end{align}
}
{with
\begin{align}
\beta_{\parallel} =& c_{\text{d},4} + c_{\text{d},5} (\alpha - 1)^{c_{\text{d},6}}\, ,\\
\beta_{\perp}=& c_{\text{d},4} + c_{\text{d},5} (\alpha - 1)^{c_{\text{d},6}}\, ,
\end{align}
where $c_{\text{d},\text{i}}$ are the fit parameters to predict $C_{\text{D},\parallel}$ and $C_{\text{D},\perp}$}, and $C_{\text{D,Stokes},\parallel}$ and $C_{\text{D,Stokes},\perp}$ are the geometric drag coefficients, defined in~\citet{Happel1981} as
\reviewerII{
\begin{align}\label{eq:Jeffery-Orbits}
C_{\text{D,Stokes},\parallel} & = \cfrac{64}{Re_\text{p}\alpha^{1/3}}
\cfrac{1}
{\cfrac{-2 \alpha}{\alpha^2-1}+
\cfrac{2\alpha^2 - 1}{\left(\alpha^2-1\right)^{3/2}}   
\ln\left(\cfrac{\alpha + \sqrt{\alpha^2-1}}{\alpha - \sqrt{\alpha^2 - 1}}\right)}\, ,\\
C_{\text{D,Stokes},\perp} & = \cfrac{64}{Re_\text{p}\alpha^{1/3}}
\cfrac{1}
{\cfrac{\alpha}{\alpha^2-1}+
\cfrac{2\alpha^2 - 3}{\left(\alpha^2-1\right)^{3/2}}
\ln\left({\alpha + \sqrt{\alpha^2-1}}\right)}\, .
\end{align}}

The fit parameters in Eq.~\eqref{eq:CD-uniform-fit}, for both the parallel and perpendicular alignment with the main locally uniform flow, are listed in table~\ref{table:dragcoefficients-uniformflow}.
They ensure the accurate estimation of the drag coefficient from \reviewerII{a particle Reynolds number of \Rep = 0.1 toward a particle Reynolds number of \Rep = 300, for a particle of aspect ratio ranging from $\alpha = 2.5$ to $\alpha = 10$.}
The maximum, mean and median relative differences between the model fit and the DNS are of~$6.17$\%, $1.76$\% and $1.40$\%, respectively.

\begin{table}
\centering
\begin{tabular}{c | c c c c c c c c}
& $c_\textup{d,1}$ & $c_\textup{d,2}$ & $c_\textup{d,3}$ & $c_\textup{d,4}$ & $c_\textup{d,5}$ & $c_\textup{d,6}$ & $c_\textup{d,7}$ & $c_\textup{d,8}$\\
\hline
\hline
$C_{\textup{D},\parallel}$ & 2.41 & 2.52 & 0.124 & 13.4 & -12.9 & 1.95 $\times10^{-3}$ & 4.71 $\times10^{-2}$ & -0.233\\
$C_{\textup{D},\perp}$ & 2.68 & 1.35 & 0.828  & 0.265 & 4.46 $\times10^{-4}$ & 1.90 & 6.23 $\times10^{-6}$ & -3.59\\
\end{tabular}
     \caption{List of the fit parameters in Eq.~\eqref{eq:CD-uniform-fit} to model the drag coefficient of the particle in a uniform flow ($C_{\textup{D}}(Re_\textup{p}, \theta, \alpha)$).}
     \label{table:dragcoefficients-uniformflow}
\end{table}

The correlation to model the change in the particle force in case of a local shear flow compared to a locally uniform flow is given by
\begin{align}\label{eq:CD-shear-fit}
C_{\textup{D},\tilde{G}}(Re_\textup{p}, \theta, \alpha, \tilde{G}) =\left\{
    \begin{array}{ll}
        0 & Re_\textup{p} < 1\, ,\\
        C_{\textup{D},\tilde{G},\alpha=1} + C_{\textup{D},\tilde{G},\alpha>1} & Re_\textup{p} \geq 1\, ,
    \end{array}
\right.
\end{align}
where the change is zero in the viscous regime~\citep{Happel1981}.
At finite particle Reynolds number, the change in the particle drag force coefficient in case of linear shear flow compared to uniform flow is divided into a term for spherical particles, $C_{\textup{D},\tilde{G},\alpha=1}$, and a term accounting for the non-sphericity of the particle $C_{\textup{D},\tilde{G},\alpha>1}$.
The term for spherical particle is given by
\begin{equation}\label{eq:CD-shear-fit-sphere}
C_{\textup{D},\tilde{G},\alpha=1} = \tilde{G} \left( 0.571 \exp{(Re_\textup{p})}\right)^{-0.0449}\, .
\end{equation}
This expression is fitted from the correlation of~\citet{Kurose1999} to predict the drag coefficient for a spherical particle subject to an unbounded linear shear flow.
{The range of validity of the correlation derived in~\citet{Kurose1999} spans from a particle Reynolds number of \Rep = 1 to 500, and a dimensionles shear rate of $\tilde{G} = 0$ to 0.4.}
The second term in Eq.~\eqref{eq:CD-shear-fit} is given by
\begin{equation}\label{eq:non-spherical-shear-drag}
C_{\textup{D},\tilde{G},\alpha>1} = \left[\left(c_{\textup{d},\tilde{G},1}\tilde{G}^{c_{\textup{d},\tilde{G},2} + c_{\textup{d},\tilde{G},3} \alpha}\right)\exp{(\text{\Rep}^{c_{\textup{d},\tilde{G},4}})}\right]\sin(\varPsi_{\textup{D},\tilde{G}})^{\beta_{D,\tilde{G}}}\, ,
\end{equation}
where the exponent of the sinusoidal term $\beta_{D,\tilde{G}}$ is given by
\begin{equation}
\beta_{D,\tilde{G}} = {2 + c_{\textup{d},\tilde{G},5}\left(c_{\textup{d},\tilde{G},6}\alpha^{c_{\textup{d},\tilde{G},7}}\right)^{-1}}\, ,
\end{equation}
and $\varPsi_{\textup{D},\tilde{G}}$ is given by
\begin{equation}\label{eq:shift-drag}
\varPsi_{\textup{D},\tilde{G}} =\pi \left(\cfrac{\theta}{\pi}\right)^{3.35 + 
\left[\left(\cfrac{\ln(\text{\Rep})}{1.93}\right)\cfrac{1}{\alpha}\right]^{0.089}}\, .
\end{equation}
The exponent on the sinusoidal term, $\beta_{D,\tilde{G}}$, and the shift function, $\varPsi_{\textup{D},\tilde{G}}$, aim to account for the influence of the aspect ratio of the particle on the change of the profile of the drag coefficient and the magnitude of the increase of the drag coefficient of the particle in case of local shear flow.
The fit parameters in Eq.~\eqref{eq:CD-shear-fit-sphere}-\eqref{eq:shift-drag}, $c_{\textup{d},\tilde{G},\text{i}}$, are given in table~\ref{table:dragcoefficients-shearrate}.

\begin{table}
\centering
\begin{tabular}{c c c c c c c}
$c_{\textup{d},\tilde{G},1}$ & $c_{\textup{d},\tilde{G},2}$ & $c_{\textup{d},\tilde{G},3}$ & $c_{\textup{d},\tilde{G},4}$ & $c_{\textup{d},\tilde{G},5}$ & $c_{\textup{d},\tilde{G},6}$ & $c_{\textup{d},\tilde{G},7}$\\
\hline
\hline
$1.21$ & $1.48$ & $-3.74 \times 10^{-2}$ & $-0.385$ & $2.75 \times 10^{-2}$ & $-9.89 \times 10^{-2}$ & $2.01$\\
\end{tabular}
     \caption{List of the fit parameters in Eq.~\eqref{eq:non-spherical-shear-drag}, used in the correlation to model the change in the drag force in case of local shear flow compared to uniform flow ($C_{\textup{D},\tilde{G}}(Re_\textup{p}, \theta, \alpha, \tilde{G})$).}\label{table:dragcoefficients-shearrate}
\end{table}

\begin{figure}[htbp!]
\includegraphics[width=0.485\columnwidth]{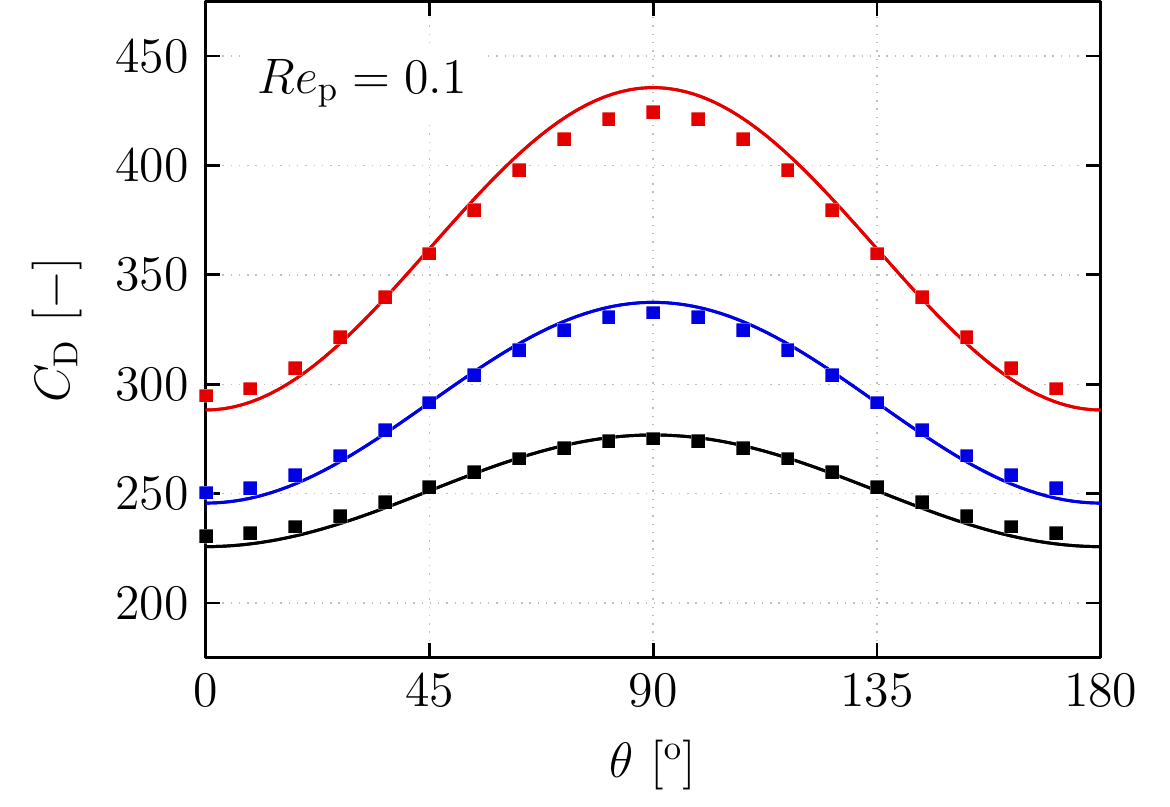}
\includegraphics[width=0.48\columnwidth]{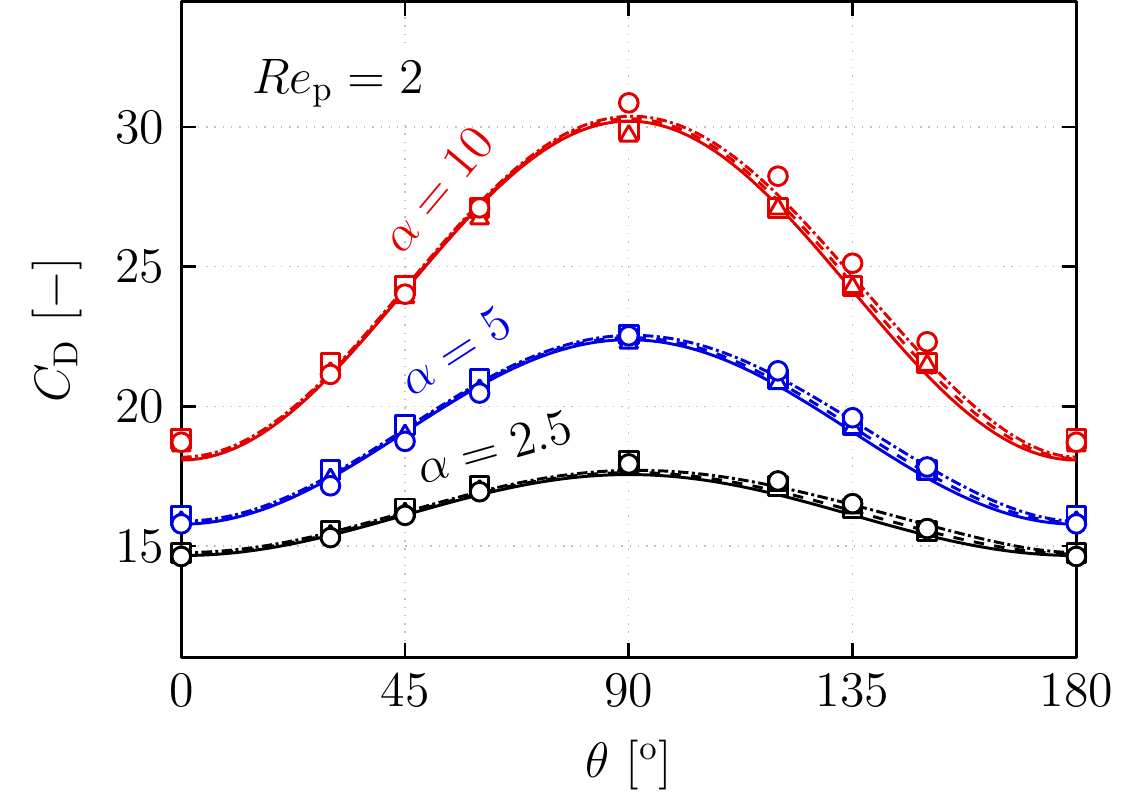}\\

\includegraphics[width=0.475\columnwidth]{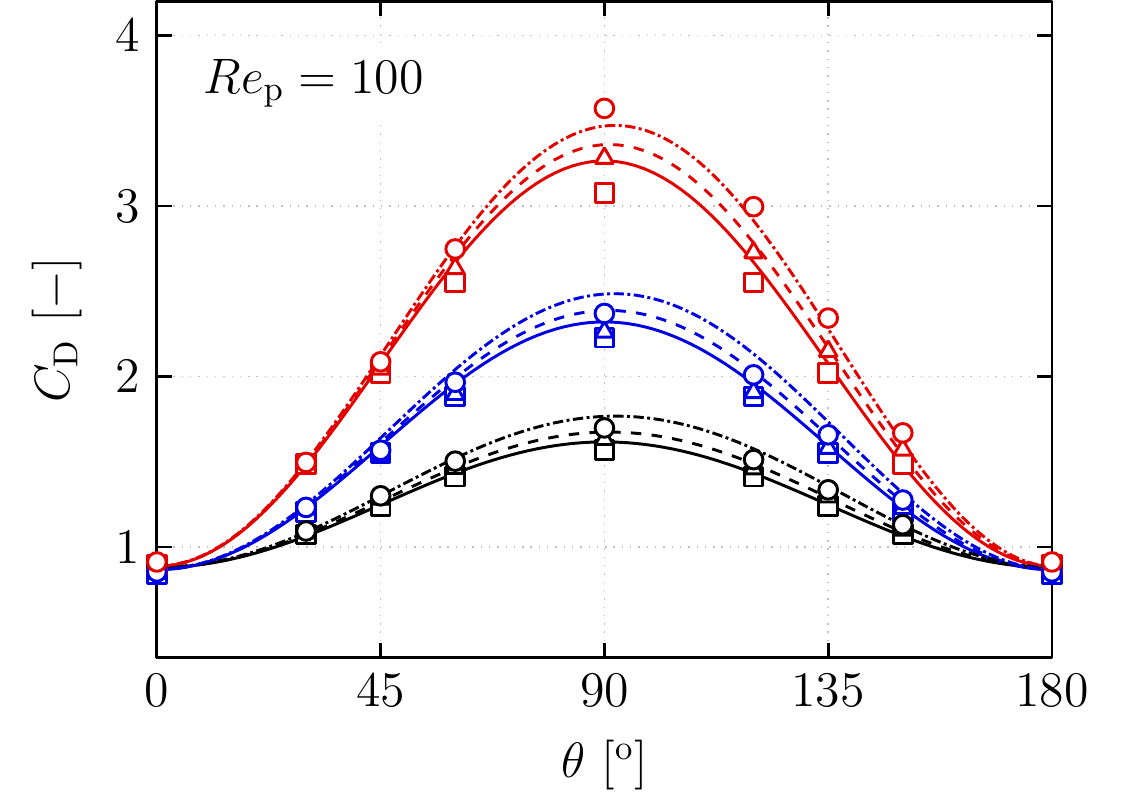}
\includegraphics[width=0.485\columnwidth]{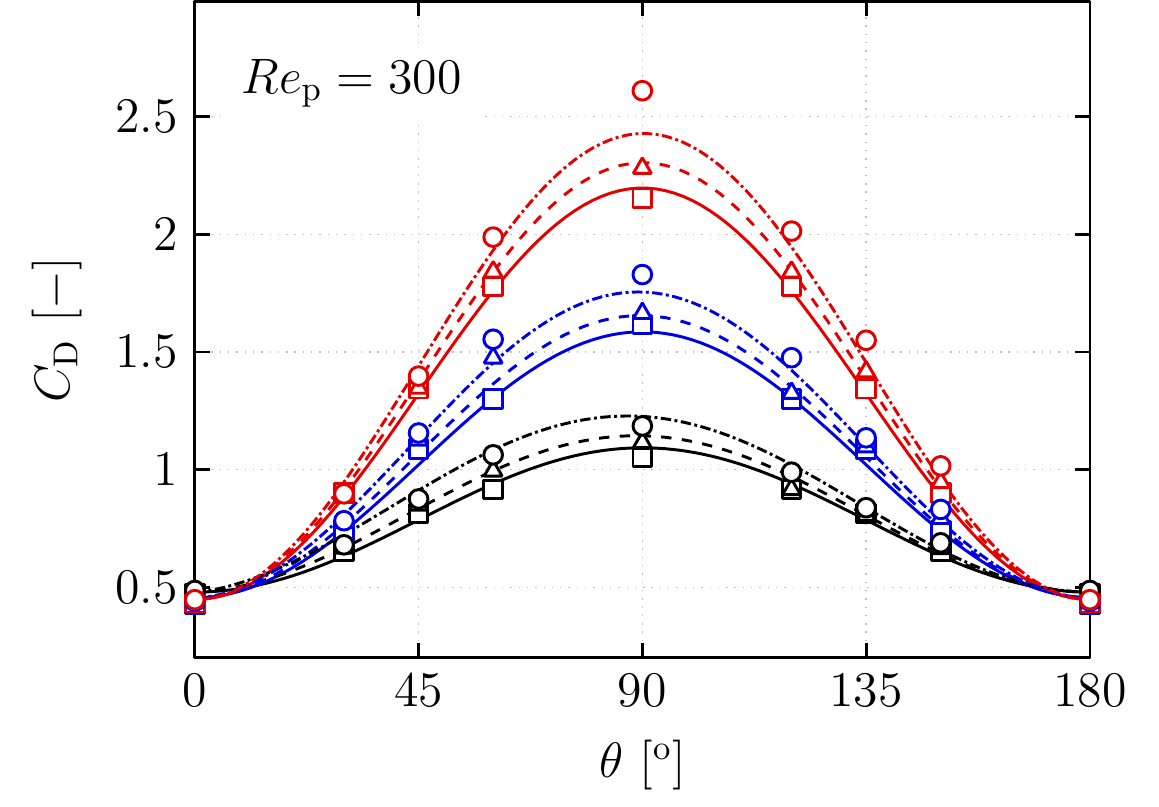}\\
    \caption{The drag coefficient, $C_\text{D}$, as a function of the orientation angle $\theta$.
    The color indicates the aspect ratio of the particle, $\alpha  =2.5$: black, $\alpha  =5$: {blue}, and $\alpha  =10$: {red}.
    The symbols represent the DNS results for the shear rate $\tilde{G} = 0$: $\square$, $\tilde{G} = 0.1$: $\triangle$, and $\tilde{G} = 0.2$: \protect\tikz{\protect\draw[thick] (0,0) circle (3pt)}.
    The correlation to model the drag coefficient $C_\text{D}$ (Eq.~\eqref{eq:CD-GeneralExpression}) is shown for all the flow regimes varying the line style, $\tilde{G} = 0$: solid line, $\tilde{G} = 0.1$: dashed line, and $\tilde{G} = 0.2$: dash dotted line.
    In the viscous regime the analytical results of~\citet{Happel1981} are shown for the uniform flow, $\tilde{G} = 0$: $\blacksquare$.}\label{fig:EvolCoeffCD}
\end{figure}

The analytical solution and the results of the DNS along with the model for the drag coefficient are shown in figure~\ref{fig:EvolCoeffCD}.
In the viscous regime, the correlation for the drag coefficient accurately recovers the analytical solution of~\citet{Brenner1961}.
For a finite Reynolds number, the correlation to predict the drag coefficient of the particle in a uniform flow accurately recovers the DNS results.

The expression derived to model the change in the drag coefficient in case of local shear flow compared to uniform flow is also in good agreement with the DNS results.
For instance, the shift of the profile of the drag coefficient, shown in figure~\ref{fig:CDvsRepvsTheta}, is accurately recovered.
The maximum, mean and median relative differences between the drag coefficient model and the DNS are of~$6.75$\%, $1.67$\%, and $1.35$\%, respectively.

\subsection{Lift coefficient\label{subsec:liftforceresults}}
\subsubsection{Results and discussion}

\begin{figure}[htbp!]
\centering
    \includegraphics[width=0.6\columnwidth, trim={0 0 0 0}, clip]{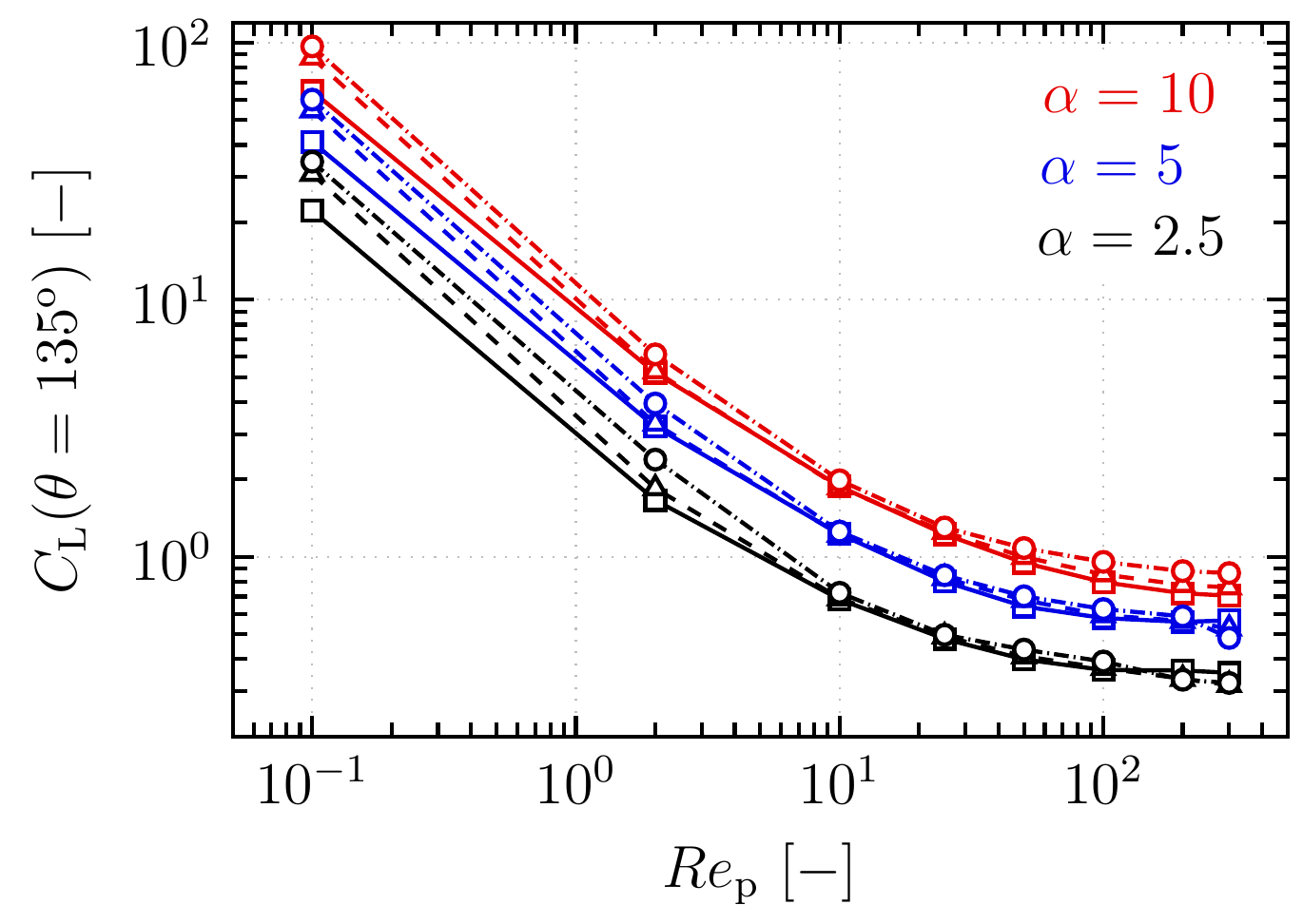}
    \caption{The lift coefficient, \cl, at an orientation angle of $\theta = 135^\text{o}$, as a function of the particle Reynolds number, \Rep, for all particles and shear rate studied.
    The color indicates the aspect ratio of the particle, $\alpha  =2.5$: black, $\alpha  =5$: {blue}, and $\alpha  =10$: {red}.
    The line style indicates the shear rate $\tilde{G} = 0$: solid line, $\tilde{G} = 0.1$: dashed line, $\tilde{G} = 0.2$: dash dotted line.}
    \label{fig:CLvsReynoldsCorrelation}
\end{figure}

In the viscous regime, the profile of the lift force of the particle in a uniform flow as a function of the orientation angle exhibits a cosine-sine profile, with absolute maxima at orientation angles of $\theta = 45^{\text{o}}$ and $135^{\text{o}}$, \reviewerII{see figure~\ref{fig:Convergence-ViscousRegime}} and~\citet{Happel1981}.
Hence, the lift coefficients at the orientation angle $\theta = 135^{\text{o}}$ are shown in figure~\ref{fig:CLvsReynoldsCorrelation} as a function of the particle Reynolds number for all the cases studied.
In the viscous regime, the lift force coefficient of the particle is always positively increased in case of local shear flow compared to uniform flow.
These differences between the lift coefficients reduce with increasing particle Reynolds number for all cases.
At a particle Reynolds number of \Rep = 200 and aspect ratio of $\alpha = 2.5$, the particle lift force is negatively increased in case of local shear flow compared to uniform flow.
This phenomenon is also observed at a particle Reynolds number of \Rep = 300 for the particle of aspect ratio $\alpha = 5$.
On the other hand, the lift force of the most elongated particle is always positively increased in case of a local shear flow compared to a locally uniform flow for this specific orientation angle.

\begin{figure}[htbp!]
\includegraphics[width=0.49\columnwidth, trim={0 25 200 90}, clip]{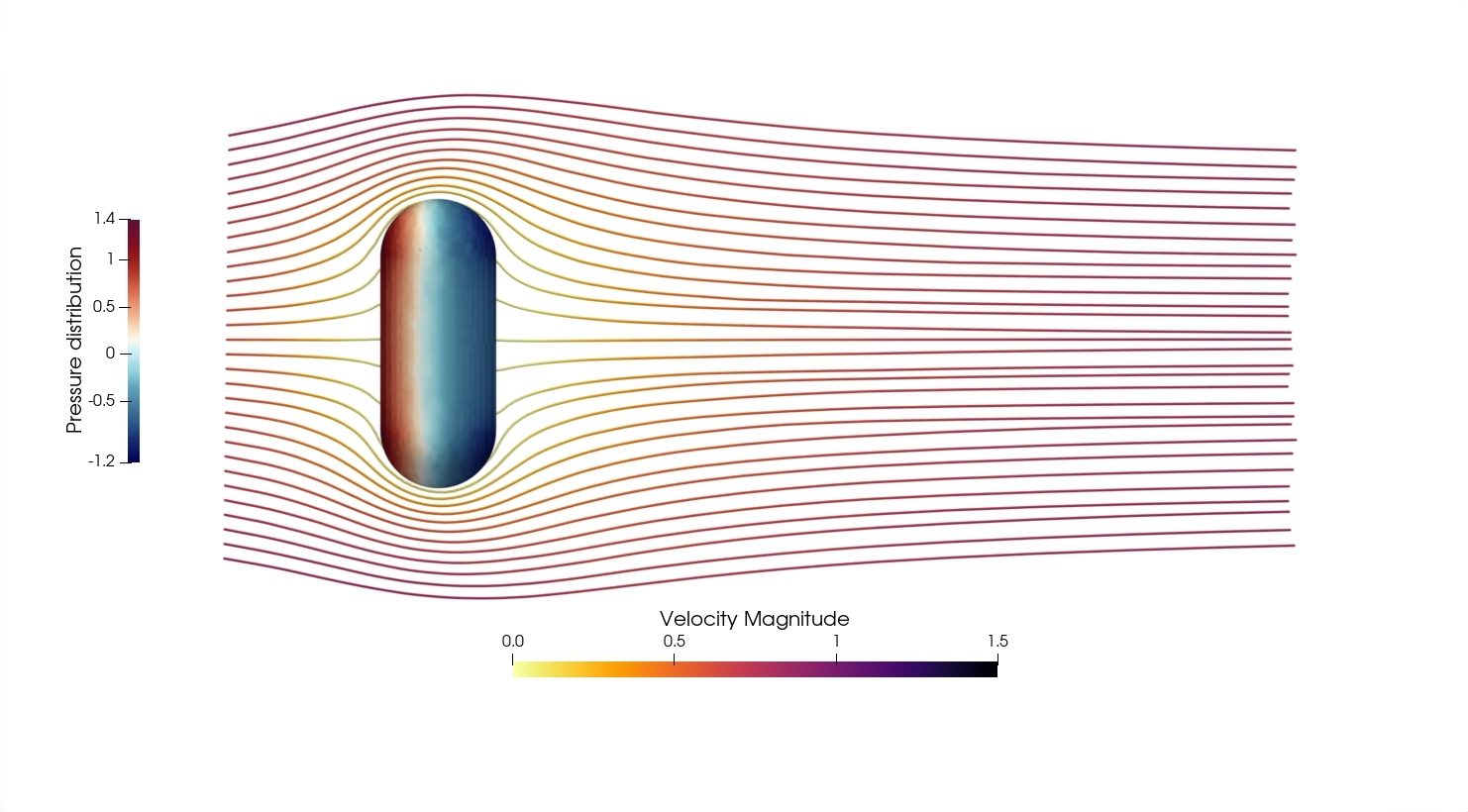}\hfill
\includegraphics[width=0.49\columnwidth, trim={0 25 200 90}, clip]{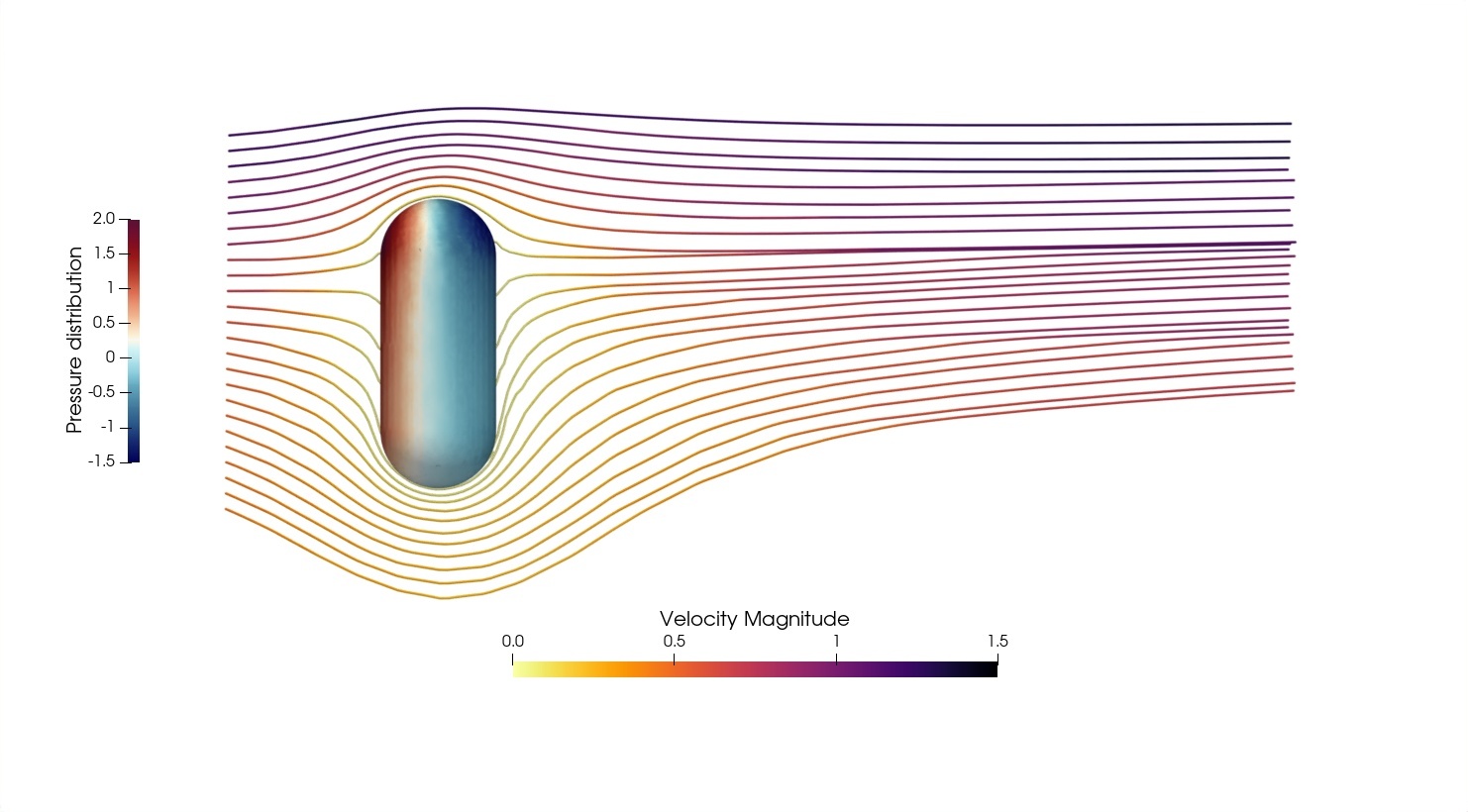}\\
\includegraphics[width=0.49\columnwidth, trim={0 25 200 90}, clip]{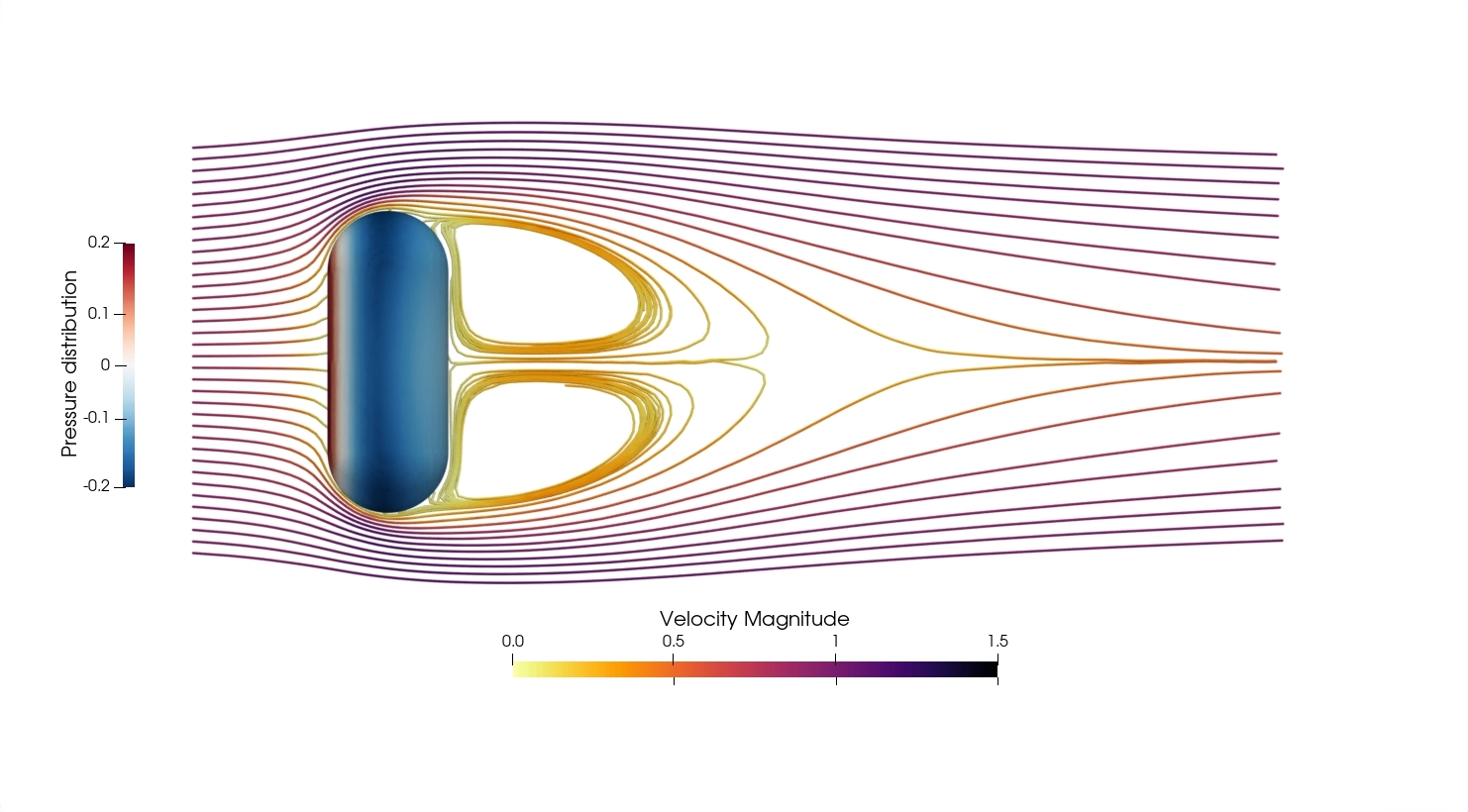}\hfill
\includegraphics[width=0.49\columnwidth, trim={0 25 200 90}, clip]{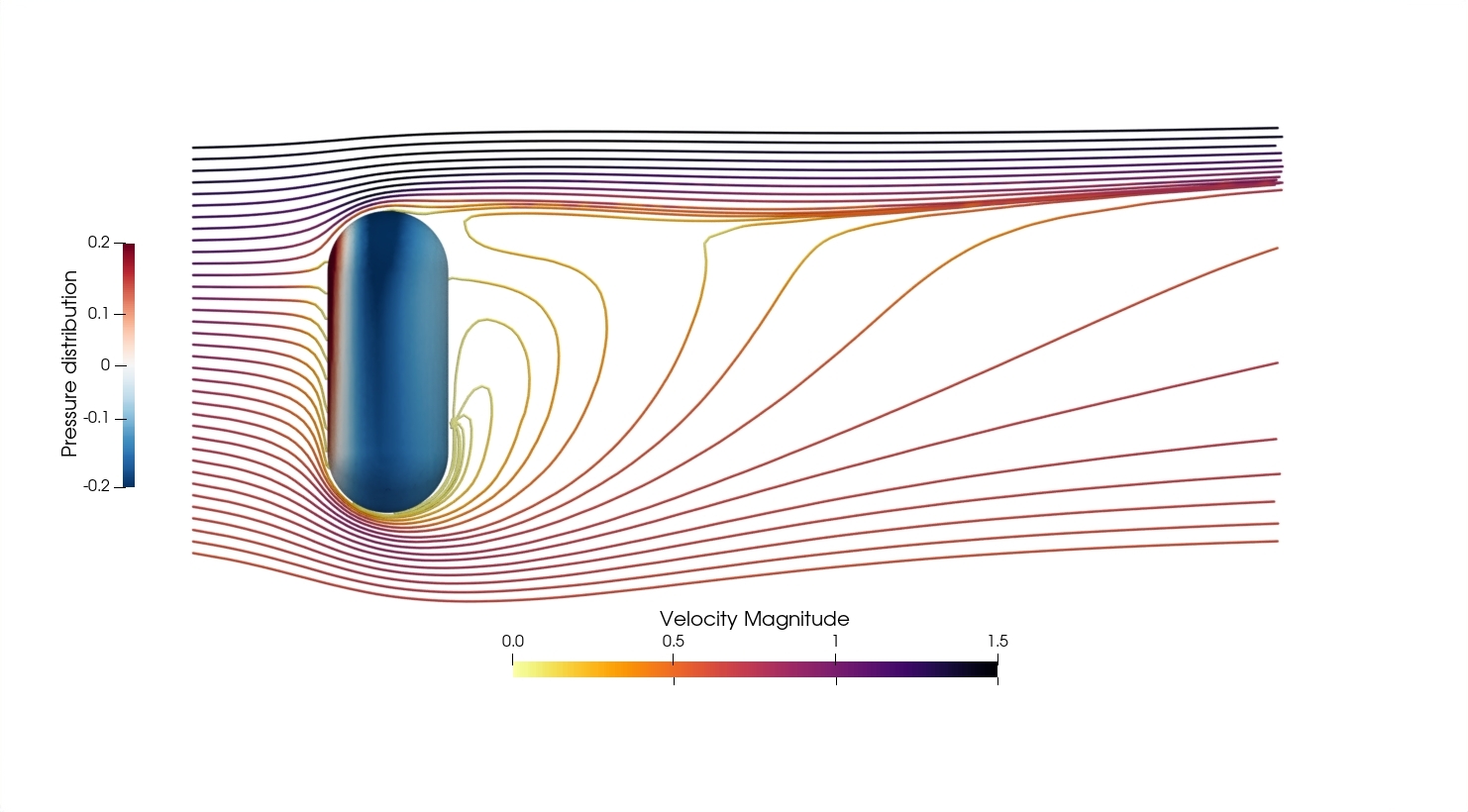}
    \caption{The pressure on the surface of the particle and the streamlines of the fluid velocity, colored by its magnitude, past a particle of aspect ratio $\alpha=2.5$, at a particle Reynolds number of \Rep = 2 (top row), and \Rep = 200 (bottom row), for uniform flow configuration (left column), and linear flow (right column).}\label{fig:StreamlinesParticles}
\end{figure}

The negative increase in the particle lift force in case of a local shear flow compared to a locally uniform flow is also observed for spherical particles~\citep{Kurose1999,Bagchi2002}, and slightly elongated prolate spheroid~\citep{Holzer2009}.
In the work of~\citet{Kurose1999}, the authors analyze the pressure and the viscous force distribution on the surface of the particle, and point out that the negative increase in the lift force in case of a local shear flow is caused by a modification of the flow recirculation in the wake of the spherical particle.
In figure~\ref{fig:StreamlinesParticles}, the streamlines of the fluid velocity past a particle of aspect ratio $\alpha = 2.5$ are shown for a particle Reynolds number of \Rep = 2 (top row), and \Rep = 200 (bottom row), and with shear rates $\tilde{G}=0$ (left column), and $\tilde{G}=0.2$ (right column).
The pressure distribution on the surface of the particle is also shown for all cases.
For both uniform flow results (left figures in figure~\ref{fig:StreamlinesParticles}), the motion of the fluid around the particle is symmetric around the particle.
The distribution of the pressure on the surface of the particle is also symmetric, thus the lift coefficient is \reviewerII{equal to 0}.

For both results with a local shear flow (right figures in figure~\ref{fig:StreamlinesParticles}), the distribution of the pressure on the surface of the particle, and the streamlines are no longer symmetric.
This results in a change in the lift force of the particle in case of local shear compared to uniform flow.
This change depends on the particle Reynolds number, for instance at a particle Reynolds number of \Rep = 2 the lift force of the particle is positively increased.
At a particle Reynolds number of~\Rep = 200, however, the specific flow recirculation in the wake of the particle yields to a negative increase of the particle lift force in case of linear shear flow.

\begin{figure}[htbp!]
\centering
    \includegraphics[width=\columnwidth, trim={40 0 0 0}]{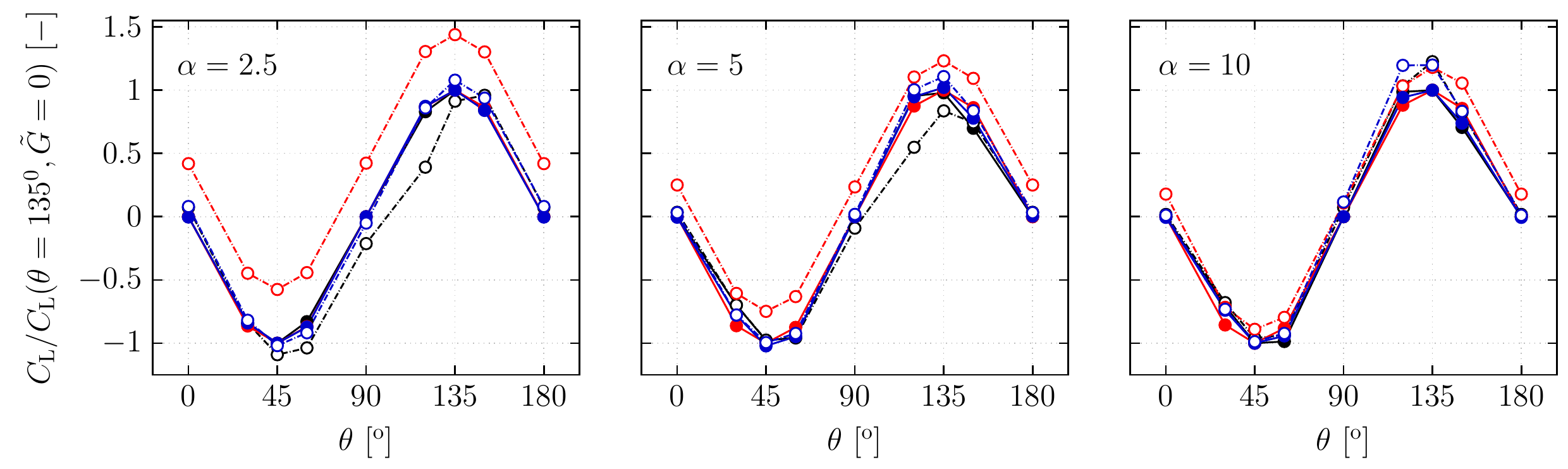}
    \caption{The lift coefficient, $C_\text{L}$, scaled by the lift coefficient of the particle in a uniform flow at an orientation angle of $\theta = 135^{\text{o}}$ $C_{\textup{L}}(\theta=135^\text{o},\tilde{G}=0)$, as a function of the orientation angle $\theta$.
    The color indicates the particle Reynolds numbers, \Rep = 2: {red}, \Rep = 100: {blue}, and \Rep = 300: black.
    The line style indicates the shear rate $\tilde{G} = 0$: solid line {with filled markers}, $\tilde{G} = 0.2$: dashed line {with empty markers}.
    The particles are shown from the left figure to the right figure: $\alpha= 2.5$, $5$, and $10$.}
    \label{fig:CLvsRepvsTheta}
\end{figure}

The profile of the lift coefficients as a function of the orientation angle, scaled by the lift coefficient of the particle in a uniform flow at an orientation angle of $\theta=135^\text{o}$, is shown in figure~\ref{fig:CLvsRepvsTheta} for the flow configurations with shear rates of $\tilde{G} = 0$ and $0.2$.
The results are shown at a particle Reynolds number of \Rep  = 2, 100 and 300.
At \Rep = 2, the lift force is increased in case of a local shear flow compared to a local uniform flow for all the particles and orientation angles.
The most significant change in the lift force is observed for the particle of aspect ratio $\alpha = 2.5$.
With the increase of the particle Reynolds number the change in the lift force coefficient of the particle is less significant, and almost reduces to 0 for the particle of aspect ratio $\alpha = 2.5$ at a particle Reynolds number of \Rep = 100.
At particle Reynolds number \Rep = 300, the lift force of the particle is negatively increased in case of local shear flow compared to uniform flow in the range of orientation angles $\theta =45^\text{o}$ to $150^\text{o}$, and $\theta =60^\text{o}$ to $135^\text{o}$, for the particles of aspect ratios $\alpha = 2.5$ and $5$, respectively.
This specific behavior of the lift force at a high particle Reynolds number in case of linear shear flow is also observed by~\citet{Holzer2009} for a prolate ellipsoid particle with an aspect ratio $\alpha=1.5$.
For the particle of aspect ratio $\alpha = 10$, the particle lift force is always positively increased in case of a local shear flow compared to a locally uniform flow, hence showing the interplay of parameters between the aspect ratio of the particle and the particle Reynolds number on the change in the values, the profile and the sign of the lift force coefficients in case of a local shear flow.

\subsubsection{Correlation for the lift coefficient}

Similarly to the correlation for the drag coefficient, the expression to predict the lift coefficient is divided into a term accounting for the local uniform flow and a term accounting for the local shear flow:
\begin{equation}\label{eq:CL-GeneralExpression}
C_\textup{L}(Re_\textup{p}, \theta, \alpha, \tilde{G}) = C_\text{L}(Re_\textup{p}, \theta, \alpha) + C_{\textup{L},\tilde{G}}(Re_\textup{p}, \theta, \alpha, \tilde{G})\, ,
\end{equation}
where $C_{\textup{L}}(Re_\textup{p}, \theta, \alpha)$ is the lift coefficient of the particle in a uniform flow, and $C_{\textup{L},\tilde{G}}(Re_\textup{p}, \theta, \alpha, \tilde{G})$ is the lift coefficient to account for the change in the lift coefficient in case of linear shear flow compared to uniform flow.
\reviewerII{The correlations for both the uniform lift coefficient and the change in the lift coefficient in case of linear shear flow are both valid in the range $\alpha = 2.5$ to $\alpha = 10$, and particle Reynolds number ranging from \Rep = 0.1 to \Rep = 300.}
Similarly to the drag coefficient, the derivation of the correlation to fit the lift coefficient for the locally uniform flow follows the work of~\citet{Frohlich2020} and~\citet{Sanjeevi2022}
\begin{align}\label{eq:CL-uniform-fit}
C_{\textup{L}}(Re_\textup{p}, \theta, \alpha) = &\left[\left(C_{\text{L},\text{Stokes}} +
\cfrac{c_{\text{l},1}(\alpha -1)^{c_{\text{l},2}}}{c_{\text{l},3}\text{\Rep}^{c_{\text{l},4} + c_{\text{l},5}(\alpha - 1)^{c_{\text{l},6}}}}\right)\left(1+\exp{(-c_{\text{l},7}\text{\Rep}^{c_{\text{l},8}})}\right)\right]\notag\\
& \cos(\varPsi_{\text{L}})\sin(-\varPsi_{\text{L}})\, ,\\
C_{\text{L},\text{Stokes}} = & \reviewerII{ C_{\text{D},\text{Stokes},\perp} - C_{\text{D},\text{Stokes},\parallel}}\, ,
\end{align}
with $C_{\text{L},\text{Stokes}}$ the lift coefficient in the Stokes regime, $c_{\text{l},i}$ are the fit parameters, and $\varPsi_{\text{L}}$ is the shift function accounting for the finite particle Reynolds number effects~\citep{Frohlich2020,Sanjeevi2022}
\begin{equation}
\varPsi_\text{L} = 
=\left\{
    \begin{array}{ll}
        \cfrac{\pi}{2} \left(\cfrac{\theta}{\nicefrac{\pi}{2}}\right)^{\text{f}_{\text{L,shift}}} & \theta < \cfrac{\pi}{2}\, ,\\
         - \cfrac{\pi}{2} \left(\cfrac{\pi - \theta}{\nicefrac{\pi}{2}}\right)^{\text{f}_{\text{L,shift}}} &  \theta \geq \cfrac{\pi}{2}\, ,
    \end{array}
\right.
\end{equation}
and $\text{f}_{\text{L,shift}}$ the so-called shifting term
\begin{align}\label{eq:CL-shifting-term}
\text{f}_{\text{L,shift}} =\left\{
    \begin{array}{ll}
        1 & Re_\textup{p} < 1\, ,\\
        1 + 9.57 \times 10^{-3} \ln{\left(\alpha\right)}^{1.70} \ln{\left(\text{\Rep}\right)}^{1.25} & \text{\Rep $\geq$ 1}\, .
    \end{array}
\right.
\end{align}
\reviewerII{The shifting term is set to 1 for particle Reynolds number values lower than $Re_\textup{p} < 1$ in order to conserve the cosine-sine profile of the lift coefficient in the viscous regime~\citep{Happel1981}.}
The fit parameters in Eq.~\eqref{eq:CL-uniform-fit} are listed in table~\ref{table:liftcoefficients-uniform}.
The maximum, mean and median relative differences between the simulations results and the model fit are of~$6.39$\%, $1.88$\%, and $1.35$\%, respectively.
\begin{table}
\centering
\begin{tabular}{c c c c c c c c}
$c_{\textup{l},1}$ & $c_{\textup{l},2}$ & $c_{\textup{l},3}$ & $c_{\textup{l},4}$ & $c_{\textup{l},5}$ & $c_{\textup{l},6}$ & $c_{\textup{l},7}$ & $c_{\textup{l},8}$\\
\hline
\hline
$87.0$ & $0.640$& 84.1 & 0.132& 0.029 &0.580& 68.8&  -0.711\\
\end{tabular}
     \caption{List of the fit parameters in Eq.~\eqref{eq:CL-uniform-fit} to model the uniform lift coefficient correlation ($C_{\textup{L}}(Re_\textup{p}, \theta, \alpha)$).}\label{table:liftcoefficients-uniform}
\end{table}

The correlation to model the change in the lift force coefficient in case of linear shear flow compared to uniform flow, see Eq.~\eqref{eq:CL-GeneralExpression}, is given by
\begin{align}\label{eq:CL-shear-fit}
C_{\textup{L},\tilde{G}}(Re_\textup{p}, \theta, \alpha, \tilde{G}) = C_{\textup{L},\tilde{G},\text{Stokes}} + C_{\textup{L},\tilde{G}}\, ,
\end{align}
where the term $C_{\textup{L},\tilde{G},\text{Stokes}}$ is fitted from the Eq.~\eqref{eq:Fan1995}, which expresses the shear-induced lift force experienced by a particle in the viscous regime~\citep{Happel1981}, and the term $C_{\textup{L},\tilde{G}}$ accounts for the finite particle Reynolds number effects.
The term $C_{\textup{L},\tilde{G},\text{Stokes}}$ is given by
\begin{align}\label{eq:CL-shear-fit-Stokes}
C_{\textup{L},\tilde{G},\text{Stokes}} = \left(\pi^2 9.02  \tilde{L}^{0.831}\right)\left[1 + \left(1- \left(1.83 \times 10^{-3} \alpha\right)\tilde{G}\right)\mathcal{S}(\text{\Rep,$\tilde{G}$})\right]\, ,
\end{align}
with $\tilde{L}$ a dimensionless coefficient obtained from Eq.~\eqref{eq:Fan1995}
\begin{align}\label{eq:CL-shear-fit-Stokes-LiftCoefficient}
\tilde{L} = \left(\cfrac{a}{D_\text{eq}}\right)^2 \cfrac{\tilde{G}}{\sqrt{\tilde{G}}} \left[\left(\tens{K}\text{ } \tens{L}\text{ } \tens{K}\right)\vect{\tilde{u}}^*\right]\vect{e}_y\, ,
\end{align}
with $\vect{\tilde{u}}^*$ a dimensionless unit column vector set to $\vect{\tilde{u}}^* = [1, 0, 0]^{T}$ to dot the matrix product $\tens{K}\text{ } \tens{L}\text{ } \tens{K}$ in the main flow direction,
and the sigmoid function $\mathcal{S}(\text{\Rep},\tilde{G})$ accounts for the decay of the change in the particle lift force in case of linear shear flow compared to uniform flow from the viscous regime to intermediate particle Reynolds number, as shown in figure~\ref{fig:CLvsReynoldsCorrelation}.
This function is given by
\begin{equation}\label{eq:lift-shear-sigmoid}
\mathcal{S}(\text{\Rep},\tilde{G}) = \cfrac{x}{1+\vert x\vert}\, ,\quad\text{with } x = 0.206 \tilde{G}^{-2.55}\left(-0.426 \text{\Rep}^{1.61}\right)\, .
\end{equation}
The finite Reynolds term in Eq.~\eqref{eq:CL-shear-fit} is given by
\begin{align}\label{eq:CL-shear-fit-Rep}
C_{\textup{L},\tilde{G}} =\left\{
    \begin{array}{ll}
        0 & Re_\textup{p} < 1\, ,\\
        \left(-2.50 \times 10^{-4} \text{\Rep} \tilde{G}^{0.309}\right)\left(\tilde{G}\alpha^{1.50}\right)^{-1}\sin(\varPsi_{\text{L},\tilde{G}})
        & \text{\Rep $\geq$ 1}\, ,
    \end{array}
\right.
\end{align}
where the shift function $\varPsi_{\text{L},\tilde{G}}$ is used to model the variation from the positive increase to the negative increase of the lift force coefficient in case of linear shear flow compared to uniform flow as a function of the aspect ratio and the particle Reynolds number.
It also aims to account for the decrease in the change of the particle lift force coefficient in case of a local shear flow compared to a locally uniform flow for the orientation angles near $\theta = 0^{\text{o}}$ and $\theta = 180^{\text{o}}$, as shown in figure~\ref{fig:CLvsRepvsTheta} for a particle Reynolds number of \Rep $\geq$ 100.
It is given by
\begin{equation}\label{eq:CL-cosine-sine-shear-shift}
\varPsi_{\text{L},\tilde{G}} =\pi \left(\cfrac{\theta}{\pi}\right)^{f_{\text{L,$\tilde{G}$}}}, \,\text{with } f_{\text{L,$\tilde{G}$}} = 1 + 6.96 \times 10^{-2} \ln{(\text{\Rep})} +  7.33 \times 10^{-2} \ln{(\text{\Rep})} \ln{(\alpha)}^{0.211}\, .
\end{equation}

\begin{figure}[htbp!]
\includegraphics[width=0.485\columnwidth]{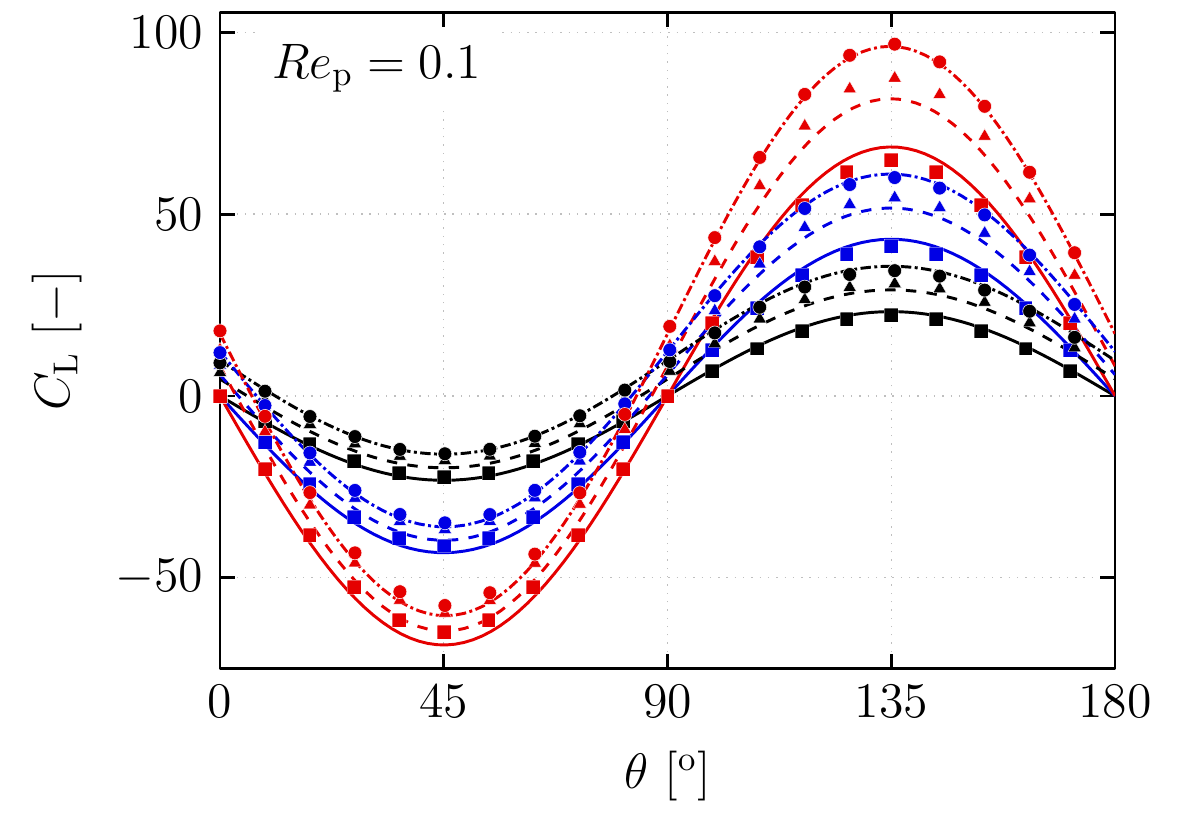}
\includegraphics[width=0.48\columnwidth]{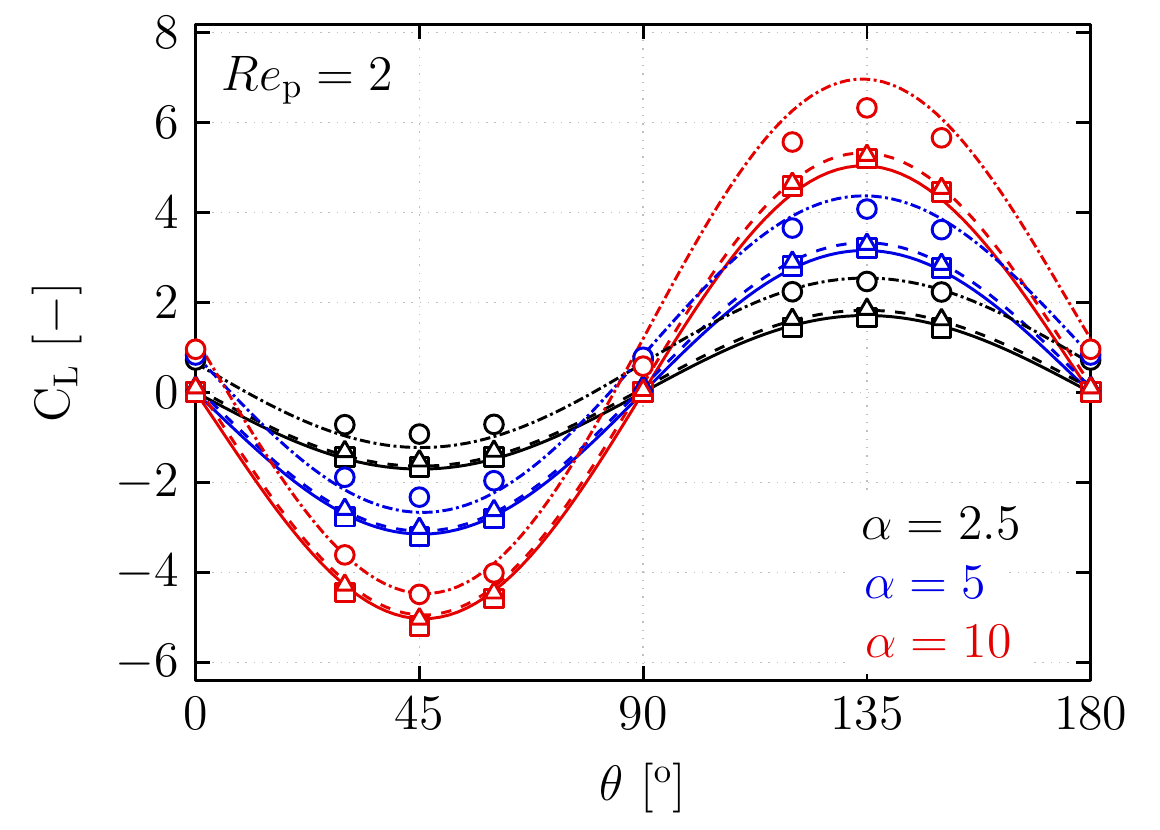}\\

\includegraphics[width=0.475\columnwidth]{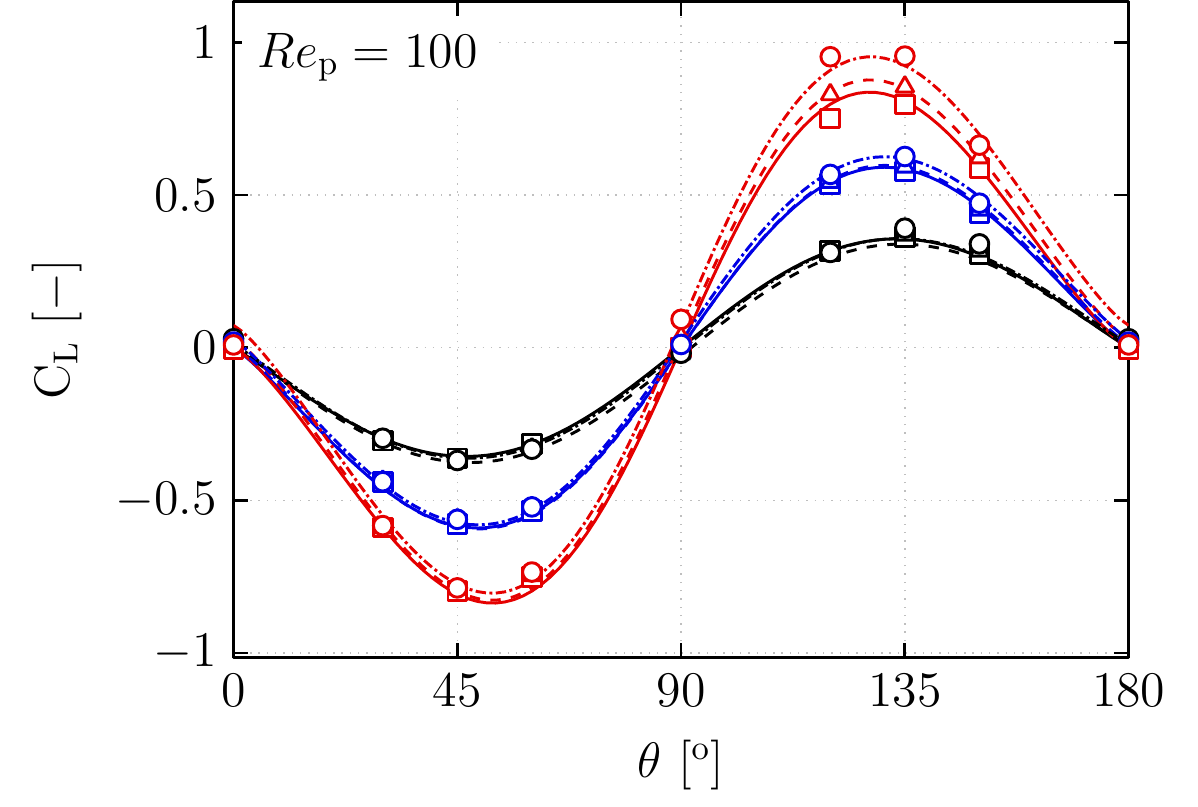}
\includegraphics[width=0.485\columnwidth]{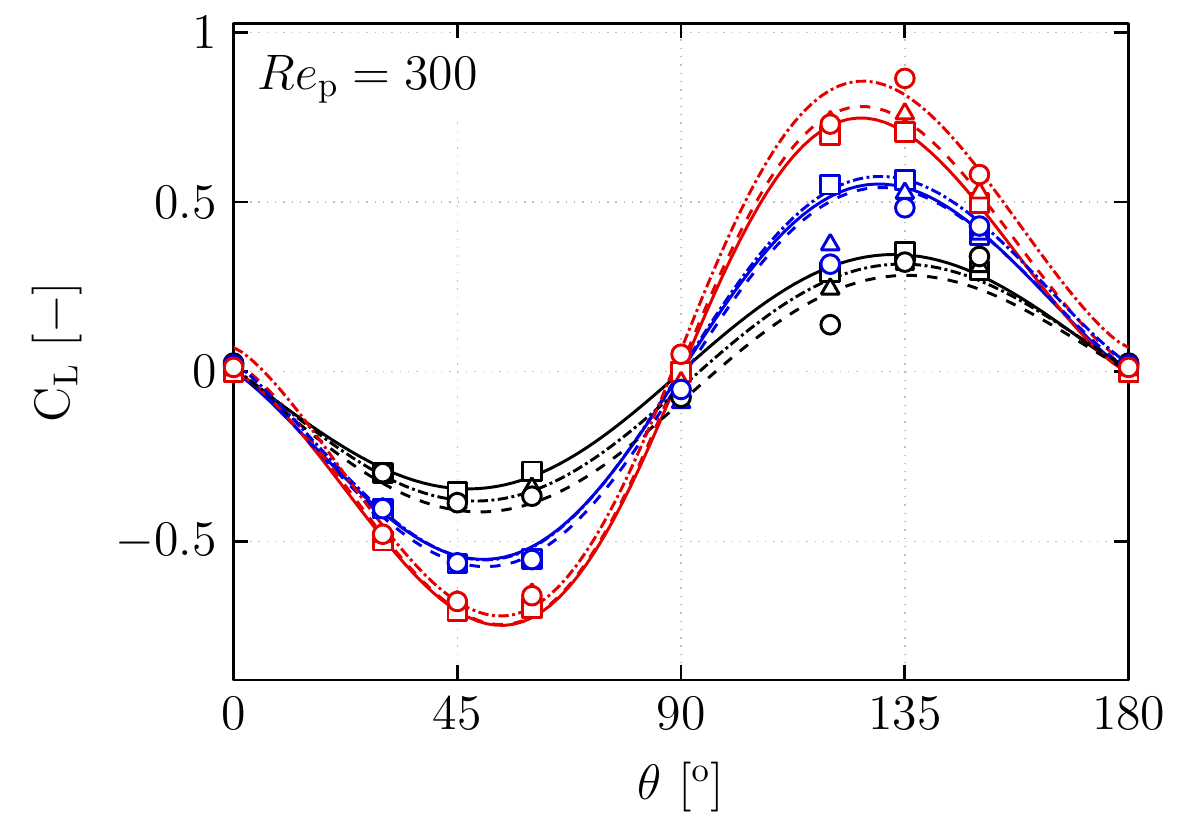}\\
    \caption{The lift force coefficient, $C_\text{L}$, as a function of the orientation angle $\theta$.
    The color indicates the aspect ratio of the particle, $\alpha  =2.5$: black, $\alpha  =5$: {blue}, and $\alpha  =10$: {red}.
    The symbols represent the DNS results for the shear rate $\tilde{G} = 0$: $\square$, $\tilde{G} = 0.1$: $\triangle$, and $\tilde{G} = 0.2$: \protect\tikz{\protect\draw[thick] (0,0) circle (3pt)}.
    The correlation to model the lift coefficient $C_\text{L}$ (Eq.~\eqref{eq:CL-GeneralExpression}) is shown for all the flow regimes varying the line style, $\tilde{G} = 0$: solid line, $\tilde{G} = 0.1$: dashed line, and $\tilde{G} = 0.2$: dash dotted line.
    In the viscous regime the analytical results of~\citet{Harper1968} are shown for the three shear rates, $\tilde{G} = 0$: $\blacksquare$, $\tilde{G} = 0.1$: $\blacktriangle$, and $\tilde{G} = 0.2$: \protect\tikz{\protect\draw[thick, fill] (0,0) circle (3pt)}.}\label{fig:EvolCoeffCL}
\end{figure}

The correlation to predict the lift coefficient, the analytical solutions in the viscous regime and the DNS results are shown in figure~\ref{fig:EvolCoeffCL} for particle Reynolds numbers of \Rep = 0.1, 2, 100 and 300.
In general, the correlation to predict the lift force of the particle in an uniform flow is in very good agreement for all the cases.
The correlation to account for the change in the particle lift force in case of a local shear flow compared to a locally uniform flow accurately recovers the analytical results~\citep{Harper1968}, and the DNS results until a particle Reynolds number of \Rep = 100.
For the particle Reynolds number of \Rep = 300, the correlation underestimates the negative increase of the particle lift force in case of local shear flow for the particles of aspect ratios $\alpha=2.5$ and $5$.
In general, the correlation for the lift coefficient provides a good agreement with the DNS results; the maximum, mean and median relative differences between the model fit and the DNS are of~$52.37$\%, $5.35$\% and $1.95$\%, respectively.
The large relative maximum error is caused by the decreasing differences in values between the lift coefficients with local shear flow compared to local uniform flow for the orientation angles $\theta = 0^{\text{o}}$, $90^{\text{o}}$, and $180^{\text{o}}$.
{The error is maximum for the configuration at shear flow $\tilde{G} = 0.2$, particle Reynolds number \Rep = 300, aspect ratio $\alpha = 5$, and orientation angle $\theta = 90^{o}$.}
At these orientation angles, the change in the lift coefficient of the particle in case of linear shear flow compared to uniform flow is almost 0, hence the large relative errors are caused by the very small lift coefficients, and the absolute error of the proposed correlation is very small.

\subsection{Torque coefficient\label{subsec:torqueforceresults}}
\subsubsection{Results and discussion}

A particle experiences a torque when the center of pressure of this particle no longer coincides with its center of mass.
For non-spherical particles subject to a locally uniform flow, a torque is generated when the particle is not aligned with or perpendicular to the main local flow direction.
For an unbounded linear shear flow, the velocity field and the pressure distribution on the particle surface is different compared to the case with local uniform flow, see figure~\ref{fig:StreamlinesParticles}.
Hence, the profile of the torque coefficient as a function of the orientation differs in case of local shear flow compared to a local uniform flow.
This results in a non-zero torque coefficient at the orientation angles $\theta = 0^{\text{o}}$, $90^{\text{o}}$ and $180^{\text{o}}$.

\begin{figure}[htbp!]
\centering
    \includegraphics[width=0.6\columnwidth, trim={0 0 0 0}, clip]{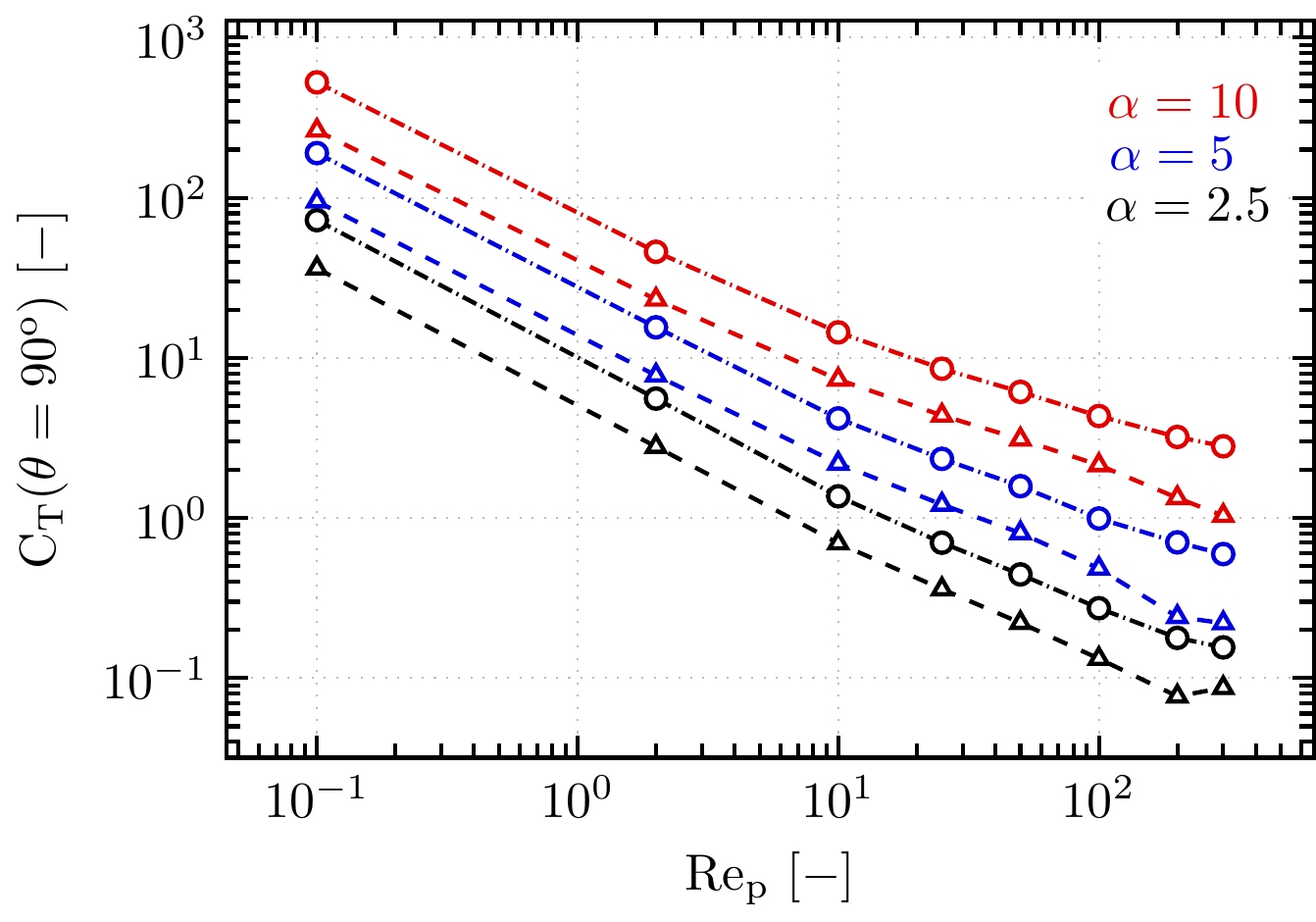}
    \caption{The torque coefficient, \ct, at the orientation angle of $\theta = 90^\text{o}$ as a function of the particle Reynolds number.
    The line color indicates the aspect ratio, $\alpha= 2.5$: black, $\alpha= 5$: {blue}, and $\alpha= 10$: {red}.
    The line style indicates the shear rate $\tilde{G} = 0.1$: dashed line, $\tilde{G} = 0.2$: dash dotted line.}
    \label{fig:CTvsReynoldsCorrelation}
\end{figure}

The change in the particle torque in case of a local shear flow compared to a locally uniform flow can be illustrated by showing the evolution of the absolute value of the torque coefficient at the orientation angle of $\theta = 90^\text{o}$, as a function of the particle Reynolds number.
This specific orientation angle is chosen as it provides the maximum torque coefficient in the viscous regime for this shear flow configuration~\citep{Jeffery1922}, and has the advantage of being equal to 0 in case of uniform flow.
In figure~\ref{fig:CTvsReynoldsCorrelation}, the torque coefficients are shown for all the considered particle aspect ratios at flow configurations with shear rates $\tilde{G} = 0.1$ and $0.2$.

At a low particle Reynolds number, the torque of the particle is significantly modified in case of a local shear flow compared to a local uniform flow, for all the linear shear rates and particles, and the more the particle is elongated the larger is the modification from the local uniform flow results.
At higher particle Reynolds number, the differences between the local uniform flow and the local shear flow configurations decrease; and the smaller is the aspect ratio of the particle, the smaller is the difference.
This observation agrees well with the results of~\citet{Holzer2009} for a prolate of aspect ratio $\alpha = 1.5$.

\begin{figure}[htbp!]

\centering
    \includegraphics[width=0.99\columnwidth, trim={0 0 0 0}, clip]{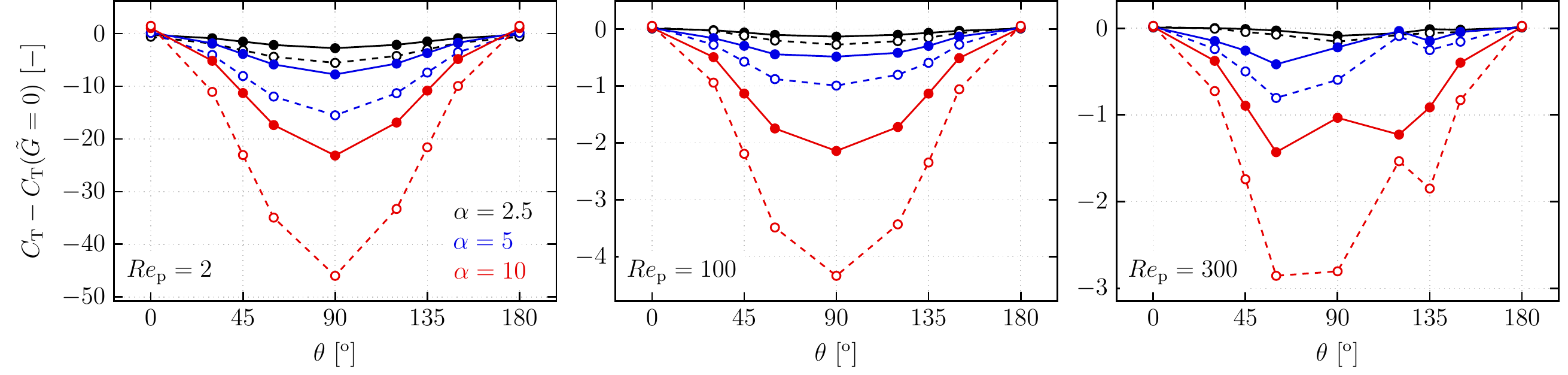}
    \caption{The torque coefficient minus the uniform torque coefficient, $C_\text{T} - C_\text{T}(\tilde{G}=0)$, as a function of the orientation angle $\theta$.
    The color indicates the aspect ratio of the particle, $\alpha  =2.5$: black, $\alpha  =5$: {blue}, and $\alpha  =10$: {red}.
    The line style indicates the shear rate, $\tilde{G} = 0.1$: solid line, $\tilde{G} = 0.2$: dashed line.}
    \label{fig:CTScaled-Theta}
\end{figure}

A consequence of the non-zero torque coefficient for the orientation angles $\theta = 0^{\text{o}}$, $90^{\text{o}}$ and $180^{\text{o}}$ in case of linear shear flow is that the cosine-sine profile of the torque, often used to model the torque coefficient of axi-symmetric particles subject to uniform flow~\citep{Zastawny2012c,Frohlich2020,Sanjeevi2022}, is no longer an adequate model to predict the torque coefficient in case of linear shear flow.
This change is studied by showing the difference between the torque coefficients at shear rates $\tilde{G} = 0$ and $\tilde{G} > 0$ as a function of the orientation angle, $C_\text{T}(\theta,\text{\Rep}) - C_\text{T}(\theta,\text{\Rep},\tilde{G}=0)$, in figure~\ref{fig:CTScaled-Theta} for all studied orientation angles and aspect ratios.
The results are shown at particle Reynolds numbers of~\Rep = 2, 100 and 300.
For the particle Reynolds numbers \Rep = 2 and 100, the change in the torque of the particle in case of local shear flow compared to uniform flow follows a sinusoidal profile for all the particles studied.
At \Rep = 300, the evolution of this coefficient is more complex and it is difficult to capture its behavior.\\
\\
{The specific behavior of the change in the torque coefficient in case of linear shear flow at \Rep = 300 is the result of an interplay between the shift of the maximum torque coefficient in case of uniform flow towards an orientation angle of~$\theta = 60^{o}$, or by symmetry $120^{o}$~\citep{Frohlich2020,Sanjeevi2022} which reduces the change in the torque coefficient at this specific angle.
However, in case of a linear shear flow, the orientation angle of the particle at which the maximum torque coefficient occurs drastically changes, even for low particle Reynolds number, as shown by~\citet{Jeffery1922}. Therefore, there are two separate mechanisms which modify at which orientation angle of the particle the maximum absolute value of the torque coefficient occurs: the Reynolds number as well as the strength of the linear shear. The interplay of these two effects results in the non-linear behavior observed at large particle Reynolds number.}

\subsubsection{Correlation for the torque coefficient}

Similarly to the correlations to predict the drag and the lift coefficients, the correlation to model the torque coefficient is divided into a term accounting for the uniform flow, and into a term describing the change in the torque in case of local shear flow compared to local uniform flow.
The general correlation is given by
\begin{equation}\label{eq:CT-GeneralExpression}
C_\textup{T}(Re_\textup{p}, \theta, \alpha, \tilde{G}) = C_\text{T}(Re_\textup{p}, \theta, \alpha) + C_{\textup{T},\tilde{G}}(Re_\textup{p}, \theta, \alpha, \tilde{G})\, ,
\end{equation}
where $C_{\textup{T}}(Re_\textup{p}, \theta, \alpha)$ is the torque coefficient for the local uniform flow, and $C_{\textup{T},\tilde{G}}(Re_\textup{p}, \theta, \alpha, \tilde{G})$ is the torque coefficient to account for the change in the torque coefficient caused by the linear shear flow with respect to the uniform flow.
\reviewerII{The correlations for the uniform torque coefficient is valid in the range from $\alpha = 2.5$ to $\alpha = 10$, and particle Reynolds number ranging from \Rep = 1 to \Rep = 300.
The correlation to predict the uniform torque coefficient is not valid for lower particle Reynolds number. 
Since the torque is known to vanish for lower values and is often consider equal to 0~\citep{Sanjeevi2018}, and is modeled accordingly to these findings in this work.
The correlation to predict the torque coefficient in case of linear shear flow, however, is valid in the range from $\alpha = 2.5$ to $\alpha = 10$, and particle Reynolds number ranging from \Rep = 0.1 to \Rep = 300.}
The derivation of the correlation to fit for the torque coefficient of the particle in a uniform flow follows the work of~\citet{Zastawny2012c} and~\citet{Frohlich2020}, and is given by
\begin{equation}\label{eq:CT-uniform-fit}
C_{\textup{T}}(Re_\textup{p}, \theta, \alpha) =
\left[\cfrac{c_{\textup{t},1} \ln{(\alpha)}^{c_{\textup{t},2}}}{c_{\textup{t},3}\text{\Rep}^{c_{\textup{t},4}}} +
\cfrac{c_{\textup{t},5} \ln{(\alpha)}^{c_{\textup{t},6}}}{c_{\textup{t},7}\text{\Rep}^{c_{\textup{t},8} + c_{\textup{t},9} \ln(\alpha)^{c_{\textup{t},10}}}}\right]\cos(\theta)\sin(\theta)^{f_{\text{T},\text{shift}}}\,
\end{equation}
where $c_{\text{t},i}$ are the fit parameters, and $f_{\text{T},\text{shift}}$ is a function used to shift the exponent of the sinusoidal term, which depends on both the particle Reynolds number and the aspect ratio of the particle.
It is given by
\begin{equation}
f_{\text{T},\text{shift}} = 1 + 8.216 \times 10^{-5} \alpha^{1.388}\ln{(\text{\Rep})}^{3.513}\, .
\end{equation}
The fit parameters, $c_{\textup{t},\text{i}}$, used in Eq.~\eqref{eq:CT-uniform-fit} are listed in table~\ref{table:torquecoefficients-uniform}.
The maximum, mean and median relative differences between the model fit and the DNS results are of~$8.31$\%, $2.15$\% and $0.77$\%, respectively.
\begin{table}
\centering
\begin{tabular}{c c c c c c c c c c}
$c_{\textup{t},1}$ & $c_{\textup{t},2}$ & $c_{\textup{t},3}$ & $c_{\textup{t},4}$ & $c_{\textup{t},5}$ & $c_{\textup{t},6}$ & $c_{\textup{t},7}$ & $c_{\textup{t},8}$& $c_{\textup{t},9}$& $c_{\textup{t},10}$\\
\hline
\hline
-0.645 & 2.34 & -12.7 & -0.359 & -5.66 & 1.88 & -1.64 & -3.72 $\times 10^{2}$ & 3.73 $\times 10^{2}$ & $8.48 \times 10^{-4}$\\
\end{tabular}
     \caption{List of the fit parameters in Eq.~\eqref{eq:CT-uniform-fit} to model the uniform torque coefficient correlation ($C_{\textup{T}}(Re_\textup{p}, \theta, \alpha)$).}\label{table:torquecoefficients-uniform}
\end{table}

The correlation to model the change in the torque of the particle in case of linear shear flow compared to uniform flow, Eq.~\eqref{eq:CT-GeneralExpression}, is fitted from the sinusoidal expression
\begin{equation}\label{eq:CT-shear-fit}
C_{\textup{T},\tilde{G}}(Re_\textup{p}, \theta, \alpha, \tilde{G}) = C_{\textup{T},\tilde{G},\parallel} + \left[C_{\textup{T},\tilde{G},\perp} - C_{\textup{T},\tilde{G},\parallel}\right] \sin(\theta)^{2}\, ,
\end{equation}
where the coefficients $C_{\textup{T},\tilde{G},\parallel}$ and $C_{\textup{T},\tilde{G},\perp}$ are described by
\begin{align}\label{eq:CT-shear-fit-Para-Perp}
C_{\textup{T},\tilde{G},\parallel} & = \reviewerII{10} \tilde{T}_{\tilde{G},\parallel}\left(\cfrac{c_{\textup{t},\tilde{G},1}}{\text{\Rep}^{c_{\textup{t},\tilde{G},2}}}\right)
+ \cfrac{c_{\textup{t},\tilde{G},3} \alpha^{c_{\textup{t},\tilde{G},4}}}{c_{\textup{t},\tilde{G},5}\text{\Rep}^{\gamma_{\parallel}}}\, ,  \\
C_{\textup{T},\tilde{G},\perp} & = 10  \tilde{T}_{\tilde{G},\perp}\left[\cfrac{c_{\textup{t},\tilde{G},1}}{\text{\Rep}^{c_{\textup{t},\tilde{G},2}}} +
\cfrac{c_{\textup{t},\tilde{G},3} \alpha^{c_{\textup{t},\tilde{G},4}}}{c_{\textup{t},\tilde{G},5}\text{\Rep}^{\gamma_{\perp}}}\right]\, ,
\end{align}
{with
\begin{align}
\gamma_{\parallel} = & \left(c_{\textup{t},\tilde{G},6} + c_{\textup{t},\tilde{G},7} \ln(\alpha)^{c_{\textup{t},\tilde{G},8}}\right)\, ,\\
\gamma_{\perp} = &  \left(c_{\textup{t},\tilde{G},6} + c_{\textup{t},\tilde{G},7} \ln(\alpha)^{c_{\textup{t},\tilde{G},8}}\right)\, ,
\end{align}}
and where $\tilde{T}_{\tilde{G},\parallel}$ and $\tilde{T}_{\tilde{G},\perp}$ are the dimensionless torque coefficients in the viscous regime, fitted from the derivations of~\citet{Jeffery1922}, and $c_{\textup{t},\tilde{G},\text{i}}$ are the fit parameters.
In the present configuration, the dimensionless torque coefficients $\tilde{T}_{\tilde{G},\parallel}$ and $\tilde{T}_{\tilde{G},\perp}$ are directly obtained from the knowledge of the flow configuration and the analytical solution derived by~\citet{Jeffery1922}.
They are given by
\begin{align}\label{eq:CT-shear-fit-TorqueCoeff}
\tilde{T}_{\tilde{G},\parallel} & = \frac{16 \pi  \left(\nicefrac{b}{D_\text{eq}}\right)^3 \alpha}{3\left(\lambda_2+\alpha^2 \lambda_1\right)}\left[\left(\alpha^2-1\right) \tilde{S}_\textup{yx, $\parallel$}^\textup{b}+\left(\alpha^2+1\right)\left(\tilde{\Omega}_\textup{yx, $\parallel$}^\textup{b}\right)\right]\, ,\\
\tilde{T}_{\tilde{G},\perp} & = \frac{16 \pi  \left(\nicefrac{b}{D_\text{eq}}\right)^3 \alpha}{3\left(\lambda_2+\alpha^2 \lambda_1\right)}\left[\left(\alpha^2-1\right) \tilde{S}_\textup{yx, $\perp$}^\textup{b}+\left(\alpha^2+1\right)\left(\tilde{\Omega}_\textup{yx, $\perp$}^\textup{b}\right)\right]\, ,
\end{align}
where $b$ is the semi-minor axis of the particle, $D_{\text{eq}}$ is the volume-based-equivalent diameter of the particle, $\alpha$ is the aspect ratio of the particle, $\lambda_1$ and $\lambda_2$ are geometric coefficients derived in~\citet{Gallily1979} and given in Eq.~\eqref{eq:Gallily-lambda}.
The fluid strain and rotation tensors, $\tilde{S}_\textup{yx}^{\text{b}}$ and $\tilde{\Omega}_\textup{yx}^{\text{b}}$ respectively, are given in the body space reference frame for a particle aligned parallel to the main local flow direction ($\theta = 0^{\text{o}}$), as indicated by the $\parallel$ subscript, and perpendicular to the main local flow direction ($\theta = 90^{\text{o}}$), as indicated by the $\perp$ subscript.
{In Eq.~\ref{eq:CT-shear-fit-TorqueCoeff}, the fluid strain rate and rotation tensors are obtained from the gradient of the fluid velocity in the body space reference frame at the center of the particle, using Eq.~\refeq{eq:Fluid strain Rate and Rotation tensor}.}
Here, the gradient of the fluid velocity is given in a dimensionless form in the Cartesian world space using the dimensionless shear rate of the flow
\reviewerII{
\begin{equation}
\nabla \mathbf{\tilde{u}} =
\begin{bmatrix}
0 & \tilde{G} & 0 \\
0& 0 & 0\\
0& 0 & 0\\
\end{bmatrix}
\, ,
\end{equation}
where the dimensionless fluid velocity gradient matrix contains only one component for the present shear flow configuration because the direction of the local shear flow is in the same plane as the particle orientation variations.
For shear flow non-aligned with one of the Cartesian component the magnitude of the dimensionless shear rate can be projected among each Cartesian components.}
The dimensionless gradient of velocity matrix is then projected to the body space reference frame {using unit-quaternions}~\citep{Zhao2013a} to determine the fluid strain and rotation tensors, $\tilde{S}_\textup{yx}^{\text{b}}$ and $\tilde{\Omega}_\textup{yx}^{\text{b}}$ for the orientation angles $\theta = 0^\text{o}$ and $90^\text{o}$.
The fit parameters derived from the construction of the correlation to account for the change in the torque coefficient caused by the local shear flow with respect to the uniform flow are listed in table~\ref{table:torquecoefficient-shearflow}.
The relatively small coefficient modeling the behavior of the dimensionless torque coefficient at $\theta = 0^{\text{o}}$ is caused by the fast decrease of the change in the torque coefficient caused by the local shear flow with respect to the uniform flow at finite particle Reynolds number for this specific angle.

\begin{table}
\centering\reviewerII{
\begin{tabular}{c | c c c c c c c c}
& $c_\textup{t,1}$ & $c_\textup{t,2}$ & $c_\textup{t,3}$ & $c_\textup{t,4}$ & $c_\textup{t,5}$ & $c_\textup{t,6}$ & $c_\textup{t,7}$ & $c_\textup{t,8}$\\
\hline
\hline
$C_{\textup{T},\tilde{G},\parallel}$ & 0.334 & 1.43 & 0.137 & 1.45  & 1.22 & 1.74 & -0.362 & 1.52\\
$C_{\textup{T},\tilde{G},\perp}$ & 1.11 & 1.05 & 0.387 & -0.498 & 0.746 & 1.75 & -1.20 & 0.298 \\
\end{tabular}
     \caption{List of the fit parameters in Eq.~\eqref{eq:CT-shear-fit-Para-Perp}, used in the correlation to predict the change in the torque coefficient caused by the linear shear flow with respect to the uniform flow ($C_{\textup{T},\tilde{G}}(Re_\textup{p}, \theta, \alpha, \tilde{G})$).}\label{table:torquecoefficient-shearflow}
     }
\end{table}

\begin{figure}[htbp!]
\includegraphics[width=0.485\columnwidth]{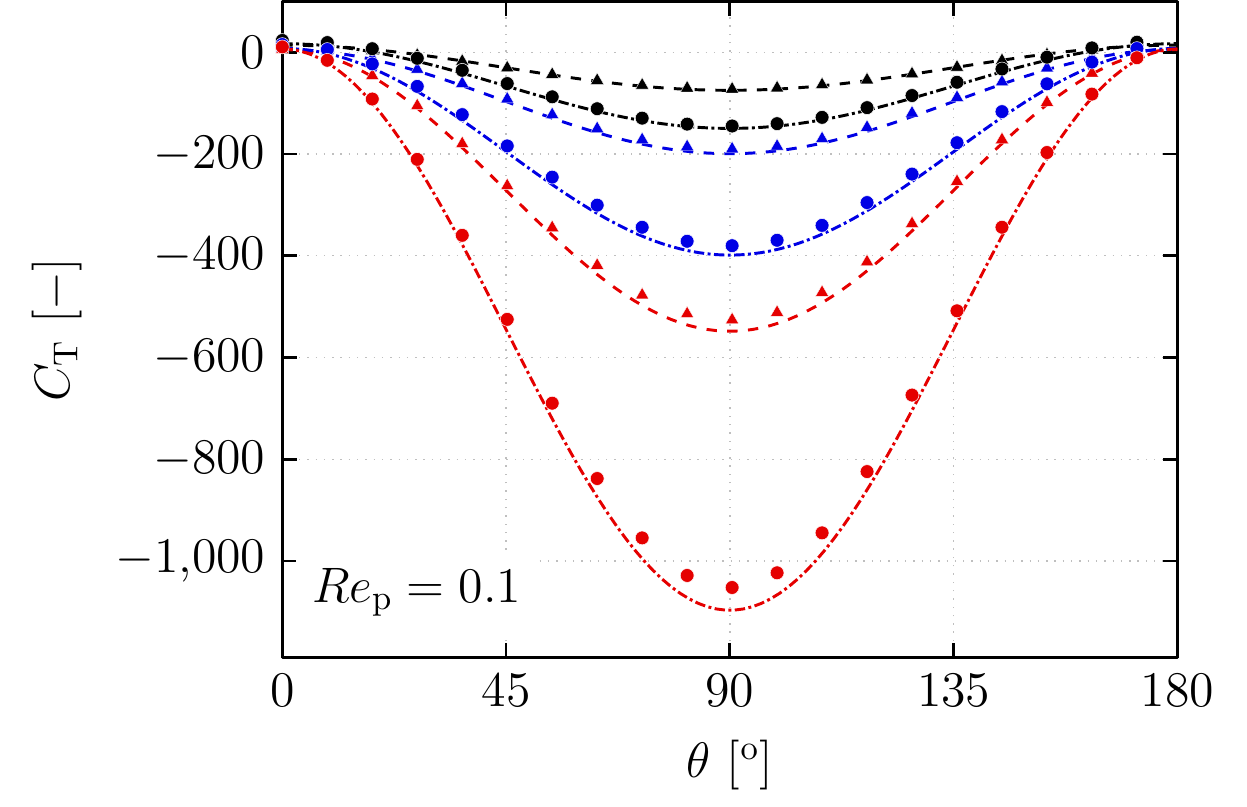}
\includegraphics[width=0.48\columnwidth]{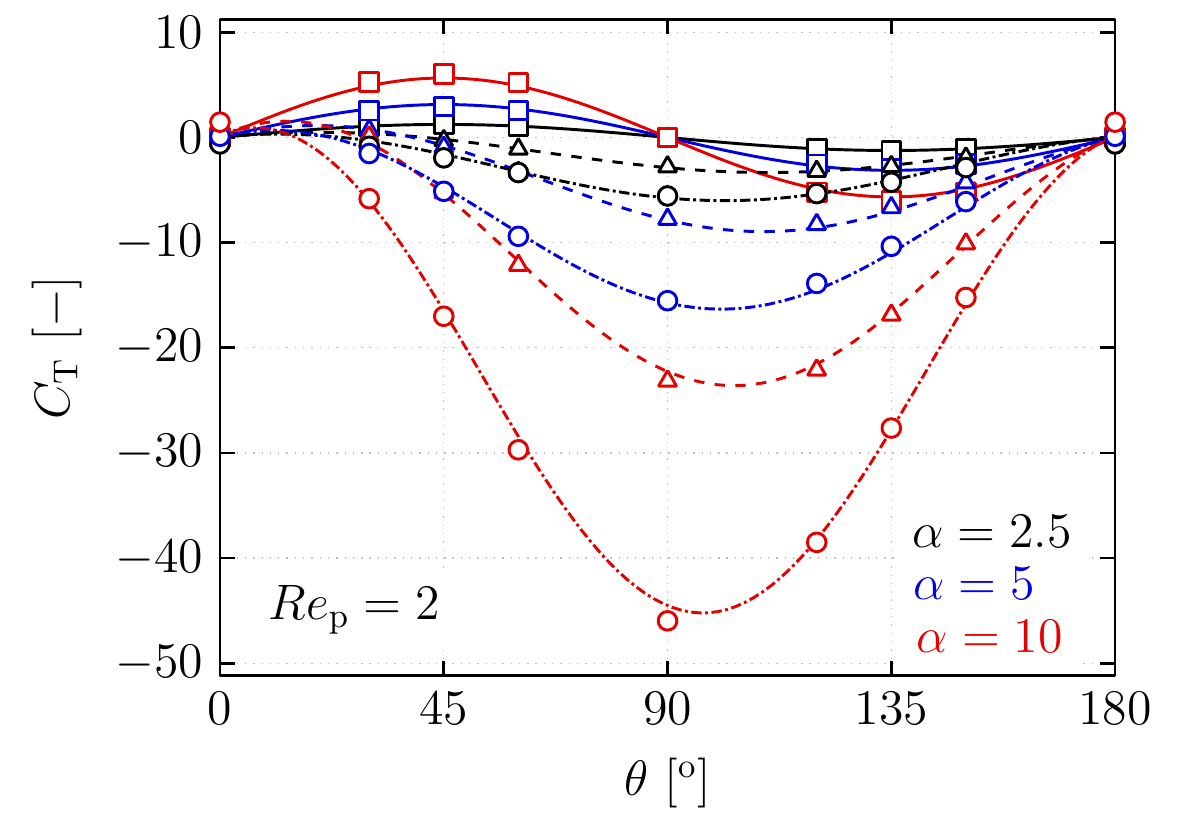}\\

\includegraphics[width=0.475\columnwidth]{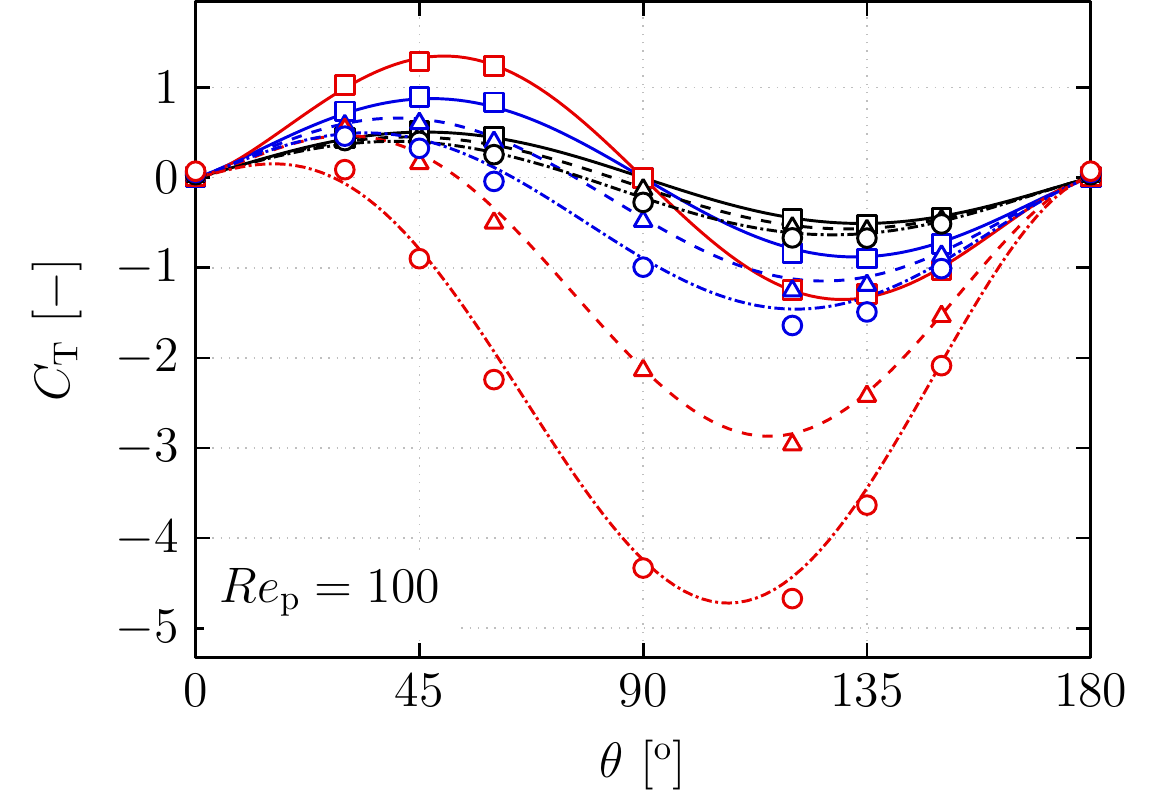}
\includegraphics[width=0.485\columnwidth]{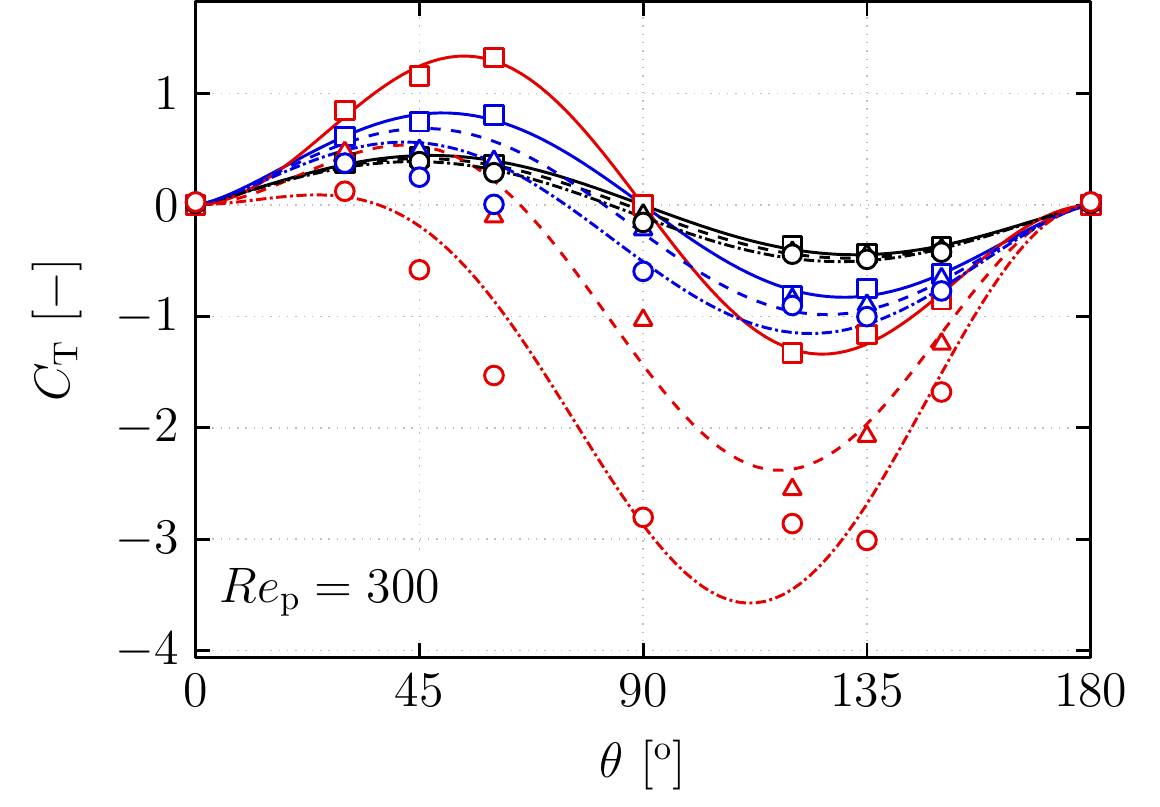}\\
    \caption{The torque coefficient, $C_\text{T}$, as a function of the orientation angle $\theta$.
    The color indicates the aspect ratio of the particle, $\alpha  =2.5$: black, $\alpha  =5$: {blue}, and $\alpha  =10$: {red}.
    The symbols represent the DNS results for the shear rate $\tilde{G} = 0$: $\square$, $\tilde{G} = 0.1$: $\triangle$, and $\tilde{G} = 0.2$: \protect\tikz{\protect\draw[thick] (0,0) circle (3pt)}.
    The correlation to model the torque coefficient $C_\text{T}$ (Eq.~\eqref{eq:CT-GeneralExpression}) is shown for all the flow regimes varying the line style, $\tilde{G} = 0$: solid line, $\tilde{G} = 0.1$: dashed line, and $\tilde{G} = 0.2$: dash dotted line.
    In the viscous regime the analytical results of~\citet{Jeffery1922} are shown for the shear rates, $\tilde{G} = 0.1$: $\blacktriangle$, and $\tilde{G} = 0.2$: \protect\tikz{\protect\draw[thick, fill] (0,0) circle (3pt)}.}\label{fig:EvolCoeffCT}
\end{figure}

The analytical and DNS results, along with the correlation for the torque coefficient, are shown in figure~\ref{fig:EvolCoeffCT}.
For a finite Reynolds number, the correlation to predict the torque coefficient of the particle in a uniform flow accurately recovers the DNS results for all the aspect ratio and studied particle Reynolds number.
The correlation to predict the change in the torque coefficient of the particle in case of local shear flow compared to uniform flow also accurately predicts the torque coefficient of the particle in the viscous regime and up to a particle Reynolds number of \Rep = 100.
At a higher particle Reynolds number, the accuracy of the correlation slightly deteriorates for the more elongated particles.
\reviewerII{The maximum, mean and median relative differences between the model fit and the DNS are of~$70.80$\%, $6.78$\% and $1.08$\%, respectively.}
\reviewerII{The large maximum relative error is caused by the decreasing differences in values between the torque coefficients with local shear flow and local uniform flow for an orientation angle of $\theta = 0^{\text{o}}$ and $180^{\text{o}}$, see for example the results at particle Reynolds number of \Rep = 2 for the particle with an aspect ratio of $\alpha = 10$, or for the specific flow configuration of shear flow $\tilde{G}$ = 0.2, particle Reynolds number \Rep = 300, aspect ratio $\alpha = 5$, and orientation angle $\theta = 120^{o}$.
Also, the notable relative error in the mean is caused by the decreasing influence of the change in the torque coefficient in case of linear shear flow compared to uniform flow for the particle with an aspect ratio of $\alpha=2.5$ at a high particle Reynolds number.
Nevertheless, the absolute error in the prediction of $C_\textup{T}(Re_\textup{p}, \theta, \alpha, \tilde{G})$, see Eq.~\refeq{eq:CT-GeneralExpression}, is small since $C_\text{T}(Re_\textup{p}, \theta, \alpha) >> C_{\textup{T},\tilde{G}}(Re_\textup{p}, \theta, \alpha, \tilde{G})$.}

\section{Conclusions\label{sec:conclusions}}

The hydrodynamic forces experienced by axi-symmetric rod-like particles in a fluid flow vary with respect to the shape of the particle, the relative velocity between the particle and the flow and properties of the fluid, the angle of attack of the particle with respect to the main local flow direction, and the profile of the local fluid velocity.
The variation of this set of parameters modifies the magnitude of the hydrodynamic forces, as well as the profile of these forces as a function of the orientation of the particle.
Several correlations exist to model the interaction forces between the fluid and the specific particle, and recent works propose models for the drag, lift and torque coefficients, varying the particle Reynolds number, the orientation angle of the particle, and the aspect ratio of the non-spherical particles~\citep{Ouchene2016,Frohlich2020,Sanjeevi2022}.
However, these correlations consider exclusively spheroidal particles, and a locally uniform fluid flow.

In this work, the immersed boundary method is used to perform many direct numerical simulations (DNS) of the flow past a rod-like particle, varying the orientation angle, $\theta$, between the main axis of the particle and the main fluid velocity direction ($0^{\text{o}} \leq \theta \leq 180^{\text{o}}$), the particle Reynolds number, \Rep, \reviewerII{($2 \leq \text{\Rep} \leq 300$)}, and the aspect ratio of the particle, \reviewerII{$\alpha$, ($\alpha = 2.5, 5$ and $10$)}.
The profile of the flow is also modified, an uniform flow and an unbounded linear shear flow are considered.
The data from our DNS combined with the expressions existing in the viscous regime~\citep{Jeffery1922,Harper1968,Happel1981}, \reviewerII{which are added to the set of data used to fit the correlations,} are used to derive the correlations to predict the drag, lift and torque coefficient for a rod-like particle.
The correlations are divided into a term accounting for the uniform flow, and a term accounting for the change in the particle forces caused by the linear shear flow.

The correlations to predict the drag, lift and torque coefficients of the rod-like particle proposed in this work, in a uniform flow are in very good agreement with both the analytical solutions and the DNS results.
The maximum relative error of the correlations always remains below $10\%$, and the median of the relative error is always close to $1\%$.
This provides accurate new correlations to transport rod-like particles in large scale simulations subject to locally uniform flows {for particle Reynolds number varying from \Rep = 0.1 to 300 and aspect ratio from $\alpha = 2.5$ to $\alpha = 10$ for the lift and drag coefficient, and valid in the range \Rep = 1 to 300 and aspect ratio from $\alpha = 2.5$ to $\alpha = 10$ for the uniform flow torque coefficient.}

The change in the particle forces caused by the locally linear shear flow compared to the uniform flow depends on the particle Reynolds number, the orientation angle of the particle, the aspect ratio of the particle and the shear rate of the flow.

The particle drag force is modified in case of a local shear flow compared to a locally uniform flow only at finite particle Reynolds number~\citep{Harper1968}.
For these flow regimes, the drag force of the particle is increased, and the magnitude of this increase is larger for high particle Reynolds number, more elongated particles, and larger shear rates.
This increase of the drag force also depends on the orientation angle, and a maximum is observed for the orientation angle of $\theta = 90^{\text{o}}$.
For the orientation angles $\theta = 0^{\text{o}}$ and $180^{\text{o}}$, the change in the drag coefficient is almost negligible.
Between these two extremes, the change in the drag coefficient in case of local shear flow is largely dependent on the particle Reynolds number and the aspect ratio of the particle.
For instance, at intermediate particle Reynolds number, the increase of the drag coefficient of the particle in case of a local shear flow compared to the uniform flow is larger for all the particles and dimensionless shear rates in the range of orientation angles $90^{\text{o}} < \theta < 180^{\text{o}}$.
At higher particle Reynolds number, however, this trend is shifted toward the orientation angles in the range of $0^{\text{o}} < \theta < 90^{\text{o}}$ for the particles of aspect ratio $\alpha = 2.5$ and $5$.

In the viscous regime, the lift force of the particle increases for all the aspect ratios and orientation angles in case of a local shear flow compared to a local uniform flow~\citep{Harper1968}.
With the increase of the particle Reynolds number, the change in the lift force of the particle reduces; this decrease occurs more quickly for the less elongated particles.
For instance, at a particle Reynolds number of~\Rep=100, the results obtained with the local shear flow and the local uniform flow almost overlap for the particles of aspect ratios $\alpha = 2.5$ and $5$.
Yet, as the particle Reynolds number increases, the asymmetric flow recirculation in the wake of the particle yields to a negative increase of the lift force of the particles of aspect ratios $\alpha = 2.5$ and $5$ in case of linear shear flow.
This increase is more significant for the orientation angles near $\theta = 120^{\text{o}}$, and for the particle of aspect ratio $\alpha = 2.5$.

Similarly to the lift force in the viscous regime, the linear shear flow modifies the torque on all the particles and for all the orientation angles considered in this work~\citep{Jeffery1922}.
Interestingly, the modification of the torque coefficient as a function of the orientation angle follows a sinusoidal profile, and is always maximum for the orientation angle $\theta = 90^{\text{o}}$.
This sinusoidal profile is observed for all the particles up to a particle Reynolds number of~\Rep=100.
With increasing particle Reynolds number, the change in the torque coefficient in case of a local shear flow compared to a local uniform flow reduces, and is almost zero for the particle of aspect ratio $\alpha = 2.5$ at a high particle Reynolds number.

The correlations to predict the change in the drag and lift forces and torque in case of a linear shear flow compared to the uniform flow requires an accurate modeling of the interplay of the particle Reynolds number, the aspect ratio of the particle, the orientation angles and the shear rate of the flow.
The correlations presented in this work are in a good agreement with the analytical results available in the literature and our DNS results.
\reviewerII{The median of the relative error remains below $2\%$ for the drag, the lift, and the torque coefficients.
The maximum relative error of the coefficients for the drag always remains below 7\%.}
Although the maximum relative error can be high in the prediction of the lift and torque coefficients, \reviewerII{this only occurs when these have a very low value for the lift and torque coefficient as compared to the uniform flow values, thus the absolute error is relatively small.}
\reviewerII{These correlations are valid for particle Reynolds number varying from \Rep = 0.1 to 300, aspect ratio from $\alpha = 2.5$ to $\alpha = 10$, and dimensionless shear rate ranging from $\tilde{G} = 0$ to $\tilde{G} = 0.2$.}

These novel correlations can be used to predict the behavior of rod-like particles of different aspect ratios in large scale particle-laden flow simulations subject to locally uniform or non-uniform flows.

\section*{
Data Availability Statement} The data that support the findings of this study are reproducible
and files to regenerate the data as well as an executable to implement the correlations to predict the drag, lift and torque coefficients are openly available in the repository 
with DOI 10.5281/zenodo.8348930 
on
\href{https://doi.org/10.5281/zenodo.8348930}{https://doi.org/10.5281/zenodo.8348930}

\section*{Acknowledgments}
This research was funded by the Deutsche Forschungsgemeinschaft (DFG, German Research Foundation) - Project-ID 448292913.
\reviewerII{The authors would like to thank the reviewers of this manuscript; with their thorough and valuable feedback we have further improved our manuscript.}

\bibliographystyle{model1-num-names}

\end{document}